\documentclass[11pt, a4paper]{article}
\usepackage{jheppub}

\usepackage{amsmath, amsfonts, amssymb}
\usepackage{hyperref}
\usepackage{tensind}
\usepackage{mathrsfs}

\tensordelimiter{?}

\newcommand {\cC}{{\cal C}}
\newcommand {\cD}{{\cal D}}
\newcommand {\cE}{{\cal E}}
\newcommand {\cF}{{\cal F}}
\newcommand {\cG}{{\cal G}}
\newcommand {\cH}{{\cal H}}

\newcommand {\cJ}{{\cal J}}

\newcommand {\cL}{{\cal L}}
\newcommand {\cM}{{\cal M}}
\newcommand {\cN}{{\cal N}}
\newcommand {\cO}{{\cal O}}

\newcommand {\cQ}{{\cal Q}}

\newcommand {\cS}{{\cal S}}

\newcommand {\cV}{{\cal V}}
\newcommand {\cW}{{\cal W}}

\def\l{\lambda}

\def\q{\theta}

\def\z{\zeta}

\def\L{\Lambda}

\newcommand{\rd}{{\rm d}}

\newcommand{\dalpha}{{\dot{\alpha}}}
\newcommand{\dbeta}{{\dot{\beta}}}
\newcommand{\dgamma}{{\dot{\gamma}}}
\newcommand{\ddelta}{{\dot{\delta}}}
\newcommand{\dmu}{{\dot{\mu}}}

\newcommand{\dphi}{{\dot{\phi}}}

\newcommand{\1}{{\underline{1}}}
\newcommand{\2}{{\underline{2}}}

\newcommand{\eps}{\epsilon}

\newcommand{\pa}{\partial}
\newcommand{\ul}{\underline}

\newcommand{\bsigma}{{\bar \sigma}}

\newcommand{\bphi}{{\bar\phi}}
\newcommand{\bpsi}{{\bar\psi}}

\newcommand{\abs}[1]{\left| #1 \right|}

\DeclareMathOperator{\sdet}{sdet}

\DeclareMathOperator{\Tr}{Tr}

\newcommand{\HC}{{\mathrm{h.c.}}}

\newcommand{\eol}{\notag \\}

\newcommand{\sym}[1]{\stackrel{\scriptstyle#1}{\mbox{\tiny $\smile$}}}

\newcommand{\bbA}{{\mathbb A}}
\newcommand{\bbD}{{\mathbb D}}
\newcommand{\bnabla}{{\bar \nabla}}
\newcommand{\trD}{{\rm \scriptscriptstyle D}}

\numberwithin{equation}{section}

\title{A new approach to curved projective superspace}
\author{Daniel Butter}
\affiliation{Nikhef Theory Group, \\
Science Park 105, 1098 XG Amsterdam, The Netherlands}
\emailAdd{dbutter@nikhef.nl}

\preprint{Nikhef-2014-014}

\arxivnumber{1406.6235}

\abstract{We present a new formulation of curved projective
superspace. The $4D$ $\cN=2$ supermanifold $\cM^{4\vert 8}$
(four bosonic and eight Grassmann coordinates)
is extended by an auxiliary $\rm SU(2)$ manifold, which involves
introducing a vielbein and
related connections on the full $\cM^{7\vert 8} = \cM^{4\vert 8} \times \rm SU(2)$.
Constraints are chosen so that it is always possible to return to the
central gauge where the auxiliary $\rm SU(2)$ manifold largely decouples
from the curved manifold $\cM^{4\vert 8}$ describing
$4D$ $\cN=2$ conformal supergravity.
We introduce the relevant projective superspace action principle in the analytic subspace
of $\cM^{7\vert 8}$ and construct its component reduction in terms of a
five-form $\cJ$ living on $\cM^4 \times \cC$, with $\cC$ a contour in
$\rm SU(2)$.
This approach is inspired by and generalizes the original approach
taken in arXiv:0805.4683 and related works, which can be identified
with a complexified version of the central gauge of the formulation
presented here.
}

\begin{document}
\maketitle

\section{Introduction}
It is well-known that theories of eight supercharges in various dimensions
possess natural on-shell representations (such as the hypermultiplet) that
do not admit off-shell representations with a finite number of auxiliary fields --
at least, not without a central charge.
In fact, a no-go theorem guarantees that the most general charged hypermultiplet
cannot be lifted to a finite off-shell representation (see e.g. 
\cite{GIOS} for a clear discussion with references).
Both harmonic and projective superspace solve this problem in the same
way: the hypermultiplet is lifted to an off-shell multiplet by introducing
an \emph{infinite} number of auxiliary fields in a controlled way.
For harmonic superspace \cite{GIKOS, GIOS},
these auxiliary fields correspond to Fourier modes
on an auxiliary $S^2$ manifold, and the hypermultiplet is associated with
a globally defined function on $S^2$.
For projective superspace \cite{KLR, LR88, LR90},
the auxiliary fields appear as components of a
Taylor (or Laurent) expansion in a coordinate $\z$ parametrizing
the space $\mathbb CP^1$. (For recent reviews, see
\cite{LR:Prop} and \cite{Kuzenko:Lectures}.)
As a result, both superspaces actually allow the direct construction
of the most general off-shell actions involving hypermultiplets.
Of equal importance is the way in which both superspaces allow
superfield gauge prepotentials for Yang-Mills theories,\footnote{The
early work in harmonic superspace \cite{GIOS:Graphs} (see also the
monograph \cite{GIOS} for references) stimulated many manifestly
supersymmetric calculations in $\cN=2$ super Yang-Mills theories.
Projective supergraphs and their applications have been discussed in 
\cite{GonzalezRey:FR1, GonzalezRey:FR2, GonzalezRey:FR3,
G-RR, JainSiegel:Hypergraphs1, JainSiegel:Hypergraphs2}.}
which are necessary for performing quantum calculations in a manifestly
supersymmetric way.

These two alternative approaches are not actually too
dissimilar and make use of the superspace introduced by
Rosly \cite{Rosly} (see also \cite{RS}).
(Hartwell and Howe have also discussed the so-called $(\cN,p,q)$
superspaces \cite{HoweHartwell:Survey, HoweHartwell:Npq},
which provide generalizations to higher $\cN$.)
Proposed relations between harmonic and projective superspaces
have also been discussed in \cite{Kuzenko:HarmProj}
and \cite{JainSiegel:HarmProj, Butter:HarmProj}.
However, our concern in this paper will strictly be
with $4D$ $\cN=2$ projective superspace.

If one is interested in supergravity effects, one must naturally determine
how to incorporate a curved supermanifold in a covariant way consistent with the
projective structure.
This was explicitly accomplished first in five dimensional projective superspace
in a series of papers by Kuzenko
and Tartaglino-Mazzucchelli \cite{KT-M:5DSugra1, KT-M:5DSugra2, KT-M:5DSugra3}.
It was subsequently extended to dimensions two through six
by various collaborations involving Kuzenko, Linch, Lindstr\"om,
Ro\v{c}ek, and Tartaglino-Mazzucchelli
\cite{KLRT-M1, KLRT-M2, KT-M:DiffReps, T-M:2DSugra, KLT-M:3DSugra, LT-M:6DSugra}.\footnote{Corresponding
constructions of harmonic superspace in other
dimensions, which preceded the projective constructions, can be found in
\cite{Bellucci:2000yx, Zupnik:3D1, Zupnik:3D2, Sokatchev:Harm6D}.}
(Because we are interested here in $4D$ $\cN=2$ supersymmetry, we will make
frequent reference to the four-dimensional references \cite{KLRT-M1, KLRT-M2, KT-M:DiffReps,Kuzenko:2008qz},
but many important features were already present in 
\cite{KT-M:5DSugra1, KT-M:5DSugra2, KT-M:5DSugra3}.) The formulation
of curved projective superspace presented in these works
we will refer to as \emph{conventional projective superspace}.

A key ingredient of the conventional approach is to understand
the role of superconformal projective
multiplets of weight $n$, which are the natural objects of interest in
projective superspace \cite{Kuzenko:5Dallthat, Kuzenko:SPH}
(see \cite{Kuzenko:Lectures} for a pedagogical discussion).
In curved space, 
such superfields $\cQ^{(n)}(z, v^i)$ are holomorphic
in $v^i$ on some open domain of $\mathbb C^{2*} \equiv \mathbb C^2  \setminus \{0\}$,
homogeneous in $v^i$ of degree $n$,
$\cQ^{(n)}(z, c v^i) = c^n \cQ^{(n)}(z, v)$, and transform under the
superconformal gauge transformations as
\begin{align}\label{eq:deltaQ1}
\delta \cQ^{(n)} = \xi^A \cD_A \cQ^{(n)} + n \L_{\trD} \cQ^{(n)}
	-\l^i{}_j \,v^j \frac{\pa}{\pa v^i} \cQ^{(n)}
\end{align}
where the covariant derivatives $\cD_A$ are built from the supervielbein
and other connections of some curved supermanifold $\cM^{4|8}$, with
$\L_{\trD}$ and $\l^i{}_j$, respectively, the dilatation and
${\rm SU}(2)_R$ gauge parameters.
The ${\rm SU(2)}_R$ transformation can be rewritten
\begin{align}\label{eq:deltaQ2}
\delta_\l \cQ^{(n)} &= -\l^{++} D^{--} \cQ^{(n)} + n \l^0 \cQ^{(n)}~, \qquad
D^{--} = \frac{u^i}{(v,u)} \frac{\pa}{\pa v^i}~, \eol
\l^{++} &= \l^{ij} v_i v_j~, \qquad \l^0 = \l^{ij} \frac{v_i u_j}{(v,u)}~, \qquad
(v,u) = v^j u_j~.
\end{align}
The parameter $u_i$ appearing in \eqref{eq:deltaQ2} is an arbitrary coordinate,
required only to obey $(v,u) \neq 0$ in the region of interest.
Given this prescription, it is consistent to impose the covariant analyticity
constraint\footnote{Such superfields $\cQ^{(n)}$ with these properties can be understood as
generalizations of complex $\cO(n)$ superfields
$\cG^{(n)} = v_{i_1} \cdots v_{i_n} \cG^{i_1 \cdots i_n}(z)$
whose components $\cG^{i_1 \cdots i_n}$ transform as symmetric tensors
of $\rm SU(2)$,
with the constraint
$\cD_\alpha^{\,(j} \cG^{i_1 \cdots i_n)} = \bar\cD_\dalpha^{\,(j} \cG^{i_1 \cdots i_n)} = 0$
\cite{Ketov:HKSigmaModels, Ketov:YMGen}.}
\begin{align}
v_i \,\cD_\alpha{}^i \cQ^{(n)} = v_i \,\bar\cD_\dalpha{}^i \cQ^{(n)} = 0~.
\end{align}
This implies that $\cQ^{(n)}$ depends on only
half the Grassmann coordinates of superspace, in much the same way as
chiral multiplets in $\cN=1$ superspace depend (essentially)
on only $\q$ and not $\bar \q$.

Once the means to minimally couple supergravity is understood, the curved
extension of many flat space results becomes possible.
This is done by generalizing the natural action principle of
flat projective superspace \cite{KLR, Siegel:ChiralActions, Kuzenko:SPH}
\begin{align}
S = -\frac{1}{2\pi} \oint_\cC v_i \,\rd v^i \int \rd^4x\, \rd^4 \q^+ \, \mathscr{L}^{++}~, \qquad
\q^{\alpha+} = \q^{\alpha i} v_i~, \quad \bar \q^{\dalpha +} = \bar\q^{\dalpha i} v_i~,
\end{align}
where $\mathscr{L}^{++}$ is a weight-two projective multiplet Lagrangian
and $\cC$ is some contour in $\mathbb CP^1$. The component form of this
action can be written
\begin{align}
S = -\frac{1}{2\pi} \oint_\cC v_i \,\rd v^i \int \rd^4x\, \cL^{--}
~, \qquad
\cL^{--} = \frac{1}{16} \frac{u_i u_j u_k u_l}{(v,u)^4} D^{ij} \bar D^{kl} \mathscr{L}^{++}~,
\end{align}
in terms of an additional coordinate $u_i$; however, the result is actually
independent of $u_i$, except for the requirement that
$(v,u) \neq 0$ along the contour $\cC$.
The extension to the curved case was given in \cite{KT-M:DiffReps} as
\begin{align}\label{eq:IntroCompAction}
S= -\frac{1}{2\pi} \oint_\cC v_i \,\rd v^i \int \rd^4x\, e\, \cL^{--}~, \qquad
\cL^{--} = \frac{1}{16} \frac{u_i u_j u_k u_l}{(v,u)^4} \cD^{ij} \bar \cD^{kl} \mathscr{L}^{++} + \cdots
\end{align}
An additional requirement of constant $u_i$ turned out to be useful to impose.
The elided terms in the above expression for $\cL^{--}$ were
determined by requiring independence under small shifts of the constant $u_i$.
Large classes of actions can then be constructed directly from \eqref{eq:IntroCompAction}
by choosing $\mathscr{L}^{++}$ to be built out of fundamental arctic, antarctic,
vector and tensor multiplets: the resulting actions include
general supergravity-matter systems \cite{Kuzenko:2008qz}.
The coupling to conformal supergravity naturally occurs automatically
because of the super-Weyl invariance of the action \cite{KLRT-M1}.

There are some curious features about this formulation.
First, as noted in \cite{KLRT-M1}, the coordinates $v^i$ are effectively invariant
under $\rm SU(2)$ transformations. Second, the manifold is effectively
$\cM^{4|8} \times \mathbb CP^1$ but the action and constraints are clearly formulated
in a \emph{central gauge} (or \emph{central basis} in the language of \cite{GIOS})
where $\cM^{4|8}$ and $\mathbb CP^1$ are largely decoupled. One is not
permitted to make $\mathbb CP^1$-dependent Lorentz transformations (for example)
or arbitrary diffeomorphisms on $\mathbb CP^1$.
Finally, an auxiliary coordinate $u_i$ must be introduced to evaluate the action,
subject only to the condition that $(v,u) \neq 0$ along $\cC$.
(Such a constant $u_i$ exists for any contour.)
In the original flat superspace approach of
\cite{KLR, Siegel:ChiralActions}, the coordinate $u_i$ could
actually be chosen to vary along the contour;
in the curved superspace approach, it was chosen constant for convenience.

In this paper, we will shed some light on these features by presenting a modified
version of curved projective superspace.
The main idea will be to introduce a supermanifold $\cM^{4|8} \times \mathrm{SU}(2)$,
that admits gauge transformations and diffeomorphisms involving both the
coordinates $z^M$ of $\cM^{4|8}$ and the coordinates $v^{i\pm}$
of $\mathrm{SU}(2)$, placing them on an equal footing.\footnote{A similar
approach was sketched by Hartwell and Howe \cite{HoweHartwell:Npq}.}
Because our fields will always be chosen to depend only on
$\mathbb CP^1 \cong \rm SU(2) / U(1)$, the supermanifold will
effectively be $\cM^{4|8} \times \mathbb CP^1$. As in harmonic superspace,
we will assume that there exists a central gauge where these two factors
largely decouple.

We will find that the coordinates $v^{i\pm}$ indeed transform under
$\rm SU(2)$ diffeomorphisms; however, upon restriction to a
central gauge they can be interpreted as inert.
This in turn explains the two curious features mentioned above.
In the new framework, the role of the coordinate $u_i$
will be played by the complex conjugate $v_i^-$ of $v^{i+}$, so that $v^{i+} v_i^- = 1$.
The conventional formulation of projective superspace
will arise after a complexification of $v^{i+} \rightarrow v^i$ and
$v_i^-\rightarrow u_i / (v,u)$, which is
always possible provided $(v, u)$ is nonzero along the contour $\cC$ of interest.

Although a full discussion would be beyond the scope of this paper,
there is a deep relationship between the harmonic superspace approach to
supergravity and the projective superspace approach we are advocating
here. Aside from the obvious similarity -- the use of the same supermanifold
$\cM^{4|8} \times {\rm SU}(2)$ (with the $\rm SU(2)$ factor effectively
$\mathbb CP^1$) -- the projective approach we present admits
diffeomorphisms on the auxiliary manifold, as in the analytic basis of
harmonic superspace.
This can potentially lead to some important conceptual advantages,
which we will discuss in the conclusion.

This paper is organized as follows.
In section \ref{sec:SU2Geo}, we review the properties of the
$\rm SU(2)$ manifold that will augment the usual supermanifold
$\cM^{4\vert 8}$. Many of the important features of the full superspace
will already be apparent when considering just the $\rm SU(2)$ manifold itself.
Section \ref{sec:CPS} presents the structure of the supermanifold
$\cM^{4|8} \times \mathrm{SU}(2)$, upon which projective superspace can be
placed.
In section \ref{sec:SuperActions}, we present three
action principles on $\cM^{4|8} \times \mathrm{SU}(2)$ involving,
respectively,
integration over all, half, or 3/4 of the Grassmann coordinates.
The most important of these is the analytic superspace action
involving half the Grassmann coordinates
(the others can always be reduced to it) so we give
its component reduction in section \ref{sec:CompAction}. This
yields an interesting surprise: in a general gauge, the
component action can always be written as the integral of a
five-form $\cJ$ living on $\cM^4 \times \cC$, where
$\cM^4$ is the spacetime manifold
and $\cC$ is the contour in $\mathrm{SU}(2)$.
When restricted to the central gauge, the five-form leads to
a component action similar to \eqref{eq:IntroCompAction} with one
intriguing difference. In the conclusion, we briefly speculate
on possible advantages of this new extended formulation.

Three appendices are included. Appendix \ref{app:Curvatures}
covers details of the superspace curvatures that are not included
in section \ref{sec:CPS}. Appendix \ref{app:Integration} briefly
reviews how to formulate invariant integrals over submanifolds,
which is necessary for constructing invariant actions over
1/2 or 3/4 of the Grassmann coordinates. Appendix \ref{app:CompDetails}
presents the details of the component reduction of the
analytic superspace action.

The notation and conventions for the $\rm SU(2)$ manifold are largely
those of \cite{GIOS} and are straightforwardly related to those
employed in \cite{KLRT-M1, KLRT-M2, KT-M:DiffReps}. The conventions
for $\cN=2$ superspace, spinors, $\sigma$-matrices, and so on
follow \cite{Butter:CSG4d_2}.

\section{Geometric properties of $\rm SU(2)$}\label{sec:SU2Geo}
The formulation of projective superspace we will introduce in subsequent
sections is based on the product of a supermanifold
$\cM^{4\vert 8}$ with an internal $\rm SU(2)$ manifold.
As in \cite{KLRT-M1}, we are not actually interested in $\rm SU(2)$
but rather the projective space $\mathbb CP^1$.
This will come about because,
as in harmonic superspace \cite{GIOS},
we will always be dealing with quantities of fixed charge under
the diagonal $\rm U(1)$ subgroup of $\rm SU(2)$. In other words,
the effective space will actually be the coset
${\rm SU}(2) / {\rm U}(1) \cong \mathbb CP^1 \cong S^2$.

It is important to emphasize that almost all of the conventions
and technology discussed below, such as the derivatives $D^{\pm\pm}$ and $D^0$,
are exactly those introduced first in the harmonic superspace literature.
As harmonic and projective superspace utilize the same auxiliary manifold,
there is no obstruction to exploiting the same technology in both;
in fact, a common notation can help accentuate the
meaningful differences between them.

In this section, we will review the structure of the auxiliary
manifold before moving on to review elements of analysis on
$\rm SU(2)$. Afterwards, we will highlight how complexifying
$\rm SU(2)$ to $\rm SL(2,\mathbb C)$
naturally recovers the formulation used in \cite{KLRT-M1}.

\subsection{The relations $\mathrm{SU}(2) / \mathrm{U}(1) \cong S^2 \cong \mathbb CP^1$}
Let us begin with the usual formulation of $\mathbb CP^1$
as $\mathbb C^{2*} / \mathbb C^*$.
Introduce two complex coordinates $v^i$ for $i=\1,\2$.
Their complex conjugates are $\bar v_i = (v^i)^*$.
Under the identification
\begin{align}\label{eq:vIsoProj}
v^i \sim c v^i~, \qquad c\in \mathbb C^*
\end{align}
the coordinates $v^i$ are the so-called
homogeneous coordinates on $\mathbb CP^1$. The north chart
of $\mathbb CP^1$ is that region where $v^\1$ is nonzero, while the
south chart of $\mathbb CP^1$ possesses nonzero $v^\2$. We denote
the point $v^i \sim (1,0)$ as the \emph{north pole}
and $v^i \sim (0,1)$ as the \emph{south pole}.\footnote{Note that
some references (e.g. \cite{Kuzenko:Lectures}) define the north pole
to lie at $v^i \sim (0,1)$ and the south pole at $v^i \sim (1,0)$.
In that convention, the north chart is generated by stereographic projection
from the north pole, and so the north pole lies outside the north chart.}

The space $\mathbb CP^1$ can alternatively be described within the space
$\rm SU(2) \cong \mathbb C^{2*} / \mathbb R_+$. The
normalized harmonic variables
\begin{align}\label{eq:HarmDefIso}
v^{i +} := \frac{v^i}{\abs v}, \qquad
v_i^- := \frac{\bar v_i}{\abs v}, \qquad \abs v^2 = (v, \bar v) \equiv v^k \bar v_k, \qquad
v^{i +} v_i^- = 1~,
\end{align}
can be used to construct a generic $\rm SU(2)$ group element
\begin{align}\label{eq:gsu2}
\mathbf g = \left(\begin{array}{rr}
v^{\1 +} & -v_{\2}^- \\
v^{\2 +} & v_{\1}^-
\end{array}\right)
= \left(\begin{array}{rr}
v^{\1 +} & -v^{\1 -} \\
v^{\2 +} & -v^{\2 -}
\end{array}\right)
~, \qquad \mathbf g^{-1} = \mathbf g^\dag~, \quad \det \mathbf g = 1~.
\end{align}
$\mathbb CP^1$ is then identified as $\rm SU(2) / U(1)$ by imposing the equivalence relation
\begin{align}\label{eq:vIsoU1}
v^{i+} \sim e^{i \alpha} v^{i+}~, \qquad e^{i \alpha} \in \rm U(1)~.
\end{align}

We can use the inhomogeneous coordinate $\zeta = v^{\2} / v^{\1}$
of $\mathbb CP^1$ to parametrize the harmonics.
The north pole corresponds to $\zeta=0$
while the south pole corresponds to $\zeta=\infty$.
The harmonics $v_i^\pm$ are given in terms of $\zeta$ and
the phase $e^{i \psi} := v^\1 / |v^\1|$ by
\begin{align}\label{eq:InhomoCoords}
v^{i+} = (v^{\1 +}, v^{\2 +}) = \frac{e^{i \psi}}{\sqrt{1+\zeta \bar \zeta}} \, (1, \zeta)~, \qquad
v_i^- = (v_{\1}^-, v_{\2}^-) = \frac{e^{-i \psi}}{\sqrt{1+\zeta \bar \zeta}} \, (1, \bar\zeta)~.
\end{align}
Because the coordinates $\z, \bar\z$ describe the north chart
of the Riemann sphere, we refer to the coordinates
$y^{\ul m} = (\z, \bar \z, \psi)$ as the
north chart of $\rm SU(2)$. In what follows, we will
frequently present quantities in terms of this chart.

Following \cite{GIOS}, we introduce a new notion of complex
conjugation corresponding to normal complex conjugation with an additional
antipodal map on $S^2$. The new complex conjugation
is denoted with a $\widetilde{\phantom{a}}$ and acts as
$\widetilde{v^{i \pm}} = -v_i^\pm$, equivalently
$\widetilde{v_i^{\pm}} = v^{i\pm}$.
This coincides with the usual smile conjugation of the conventional
formulation of projective superspace \cite{KLRT-M1}.

\subsection{Vielbeins and covariant derivatives of $\rm SU(2)$}
We introduce three derivative operations, conventionally denoted $D^{++}$, $D^{--}$
and $D^0$, corresponding to the right action of $\rm SU(2)$ on $\mathbf g$.
They are conventionally defined on the harmonic coordinates as\footnote{Note
these derivatives preserve the constraint
$v^{i+} v_i^-=1$ and so the partial derivatives with respect
to the constrained variables $v^{i\pm}$ are sensible.}
\begin{align}
D^{++} := v_i^+ \frac{\partial}{\partial v_i^-}~,\quad
D^{--} := v^{i-} \frac{\partial}{\partial v^{i+}}~, \quad
D^0 := v^{i +} \frac{\partial}{\partial v^{i+}} -  v_i^- \frac{\partial}{\partial v_i^-} ~,
\end{align}
but can also be written in terms of the homogeneous coordinates $v^i$ and $\bar v_i$,
\begin{align}
D^{++} = v_i \frac{\partial}{\partial \bar v_i}~,\qquad
D^{--} = \bar v^i \frac{\partial}{\partial v^i}~, \qquad
D^0 = v^i \frac{\partial}{\partial v^i} - \bar v_i \frac{\partial}{\partial \bar v_i}~,
\end{align}
or in terms of the inhomogeneous coordinate $\zeta$ and the phase $\psi$,
\begin{align}
D^{++} = e^{2i \psi} \Big(
	(1+\zeta \bar \zeta) \pa_{\bar\zeta} -\frac{i}{2} \zeta \pa_\psi
	\Big)~, \quad
D^{--} = - e^{-2i \psi} \Big(
	(1+\zeta \bar \zeta) \pa_\zeta + \frac{i}{2} \bar \zeta \pa_\psi\Big)~, \quad
D^0 = -i \,\pa_\psi~.
\end{align}
They possess the commutation relations
\begin{align}
[D^{++}, D^{--}] = D^0~, \qquad [D^0, D^{++}] = 2 D^{++}~, \qquad
[D^0, D^{--}] = -2 D^{--}~,
\end{align}
and one can interpret $D^0$ as a charge generator, with $D^{++}$ and $D^{--}$
respectively carrying charge $+2$ and $-2$.

It will be convenient to denote the charges $++$, $--$ and $0$ on the
derivatives by an index $\ul a$ and to introduce a convention for
lowering this index. A convenient definition is
\begin{align}\label{eq:SU2LowerConv}
D_{\ul a} = (D_{++}, D_{--}, D_0)~, \qquad
D_{++} := - D^{--}~,\quad
D_{--} := D^{++}~,\quad
D_0 := D^0~.
\end{align}
Then the algebra of these covariant derivatives can be written as
$[D_{\ul a}, D_{\ul b}] = - T_{\ul a \ul b}{}^{\ul c} D_{\ul c}$
for a constant torsion tensor.

Associated with these covariant derivatives are three vielbeins
$\cV^{\ul a} = \rd y^{\ul m} \cV_{\ul m}{}^{\ul a}$, which we denote
(using different conventions than \cite{GIOS})
\begin{align}\label{eq:cVharm}
\cV^{++} = v_i^+ \rd v^{i+}~, \qquad
\cV^{--} = v_i^- \rd v^{i-}~, \qquad
\cV^{0} = v_i^- \rd v^{i+} = v_i^+ \rd v^{i-}~.
\end{align}
In the homogeneous coordinate system, these are given by
\begin{align}
\cV^{++} = \frac{1}{(v,\bar v)} v_i \,\rd v^{i}~, \qquad
\cV^{--} = \frac{1}{(v,\bar v)} \bar v_i \,\rd \bar v^{i}~, \qquad
\cV^{0} = \frac{1}{(v, \bar v)} \frac{1}{2} \big( \bar v_i \,\rd v^i - v^i \,\rd \bar v_i\big)
\end{align}
and in the inhomogeneous coordinate system by
\begin{align}
\cV^{++} = \frac{e^{2 i \psi}}{1+\z\bar\z} \,\rd \z~,\qquad
\cV^{--} = \frac{e^{-2 i \psi}}{1+\z\bar\z} \,\rd \bar\z~, \qquad
\cV^0 = i \rd \psi + \frac{1}{2} \frac{1}{1+\z \bar \z} \left(\bar\z \rd \z - \z \rd \bar\z \right)~.
\end{align}
The Cartan structure equations are\footnote{We use the superspace
conventions for forms (see e.g. \cite{WessBagger}). This means that
exterior derivatives act from the right, i.e.
$\rd (\alpha_p \wedge \beta_q) = \alpha_p \wedge \rd \beta_q + (-1)^q \rd \alpha_p \wedge \beta_q$
for $p$-form $\alpha_p$ and $q$-form $\beta_q$.}
\begin{align}
\rd \cV^{++} = 2 \cV^{++} \wedge \cV^0~, \qquad
\rd \cV^{--} = -2 \cV^{--} \wedge \cV^0~, \qquad
\rd \cV^0 = \cV^{++} \wedge \cV^{--}~.
\end{align}
The covariant derivative can be written in the usual way, 
$D_{\ul a} = \cV_{\ul a}{}^{\ul m} \pa_{\ul m}$,
in terms of the inverse vielbein.
One can verify these relations by checking that
$\rd = \cV^{\ul a} D_{\ul a} = \rd y^{\ul m} \pa_{\ul m}$
acts as an exterior derivative on any function of the $\rm SU(2)$ coordinates $y^{\ul m}$.
We normalize the metric on $\rm SU(2)$ as
\begin{align}
\rd s^2_{\rm SU(2)} = \Tr (\rd \mathbf g^{-1} \otimes \rd \mathbf g)
	= 2 \cV^{++} \otimes \cV^{--} - 2\cV^0 \otimes \cV^0
	= 2\rd v^{i+} \otimes \rd v_i^-~,
\end{align}
although we will not use it explicitly in what follows.
Note that under the $\widetilde{\phantom{a}}$ conjugation,
the derivatives and vielbeins
are real, $\widetilde {D_{\ul a}} = D_{\ul a}$ and
$\widetilde {\cV^{\ul a}} = \cV^{\ul a}$.

The isometries of $\rm SU(2)$
correspond to the left action on the group element $\mathbf g$.
These can be denoted by generators $\hat I^i{}_j$ which act
as
\begin{align}
\hat I^i{}_j = - v^{i+} \frac{\pa}{\pa v^{j+}}
	+ v_j^- \frac{\pa}{\pa v_i^-}
	+ \frac{1}{2} \delta^i{}_j \Big(
	v^{k+} \frac{\pa}{\pa v^{k+}}
	- v_k^{-} \frac{\pa}{\pa v_k^{-}}
	\Big)~.
\end{align}
One can verify that these do indeed leave the covariant derivatives invariant,
\begin{align}
[\hat I^i{}_j, D^{++}] = [\hat I^i{}_j, D^{--}] = [\hat I^i{}_j, D^0] = 0~.
\end{align}
One can further verify that an isometry with constant parameters $\l^j{}_i$ can be
rewritten as
\begin{align}\label{eq:LItoLD}
\delta_I = \l^j{}_i \hat I^i{}_j = \l^{\ul a} D_{\ul a} =
	- \l^{++} D^{--} + \l^{0} D^0 + \l^{--} D^{++}
\end{align}
where $\l^{\pm \pm}$ and $\l^{0}$ are coordinate-dependent transformations given by
\begin{align}\label{eq:KillingLambda}
\l^{\pm\pm} := v_i^\pm v_j^\pm \l^{ij}~, \qquad
\l^{0} := v_i^+ v_j^- \l^{ij}~.
\end{align}
(It is sometimes convenient to denote $\l^0 = \l^{+-}$ in analogy with $\l^{\pm\pm}$.)
The appearance of the minus sign in \eqref{eq:LItoLD} was the reason for
introducing the sign in \eqref{eq:SU2LowerConv}.

If we now restrict to the space $S^2 \cong \rm SU(2) / U(1)$, then the
covariant derivatives $D_{\ul a}$ possess a different interpretation.
$D^0$ can be identified with the rotation generator on the tangent space
of $S^2$, while $D^{++}$ and $D^{--}$ can be identified with the
covariant holomorphic and antiholomorphic derivatives.
Then a scalar function $f^{(q)}$ of fixed
$D^0$ charge on $\rm SU(2)$ is reinterpreted as a tensor of
spin $|q|/2$ on $S^2$. In what follows, although we will always remain
with an explicit $\rm SU(2)$ manifold, we will only be dealing with
such functions $f^{(q)}$, and so it will always be possible to
reinterpret our calculations as being performed on the space
$\mathrm{SU}(2) / \mathrm{U}(1) \cong S^2 \cong \mathbb CP^1$.

\subsection{Harmonic and holomorphic tensors on $\mathbb CP^1$}
There are two interesting classes of tensors on $\mathbb CP^1 \cong S^2$.
The first are the so-called harmonic functions, which are globally defined
functions on $\rm SU(2)$ with fixed $D^0$ charge. These are given by
\begin{align}
f^{(q)} = \sum_{n=0}^\infty f^{(i_1 \ldots i_{n+q} j_1 \ldots j_n)}
	v_{i_1}^+ \cdots v_{i_{n+q}}^+ v_{j_1}^- \cdots v_{j_n}^-~, \qquad
D^0 f^{(q)} = q f^{(q)}~,
\end{align}
(assuming $q\geq 0$, but similarly for $q<0$)
and are extensively discussed in \cite{GIOS}. As already mentioned,
the requirement that they possess fixed $D^0$ charge ensures that they
describe tensor fields of spin $|q|/2$ on $S^2$.

The second interesting class are the functions $\cQ^{(q)}$
with fixed $D^0$ charge but annihilated by $D^{++}$,
\begin{align}\label{eq:Qholo}
D^0 \cQ^{(q)} = q \cQ^{(q)}~, \qquad D^{++} \cQ^{(q)} = 0~.
\end{align}
The most general class of such functions is not
globally defined on $\rm SU(2)$.
If the functions are non-singular near the north pole,
they are called \emph{arctic} and possess an expansion\footnote{Superfields
in projective superspace with such expansions were introduced in \cite{LR88}.
The arctic/antarctic nomenclature appeared later in \cite{GonzalezRey:FR1}.}
\begin{align}
\cQ^{(q)} = (v^{\1+})^q \cQ(\z) = (v^{\1+})^q \sum_{n=0}^\infty \cQ_n \z^n~.
\end{align}
Their conjugates $\widetilde \cQ^{(q)}$ are non-singular near the
south pole and are called \emph{antarctic}. They possess an expansion
\begin{align}
\widetilde\cQ^{(q)} = (v^{\2+})^q \widetilde \cQ(\z) = (v^{\2+})^q \sum_{n=0}^\infty (-1)^n \bar \cQ_n \z^{-n}~.
\end{align}
It will be convenient to refer to functions $\cQ^{(q)}$ satisfying
\eqref{eq:Qholo} as \emph{holomorphic} although strictly speaking
they are generically holomorphic only on an open domain of $\rm SU(2) / U(1)$.

Of course, it is possible for such functions to be both
holomorphic and globally defined.
These generally have an expansion of the form
\begin{align}\label{eq:GExpansion}
\cG^{(q)} = \cG^{(i_1 \ldots i_{q})} v_{i_1}^+ \cdots v_{i_{q}}^+~.
\end{align}
Such functions are both arctic and antarctic. They can be
real under conjugation only if $q$ is even.

\subsection{Integration measures and global $\rm SU(2)$ invariance}
The most straightforward integration over the auxiliary manifold
$\rm SU(2)$ is accomplished using the usual Haar measure. Given some globally
defined function $f^{(0)}(v^+, v^-)$, we may define the action integral
\begin{align}
S = \int \rd v\, f^{(0)} = \frac{i}{4 \pi^2} \int_0^{2\pi} \rd\psi \int \frac{\rd\zeta \wedge \rd\bar\zeta}{(1 + \zeta \bar\zeta)^2} \, f^{(0)}
	= \frac{i}{2 \pi} \int \frac{\rd\zeta \wedge \rd\bar\zeta}{(1 + \zeta \bar\zeta)^2} \, f^{(0)}~,
\end{align}
normalized so that $\int \rd v = 1$.
Because we have assumed $f^{(0)}$ to have vanishing $D^0$ charge, it must
be independent of $\psi$ and so the integral over $\rm U(1)$ is trivial.
Suppose now that the function $f^{(0)}$ transforms as a scalar field under
an $\rm SU(2)$ isometry,
\begin{align}
\delta_I f^{(0)} = -\l^{++} D^{--} f^{(0)} + \l^{--} D^{++} f^{(0)}~.
\end{align}
It is easy to see that the action is invariant.
Integrating by parts using
\begin{align}
\int \rd v\, D^{++} f^{--} = 0~, \qquad
\int \rd v\, D^{--} f^{++} = 0~,
\end{align}
one finds
\begin{align}
\delta_I S = \int \rd v\, \Big(D^{--} \l^{++} - D^{++} \l^{--}\Big) \,f^{(0)}(v)
	= \int \rd v\, \Big(2 \l^{0} - 2 \l^{0}\Big) f^{(0)}(v) = 0~.
\end{align}

Integrals of the above type are encountered when using harmonic superspace,
which is concerned with globally defined functions. Since we will be
dealing more with holomorphic functions, the natural integration principle
will involve a one-dimensional contour integral on $\rm SU(2)$,
with the contour avoiding regions where the functions become singular.
The natural integrand is a one-form
$\omega = \rd y^{\ul m} \omega_{\ul m} = \cV^{\ul a}\, \omega_{\ul a}$
and the corresponding integral is
\begin{align}\label{eq:contourAction}
S = \frac{1}{2\pi} \oint_\cC \cV^{\ul a} \omega_{\ul a}~.
\end{align}
Because we are actually interested in contours in $\mathbb CP^1 \cong S^2$,
we will always assume $\omega_0 = 0$ so that the resulting action is given by
\begin{align}\label{eq:contourAction2}
S = - \frac{1}{2\pi} \oint_\cC \cV^{++} \omega^{--}
	+ \frac{1}{2\pi} \oint_\cC \cV^{--} \omega^{++}
	= - \frac{1}{2\pi} \oint_\cC v_i^+ \rd v^{i+} \,\omega^{--}
	+ \frac{1}{2\pi} \oint_\cC v_i^- \rd v^{i-} \omega^{++}~.
\end{align}
For later convenience we have ``raised'' the indices on $\omega_{\ul a}$,
using the same convention as in \eqref{eq:SU2LowerConv}, so that the $D^0$ charges
of the integrands are clear.

A natural question to ask is whether the contour action \eqref{eq:contourAction2} is
invariant under $\rm SU(2)$ isometries. It turns out that the answer is yes, provided
the integrands $\omega^{--}$ and $\omega^{++}$ obey a certain condition.
First let us establish a version of Stokes' theorem.
Suppose $\omega = \rd \L^{(0)}$ for some function $\L^{(0)}$. Then
we must have
\begin{align}\label{eq:ContourStokes}
0 = - \frac{1}{2\pi} \oint_\cC \cV^{++} \,D^{--} \L^{(0)}
	+ \frac{1}{2\pi} \oint_\cC \cV^{--} D^{++} \L^{(0)}~.
\end{align}
If $\L^{(0)}$ is holomorphic, this reduces to the simpler equation
\begin{align*}
0 = -\frac{1}{2\pi} \oint_\cC \cV^{++} \,D^{--} \L^{(0)} \qquad  \text{if} \,\,\,D^{++} \L^{(0)} = 0~.
\end{align*}
These two results are quite important, so let us discuss their form in
an explicit coordinate basis. If $\tau$ is the coordinate parametrizing the contour,
one can show that
\begin{align}
\frac{1}{2\pi} \oint_\cC \cV^{++} \, D^{--} \L^{(0)}
	= -\frac{1}{2\pi} \oint_\cC \rd \tau\, \frac{\rd\zeta}{\rd \tau} \,\frac{\partial \L^{(0)}}{\partial \zeta}
\end{align}
If $\L^{(0)}$ is holomorphic, then the right-hand side vanishes as a
total derivative. If not, we find that
\begin{align}
-\frac{1}{2\pi} \oint_\cC \rd \tau\, \frac{\rd\zeta}{\rd \tau} \,\frac{\partial \L^{(0)}}{\partial \zeta}
	= \frac{1}{2\pi} \oint_\cC \rd \tau\, \frac{\rd\bar\zeta}{\rd \tau} \,\frac{\partial \L^{(0)}}{\partial \bar\zeta}
	= \frac{1}{2\pi} \oint_\cC \cV^{--} \, D^{++} \L^{(0)}~.
\end{align}
This establishes \eqref{eq:ContourStokes}.

Now let us calculate $\delta_I S$. The vielbein one-forms are necessarily
invariant under the isometry while $\omega_{\ul a}$ transforms as
$\delta_I \omega_{\ul a} = \l^{\ul b} D_{\ul b} \omega_{\ul a}$.
This implies, using the explicit form \eqref{eq:KillingLambda}
of the parameters $\l^{\ul a}$,
\begin{align}
\delta_I \omega^{--} 
	&= - D^{--} \Big( \l^{++} \omega^{--}\Big) + \l^{--} D^{++} \omega^{--}~, \eol
\delta_I \omega^{++} 
	&= D^{++} \Big( \l^{--} \omega^{++}\Big) - \l^{++} D^{--} \omega^{++}~.
\end{align}
This leads, using \eqref{eq:ContourStokes}, to
\begin{align}
\delta_I \int \cV^{++} \omega^{--}
	= - \int \cV^{--} D^{++} \Big( \l^{++} \omega^{--}\Big) + \int \cV^{++} \l^{--} D^{++} \omega^{--}~, \eol
\delta_I \int \cV^{--} \omega^{++}
	= \int \cV^{++} D^{--} \Big( \l^{--} \omega^{++}\Big) - \int \cV^{--} \l^{++} D^{--} \omega^{++}~,
\end{align}
and one can see that the difference between these two terms vanishes (and so
$\delta_I S = 0$) precisely when\footnotemark
\begin{align}\label{eq:omegaClosed}
D^{++} \omega^{--} = D^{--} \omega^{++}~.
\end{align}
This is merely the tangent space version of the condition that $\omega$ is closed.

\footnotetext{Note that it is possible to have purely holomorphic one-forms $\omega$
that obey $\omega^{++} = 0$ and $D^{++} \omega^{--} = 0$.
It will turn out that the one-forms $\omega$ we consider in projective superspace
will generally not be purely holomorphic, but will instead carry some small
non-vanishing $\omega^{++}$ piece.}

\subsection{Extension to local $\rm SU(2)$ transformations}
Up until now, we have been restricting our attention to $\rm SU(2)$ isometries.
These preserved the form of the $\rm SU(2)$ vielbein $\cV$ and were generated by
constant parameters $\l^i{}_j$. In principle, there is no reason why we
cannot perform \emph{local} $\rm SU(2)$ transformations
of the form \eqref{eq:LItoLD} but with parameters $\xi^{++}$, $\xi^{--}$ and $\xi^0$
subject only to the condition that $\xi^{\pm\pm}$ and $\xi^0$ have
$D^0$ charges $\pm 2$ and $0$, respectively. That is, we can take
\begin{align}
\delta = \xi^{\ul a} D_{\ul a} = -\xi^{++} D^{--} + \xi^{0} D^0 + \xi^{--} D^{++}
\end{align}
but with e.g. $\xi^{++}$ not necessarily of the form $\xi^{ij} v_i^+ v_j^+$.
Such $\rm SU(2)$ diffeomorphisms 
can be interpreted as diffeomorphisms on $S^2$ (generated by $\xi^{\pm\pm}$)
along with local $\rm U(1)$ frame rotations (generated by $\xi^0$).

Under such a local transformation, the vielbeins transform in the usual way
\begin{align}
\delta \cV^{\ul a} = \rd \xi^{\ul a} + \cV^{\ul b} \xi^{\ul c} T_{\ul c \ul b}{}^{\ul a}~,
\end{align}
leading to
\begin{subequations}\label{eq:deltacV}
\begin{align}
\delta \cV^{++} &= \rd \xi^{++} - 2 \cV^0 \xi^{++} + 2 \cV^{++} \xi^0 ~, \\
\delta \cV^{--} &= \rd \xi^{--} + 2 \cV^0 \xi^{--} - 2 \cV^{--} \xi^0 ~, \\
\delta \cV^0 &= \rd \xi^0 + \cV^{++} \xi^{--} - \cV^{--} \xi^{++}~.
\end{align}
\end{subequations}
One can check that the above transformations are consistent with the definitions
\eqref{eq:cVharm}, using, e.g.
$\delta v^{i+} = \xi^{\ul a} D_{\ul a} v^{i+}
	= -\xi^{++} v^{i-} + \xi^0 v^{i+}$.
Now let us investigate the consequences of requiring that the contour action \eqref{eq:contourAction2}
remain invariant under such diffeomorphisms. Viewing $\omega$ as a one-form on $\rm SU(2)$,
a general diffeomorphism can always be written as
\begin{align}
\delta_\xi \omega = \rd (\imath_\xi \omega) + \imath_\xi \rd \omega~.
\end{align}
The first term vanishes along the contour integral so we conclude
that $\omega$ must be closed. In the tangent
frame, with the condition $\omega_0 = 0$, we recover
\eqref{eq:omegaClosed}. The condition for
invariance under $\rm SU(2)$ isometries is the same condition
as for full diffeomorphism invariance.

There is an obvious geometric interpretation of the requirement that $\omega$ be closed.
An arbitrary diffeomorphism of an integral $\oint_\cC \omega$ can be interpreted
as a small deformation of the contour $\cC$. For the integral to be
stationary, the flux of $\rd \omega$ through
the closed contour $\delta \cC = \cC' - \cC$ must vanish, implying
that $\omega$ is closed in the vicinity of the contour.
This means that the integrals $\oint_\cC \omega$ we are
considering depend only upon the topological nature of the contour
within $\rm SU(2)$. In other words, for the integral to be non-vanishing,
there must be regions where $\omega$ becomes singular (or $\rd \omega$
ceases to vanish), and the contour must
enclose such a region to give a non-vanishing result.

\subsection{The complexified $\rm SU(2)$ and the emergence of a projective structure}\label{sec:SU2Geo.complex}
Our final topic in this opening section is to address how the $\rm SU(2)$
framework we have been discussing can be related to the $\mathbb CP^1$ framework
that one encounters in the conventional formulation of projective superspace
coupled to supergravity. It is after all the $\mathbb CP^1$ framework that gives
projective superspace its name.

The key idea is to complexify $\rm SU(2)$ and to treat $v^i$ and $\bar v_i$
as independent coordinates.
Beginning with the representation \eqref{eq:HarmDefIso} for the harmonic coordinates,
let us replace $\bar v_i \rightarrow u_i$, relaxing the condition that
it is the complex conjugate of $v^i$. In doing so, it will be convenient to modify the
definitions of the harmonics so that
\begin{align}
v^{i+} := v^i~, \qquad v_i^- = \frac{u_i}{(v,u)}~, \qquad
v^{i+} v_i^- = 1~, \qquad v_i^- \neq (v^{i+})^*~.
\end{align}
We have shifted the entirety of the $(v,u)$ factor into the second harmonic
because $\sqrt{(v,u)}$ is not well-defined. This shift can
be interpreted as a local complex $D^0$ gauge transformation.
This local gauge transformation has the effect of converting all quantities
of fixed $D^0$ charge $q$ into quantities of degree $q$ in $v^i$ and degree
$0$ in $u_i$. In other words, the $+$ and $-$ labels on the harmonics
(as well as any other quantities) now denote their homogeneity
under the projective transformation
\begin{align}\label{eq:ProjV}
v^i \rightarrow c \,v^i~, \qquad c \in \mathbb C~.
\end{align}
The resulting group element $\mathbf g$ given in \eqref{eq:gsu2} still
obeys $\det \mathbf g = 1$ but is no longer unitary.
In other words, we have complexified $\rm SU(2)$ to  $\rm SL(2,\mathbb C)$.

It is straightforward to extend the entirety of the previous discussion
to $\rm SL(2, \mathbb C)$. Instead of dealing with operators
and functions of fixed $D^0$ charge, we have fixed homogeneity
under \eqref{eq:ProjV} and \emph{invariance} under
$u_i \rightarrow c\, u_i$. One can introduce derivatives
\begin{align}
D^{++} = (v,u) \, v_i \frac{\pa}{\pa u_i}~, \qquad
D^{--} = \frac{u_i}{(v,u)} \frac{\pa}{\pa v^i}~, \qquad
D^0 = v^i \frac{\pa}{\pa v^i} - u_i \frac{\pa}{\pa u_i}~,
\end{align}
and their corresponding vielbeins
\begin{align}\label{eq:VielbeinsComplex}
\cV^{++} = v_i \,\rd v^i~, \qquad
\cV^{--} = \frac{u_i \,\rd u^i}{(v,u)^2} ~, \qquad
\cV^0 = \frac{u_i \,\rd v^i}{(v,u)}~.
\end{align}
Note that $\cV^0$ has changed its functional form in the complexification
more than $\cV^{\pm\pm}$ have;
this arises from the local complex $D^0$ transformation mentioned above.
The expressions for the vielbeins \eqref{eq:VielbeinsComplex} can be derived
by calculating
\begin{align}
\rd \cF^{(q)} = -\cV^{++} D^{--} \cF^{(q)}
	+ \cV^{--} D^{++} \cF^{(q)}
	+ \cV^0 D^0 \cF^{(q)}
\end{align}
on any function $\cF^{(q)}(v,u)$ of degree $q$ in $v^i$ and degree zero in $u_i$.

It is natural to convert all holomorphic functions $\cQ^{(q)}$ to new
quantities $\cQ'^{(q)} = \cQ'^{(q)}(v)$ of
definite homogeneity in $v^i$ and independent of $u_i$,
$\cQ'^{(q)}(cv) = c^q \,\cQ'^{(q)}(v)$.
These are related to the original $\cQ^{(q)}$ functions by the same complex
$D^0$ transformation, and we will drop the primes when it is clear from context
which quantities we are discussing.

Finally, the complex version of the contour integral \eqref{eq:contourAction2} takes the form
\begin{align}\label{eq:contourActionComplex}
S = - \frac{1}{2\pi} \oint_\cC \cV^{++} \omega^{--}
	+ \frac{1}{2\pi} \oint_\cC \cV^{--} \omega^{++}
\end{align}
where $\omega^{--}$ and $\omega^{++}$ are respectively
degrees $-2$ and $+2$ in $v^i$, degree zero in $u_i$, and
related by the complex version of \eqref{eq:omegaClosed}.
Under a local complex $\rm SU(2)$ (i.e. $\rm SL(2,\mathbb C)$) diffeomorphism, the coordinates
$v^i$ and $u_i$ transform as
\begin{align}\label{eq:VUgaugeT}
\delta v^i = \xi^0 v^i - \frac{\xi^{++}}{(v,u)} u^i~, \qquad
\delta u_i = -\xi^0 u_i + (v,u) \,\xi^{--}  v_i~,
\end{align}
while $\omega^{++}$ and $\omega^{--}$ transform as
\begin{align}
\delta \omega^{--} = - \xi^{++} D^{--} \omega^{--} -2 \xi^0 \omega^{--} + \xi^{--} D^{++} \omega^{--}~, \eol
\delta \omega^{++} = - \xi^{++} D^{--} \omega^{++} + 2 \xi^0 \omega^{++} + \xi^{--} D^{++} \omega^{++}~.
\end{align}
The parameters $\xi^{\pm \pm}$ and $\xi^0$ are each assumed to
be of degree zero in $u_i$ while possessing homogeneity of the
indicated degree in $v^i$.

The major advantage of the complexified $\rm SU(2)$ is that we may
choose $v^i$ and $u_i$ to have entirely uncorrelated behavior along
the contour. In particular, one can take $u_i$ to be \emph{fixed},
subject only to the condition that $(v,u) \neq 0$ along the contour.
This can be interpreted as deforming the contour $\cC$ within
$\rm SL(2,\mathbb C)$. When one adopts such a choice, the
gauge freedom \eqref{eq:VUgaugeT} is no longer arbitrary, but is restricted
by the requirement that $\delta u_i$ is similarly constant along the contour.
This implies certain constraints on the functions $\xi^{--}$ and $\xi^0$.
(This residual freedom was discussed in the context of
projective superspace in \cite{KT-M:DiffReps}.)
The advantage of adopting this choice is that the second contour integral
in \eqref{eq:contourActionComplex} \emph{automatically} vanishes even if
$\omega^{++}$ is nonzero. This is a consequence of the property that
the form of a total
contour derivative is simplified from \eqref{eq:ContourStokes} to
\begin{align}
0 = -\frac{1}{2\pi} \oint_\cC \cV^{++} D^{--} \L^0(v,u)
\end{align}
where we emphasize that $\L^0(v,u)$ may depend on $u_i$ (with degree zero).

Although taking $u_i$ to be constant can simplify
the contour integrals, we have found it useful to remain
with a real $\rm SU(2)$ manifold in defining our formulation of projective
superspace. This guarantees, for example, that
the harmonics are always well-defined; there is no requirement that
the contour avoid the location where $(v,u) = 0$. It also permits full
$\rm SU(2)$ diffeomorphisms, rather than the restricted
$\rm SL(2,\mathbb C)$ diffeomorphisms that leave $u_i$ constant.
Nonetheless, starting from a real $\rm SU(2)$ manifold
it is always possible to complexify to
$\rm SL(2,\mathbb C)$ and then to adopt the choice of
constant $u_i$ where needed.

\section{Projective superspace and $\cM^{4|8} \times {\rm SU}(2)$}\label{sec:CPS}
In this section, we will describe how to construct a covariant
projective superspace generalizing the work of
\cite{KLRT-M1, KLRT-M2, KT-M:DiffReps}.
We will do this first by constructing a
direct product of the supermanifold $\cM^{4|8}$ and $\rm SU(2)$,
and then splicing together the tangent space action of $I^i{}_j$
on $\cM^{4|8}$ with the isometry transformation on $\rm SU(2)$.
The resulting construction will correspond to that given in
the usual version of projective superspace. We will then
show how to lift to a general gauge. Finally, we will
comment briefly on the admissible types of primary analytic superfields.

\subsection{Conformal superspace on $\cM^{4|8} \times {\rm SU}(2)$: A bottom-up construction}
Let us begin with a conventional supermanifold $\cM^{4|8}$ with local
coordinates $z^M = (x^m, \q^\mu{}_\imath, \bar \q_\dmu{}^\imath)$
with $m=0,1,2,3$, $\mu=1,2$, $\dmu=1,2$ and $\imath=\1,\2$.
The associated superspace vielbein is given by
$E_M{}^A = (E_M{}^a, E_M{}^\alpha{}_i, E_M{}_\dalpha{}^i)$.
We will assume we are working with conformal superspace \cite{Butter:CSG4d_2},
so that the supermanifold possesses the full superconformal structure group,
but the framework we present here would work equally well with
$\rm SU(2)$ or $\rm U(2)$ superspace where the superconformal transformations
take the form of super-Weyl transformations \cite{KLRT-M1, Howe:CSG}.

In conformal superspace, the covariant derivative
$\nabla_A = (\nabla_a, \nabla_\alpha{}^i, \bar\nabla^\dalpha{}_i)$
is defined implicitly by\footnote{We have relabeled the $\rm SU(2)$
connection $\Phi_M{}^i{}_j$ of \cite{Butter:CSG4d_2} to $\cV_M{}^i{}_j$.}
\begin{align}\label{eq:defNabla}
\pa_M &= E_M{}^A \nabla_A + \cV_M{}^j{}_i I^i{}_j
	+ \frac{1}{2} \Omega_M{}^{ab} M_{ba}
	+ A_M \bbA
	+ B_M \bbD
	\eol & \quad
	+ F_M{}^{\alpha i} S_{\alpha i}
	+ F_M{}_{\dalpha i} \bar S^{\dalpha i}
	+ F_M{}^a K_a~,
\end{align}
which can be inverted to give
\begin{align}\label{eq:defNabla2}
\nabla_A &= E_A{}^M \Big(
	\pa_M - \cV_M{}^k{}_j I^j{}_k
	- \frac{1}{2} \Omega_M{}^{bc} M_{cb}
	- A_M \bbA
	- B_M \bbD
	\eol & \qquad\qquad
	- F_M{}^{\beta j} S_{\beta j}
	- F_M{}_{\dbeta j} \bar S^{\dbeta j}
	- F_M{}^b K_b \Big)~.
\end{align}
Obviously, \eqref{eq:defNabla} and \eqref{eq:defNabla2} are equivalent.
$M_{ab}$ is the Lorentz generator, $\bbA$ and $I^i{}_j$ are the $\rm U(1)$
and $\rm SU(2)$ $R$-symmetry generators, $\bbD$ is the dilatation generator,
$S_{\alpha i}$ and $\bar S^{\dalpha i}$ are $S$-supersymmetry generators,
and $K_a$ is the special conformal generator. Their algebra is summarized
in \cite{Butter:CSG4d_2}.

Now we wish to combine this structure with the $\rm SU(2)$ manifold
with covariant derivatives $D^{++}$, $D^{--}$, and $D^0$. The only nontrivial
step is to decide how the action of $I^i{}_j$ should be manifested
on functions $\cF(z, v^+, v^-)$ depending also on the $\rm SU(2)$ coordinates:
\begin{align}
I^i{}_j \cF &= - v^{i+} \frac{\pa}{\pa v^{j+}} \cF
	+ v_j^- \frac{\pa}{\pa v_i^-} \cF
	+ \frac{1}{2} \delta^i{}_j \Big(
	v^{k+} \frac{\pa}{\pa v^{k+}}
	- v_k^{-} \frac{\pa}{\pa v_k^{-}}
	\Big) \cF \eol
	&= v^{i+} v_j^+ D^{--} \cF
	- v^{i-} v_j^- D^{++} \cF
	- \Big(v^{i+} v_j^- - \frac{1}{2} \delta^i{}_j\Big) D^0 \cF~.
\end{align}
The operator $I^i{}_j$ acts as the isometry generator on the $\rm SU(2)$ manifold.
At this stage, we immediately recover the construction of
\cite{KLRT-M1, KLRT-M2}, since a general supergravity $\mathrm{SU}(2)_R$
transformation is given by
\begin{align}\label{eq:SU2Action}
\l^j{}_i I^i{}_j \cF = -\l^{++} D^{--} \cF + \l^{--} D^{++} \cF + \l^0 D^0 \cF~,
\end{align}
for arbitrary local $\l^i{}_j(z)$ independent of the harmonics.
Specializing this equation to holomorphic functions
$\cQ^{(n)}(z, v^+)$ of fixed $D^0$ charge $n$ recovers the transformation
law \eqref{eq:deltaQ2}, up to the complexification of $\rm SU(2)$ to $\rm SL(2,\mathbb C)$
discussed in section \ref{sec:SU2Geo.complex}.

At this stage, we have two different ways in which $I^i{}_j$ can act.
It can act on a function $\cF(z, v^+, v^-)$ as an $\rm SU(2)$ isometry,
or it can act on an $\rm SU(2)$ tensor independent of $v^{i\pm}$, such as $E_M{}^\alpha{}_i(z)$, as
a tangent space rotation. Now we wish to eliminate the latter in favor
of the former so that the operator acts in only one way. 
Consider for definiteness some superfield
$q^i$ with a single $\rm SU(2)$ index, independent of $v^{i\pm}$ and transforming
covariantly under $\mathrm{SU}(2)_R$ transformations. (For example,
$q^i$ could be $E_M{}^{\alpha i}$.) If we interpret $q^i$ as a component
of $q^+ = q^i v_i^+$, then the the action of $\mathrm{SU}(2)_R$ on
$q^i$, treating $v_i^+$ as invariant, is given by
\begin{align}
\delta_\l q^+ = \l^i{}_j q^j\, v_i^+ = -\l^{++} D^{--} q^+ + \l^0 q^+~.
\end{align}
This is exactly the same transformation rule as \eqref{eq:SU2Action},
corresponding to an isometry transformation on the $\rm SU(2)$ manifold.
If we uniformly exchange all quantities with $\mathrm{SU}(2)_R$ indices
for scalar functions on the $\rm SU(2)$ manifold, e.g.
\begin{align}
E_M{}^{\alpha i} ~\Longrightarrow~ E_M{}^{\alpha \pm}~, \qquad
E_M{}_\dalpha{}^i ~\Longrightarrow~ E_M{}_\dalpha{}^\pm~,
\end{align}
then $I^i{}_j$ can be interpreted as \emph{always} acting as
\eqref{eq:SU2Action}. In particular, the $\rm SU(2)$ connection
can be nicely rewritten as
\begin{align}
\cV_M{}^i{}_j I^j{}_i &= -\cV_M{}^{++} D^{--} + \cV_M{}^0 D^0 + \cV_M{}^{--} D^{++}~, \eol
\cV_M{}^{\pm\pm} &:= v_i^\pm v_j^\pm \cV_M{}^{ij}~, \qquad
\cV_M{}^{0} := v_i^+ v_j^- \cV_M{}^{ij}~.
\end{align}
Note that $\cV_M{}^{\pm\pm}$ and $\cV_M{}^0$ do not transform
as scalar functions under the $\rm SU(2)$ isometry, but rather as connections,
\begin{subequations}
\begin{align}
\delta_\l \cV_M{}^{++} &= \pa_M \l^{++} - 2 \cV_M{}^0 \l^{++} + 2 \cV_M{}^{++} \l^0 ~, \\
\delta_\l \cV_M{}^{--} &= \pa_M \l^{--} + 2 \cV_M{}^0 \l^{--} - 2 \cV_M{}^{--} \l^0 ~, \\
\delta_\l \cV_M{}^0 &= \pa_M \l^0 + \cV_M{}^{++} \l^{--} - \cV_M{}^{--} \l^{++}~.
\end{align}
\end{subequations}
This is exactly how the $\rm SU(2)$ vielbeins
$\cV_{\ul m}{}^{\pm\pm}$ and $\cV_{\ul m}{}^0$
transform under $\rm SU(2)$ diffeomorphisms (see \eqref{eq:deltacV}) but with
the arbitrary $\xi^{\pm\pm}$, $\xi^0$ parameters replaced with
$\l^{\pm\pm}$, $\l^0$.
Before interpreting this further, let us make a few additional comments.

The implicit expression \eqref{eq:defNabla} for the covariant
derivative can be rewritten
\begin{align}\label{eq:defNabla3}
\pa_M &= E_M{}^{\ul\alpha -} \nabla_{\ul\alpha}^+
	- E_M{}^{\ul\alpha +} \nabla_{\ul\alpha}^-
	+ E_M{}^a \nabla_a
	- \cV_M{}^{++} D^{--}
	+ \cV_M{}^{--} D^{++}
	+ \cV_M{}^0 D^0
	\eol & \quad
	+ \frac{1}{2} \Omega_M{}^{ab} M_{ba}
	+ A_M \bbA
	+ B_M \bbD
	+ F_M{}^{\ul\alpha +} S_{\ul\alpha}^-
	- F_M{}^{\ul\alpha -} S_{\ul\alpha}^+
	+ F_M{}^a K_a~,
\end{align}
where we use
\begin{align}\label{eq:CentralDefsEFOps}
E_M{}^{\ul\alpha \pm} = E_M{}^{\ul \alpha i} v_i{}^{\pm}~, \qquad
\nabla_{\ul\alpha}^\pm = v_i^\pm \nabla_{\ul\alpha}{}^i~, \qquad
F_M{}^{\ul\alpha \pm} = F_M{}^{\ul\alpha i} v_i^\pm~, \qquad
S_{\ul\alpha}^\pm = v_i^\pm S_{\ul\alpha}{}^i ~.
\end{align}
for the spinor vielbeins, $S$-supersymmetry connections, and their
corresponding operators. We have introduced a new compact notation
\begin{align}
\psi^{\ul \alpha} = (\psi^\alpha, \bar\psi^\dalpha)~, \qquad
\psi_{\ul\alpha} = (\psi_\alpha, \bar\psi_\dalpha)
\end{align}
to deal collectively with the left and right-handed vielbeins, spinor derivatives, etc.
It is helpful to introduce some further notation to simplify the
first line of \eqref{eq:defNabla3}. As in the previous section, we wish to
treat the $\pm\pm$ and $0$ indices of the $\rm SU(2)$ derivatives as
tangent space indices and to lower them using the same
conventions \eqref{eq:SU2LowerConv}, with $D_{\ul a} := (D_{++}, D_{--}, D_0)$.
It will also be useful to introduce a convention for
lowering the $\pm$ on $\nabla_{\ul\alpha}^\pm$, and similarly for the
$S$-supersymmetry generator:
\begin{align}
\nabla_{\ul\alpha -} := \nabla_{\ul\alpha}^+~, \qquad
\nabla_{\ul\alpha +} := -\nabla_{\ul\alpha}^-~, \qquad
S_{\ul\alpha -} := -S_{\ul\alpha}^+~, \qquad
S_{\ul\alpha +} := S_{\ul\alpha}^-~.
\end{align}
Now introducing $\nabla_A  = (\nabla_a, \nabla_{\ul\alpha \pm})$
and $K_A = (K_a, S_{\ul\alpha \pm})$, we can rewrite \eqref{eq:defNabla3} as
\begin{align}\label{eq:defNabla2.5}
\pa_M &= E_M{}^{A} \nabla_A
	+ \cV_M{}^{\ul a} D_{\ul a}
	+ \frac{1}{2} \Omega_M{}^{ab} M_{ba}
	+ A_M \bbA
	+ B_M \bbD
	+ F_M{}^A K_A~.
\end{align}
Recalling that the partial derivatives $\pa_{\ul m}$ 
can be written in a similar way,
\begin{align}\label{eq:defHarm2.5}
\pa_{\ul m} = \cV_{\ul m}{}^{\ul a} D_{\ul a}
	= - \cV_{\ul m}{}^{++} D^{--} + \cV_{\ul m}{}^0 D^0 + \cV_{\ul m}{}^{--} D^{++}~,
\end{align}
a new unified notation becomes apparent. Let $z^{\ul M}$ denote the full
set of coordinates $z^{\ul M} = (z^M, y^{\ul m})$ and introduce a
unified covariant derivative $\nabla_{\ul A} = (\nabla_A, D_{\ul a})$.
Then \eqref{eq:defNabla2.5} and \eqref{eq:defHarm2.5} can be written
\begin{align}\label{eq:defNabla4}
\pa_{\ul M} &= E_{\ul M}{}^{\ul A} \nabla_{\ul A}
	+ \frac{1}{2} \Omega_{\ul M}{}^{ab} M_{ba}
	+ A_{\ul M} \bbA
	+ B_{\ul M} \bbD
	+ F_{\ul M}{}^A K_A~,
\end{align}
where the full supervielbein is given by
\begin{align}\label{eq:vielbeinCentral}
E_{\ul M}{}^{\ul A}
	&=
\begin{pmatrix}
E_M{}^A & \cV_M{}^{++} & \cV_M{}^{--} & \cV_M{}^0 \\[0.2em]
0 & \cV_\z{}^{++} & \cV_\z{}^{--} & \cV_\z{}^0 \\[0.2em]
0 & \cV_{\bar\z}{}^{++} & \cV_{\bar\z}{}^{--} & \cV_{\bar\z}{}^0 \\[0.2em]
0 & \cV_\psi{}^{++} & \cV_\psi{}^{--} & \cV_\psi{}^0
\end{pmatrix}
\end{align}
and the other connections live purely on $\cM^{4|8}$,
\begin{align}\label{eq:connectionsCentral}
\Omega_{\ul M}{}^{ab} &= (\Omega_M{}^{ab}, 0, 0, 0)~, \quad
A_{\ul M} = (A_M, 0, 0, 0)~, \eol
B_{\ul M} &= (B_M, 0, 0, 0)~, \quad
F_{\ul M}{}^A = (F_{M}{}^A, 0, 0, 0)~.
\end{align}
This rearrangement is equivalent to that proposed in \cite{HoweHartwell:Npq}.

These identifications \eqref{eq:vielbeinCentral} are completely consistent
so long as two conditions are obeyed.
First, the only $\rm SU(2)$ diffeomorphisms that we may perform are
those that are isometries on the $\rm SU(2)$ manifold. Then the full
$\rm SU(2)$ vielbeins $\cV^{\ul a} = \rd z^{\ul M} \cV_{\ul M}{}^{\ul a}$
transform as \eqref{eq:deltacV} with the special choice of
$\xi^{\ul a} = \l^{\ul a}$ with $\l^i{}_j$ depending on $z^M$ alone.
Second, the only $z^M$ diffeomorphisms and other gauge transformations
(i.e. Lorentz, ${\rm U}(1)_R$, $S$-supersymmetry and special conformal) that
are allowed are those that do not depend on $v^{i\pm}$. This ensures
the zeros in the identifications \eqref{eq:vielbeinCentral} and \eqref{eq:connectionsCentral}
as well as the decompositions \eqref{eq:CentralDefsEFOps}.

As a final check, we can invert \eqref{eq:defNabla4} to find the covariant derivative $\nabla_{\ul A}$:
\begin{subequations}
\begin{align}
\nabla_{A} &= E_A{}^M \Big(
	\pa_M - \cV_M{}^{\ul a} D_{\ul a}
	- \frac{1}{2} \Omega_M{}^{bc} M_{cb}
	- A_M \bbA
	- B_M \bbD
	- F_M{}^B K_B \Big)~, \\
\nabla_{\ul a} &\equiv D_{\ul a} = \cV_{\ul a}{}^{\ul m} \pa_{\ul m}~.
\end{align}
\end{subequations}
It is easy to see that the first equation exactly matches expression \eqref{eq:defNabla2}

The algebra of the redefined operators retains its original form but with minor modifications
involving the exchange of e.g. $I^i{}_j$ for $D^{\pm\pm}$ and $D^0$ and
$S_\beta{}^i$ for $S_\beta{}^\pm$. For example,
the algebra of the $S$-supersymmetry generators with the spinor derivatives is now
given by
\begin{align}
\{S_\beta^\pm, \nabla_\alpha^\pm\} &= \pm 4 \eps_{\beta \alpha} D^{\pm\pm}, \qquad
\{\bar S^{\dbeta \pm}, \bar \nabla^{\dalpha \pm}\} = \mp 4 \eps^{\dbeta \dalpha} D^{\pm\pm}~, \eol
\{S_\beta^\mp, \nabla_\alpha^\pm\} &= \pm (2 \eps_{\beta \alpha} \bbD - 2 M_{\beta \alpha}
     - i \eps_{\beta \alpha} \bbA) - 2 \eps_{\beta \alpha} D^0~, \eol
\{\bar S^{\dbeta \mp}, \bar \nabla^{\dalpha \pm}\} &= \mp (2 \eps^{\dbeta \dalpha} \bbD - 2 M^{\dbeta \dalpha}
     + i \eps^{\dbeta \dalpha} \bbA) + 2 \eps^{\dbeta \dalpha} D^0~.
\end{align}
The full algebra of operators will be given in a general gauge in the next subsection.

It is evident that starting from this formulation of conformal superspace on
$\cM^{4|8} \times {\rm SU}(2)$, there is no intrinsic barrier to performing $v^{i\pm}$-dependent
gauge transformations and diffeomorphisms. These will move us away from the
original gauge where \eqref{eq:CentralDefsEFOps}, \eqref{eq:vielbeinCentral},
and \eqref{eq:connectionsCentral}
hold and where the $\rm SU(2)$ vielbein $\cV_{\ul m}{}^{\ul a}$ takes the
simple form \eqref{eq:cVharm}.
Of course, we can always return to this gauge. We refer to it as
the \emph{central gauge} (or central basis) in analogy with the terminology
employed within the harmonic superspace literature.\footnote{Note that the central
gauge is not unique; any harmonic-independent gauge transformation, $z^M$-diffeomorphism
or $\rm SU(2)$ isometry will take us from one central gauge to another.} In the next section,
we will extend this construction to a completely general gauge.

\subsection{Conformal superspace on $\cM^{4|8} \times {\rm SU}(2)$: The top-down construction}
In contrast to the preceding treatment where we spliced together $\rm SU(2)$ with
the supermanifold $\cM^{4|8}$ of conformal superspace, we can simply postulate the
structure of the new superspace $\cM^{7|8} = \cM^{4|8} \times {\rm SU}(2)$ and impose all
the relevant constraints. This will have the benefit of not requiring that we
begin in central gauge, although central gauge always remains a possibility.

The supermanifold $\cM^{7|8} = \cM^{4|8} \times {\rm SU}(2)$ possesses local coordinates
$z^{\ul M} = (z^M, y^{\ul m}) = (x^m, \q^{\ul\mu \pm}, \z, \bar \z, \psi)$. For convenience,
we have chosen to label the Grassmann coordinates $\q^{\ul \mu \,\imath}$ by $\imath = \pm$
to facilitate a later discussion of analytic gauge. (We emphasize that $\imath$ is
a world index and so does not correspond to any notion of charge; we could
just as well have labeled the coordinates $\q^{\ul \mu 1}$ and $\q^{\ul \mu 2}$.)

The covariant derivatives
$\nabla_{\ul A} = (\nabla_A, \nabla_{\ul a}) = (\nabla_a, \nabla_{\ul \alpha \pm}, \nabla_{\pm\pm}, \nabla_0)$
are defined implicitly by
\begin{align}
\pa_{\ul M} = E_{\ul M}{}^{\ul A} \nabla_{\ul A}
	+ \frac{1}{2} \Omega_{\ul M}{}^{ab} M_{ba}
	+ A_{\ul M} \bbA
	+ B_{\ul M} \bbD
	+ F_{\ul M}{}^A K_A~.
\end{align}
The supervielbein is required to be invertible and its components can be labeled as
\begin{align}
E_{\ul M}{}^{\ul A}
	&=
\begin{pmatrix}
E_M{}^A & E_M{}^{++} & E_M{}^{--} & E_M{}^0 \\[0.2em]
E_\z{}^A & E_\z{}^{++} & E_\z{}^{--} & E_\z{}^0 \\[0.2em]
E_{\bar\z}{}^A & E_{\bar\z}{}^{++} & E_{\bar\z}{}^{--} & E_{\bar\z}{}^0 \\[0.2em]
E_\psi{}^A & E_\psi{}^{++} & E_\psi{}^{--} & E_\psi{}^0
\end{pmatrix}~.
\end{align}
We make no assumptions about whether the vielbeins and connections are
globally defined on $\cM^{4|8} \times \rm SU(2)$.
In fact, we generically need (at least) two charts
for $\rm SU(2)$.

We introduce a prescription for raising the $\pm$ tangent space indices,
\begin{align}
\nabla_{\ul\alpha \mp} = \pm \nabla_{\ul\alpha}^\pm~, \qquad
\nabla_{\mp\mp} = \pm \nabla^{\pm \pm}~, \qquad
S_{\ul\alpha \mp} = \mp S_{\ul\alpha}^\pm~, \qquad
\nabla_0 = \nabla^0~,
\end{align}
so that they correspond to the $\nabla^0$ charge of the operator.
Now let us summarize the algebra of the various operators.
The Lorentz generator is normalized to obey
\begin{subequations}\label{eq:LorentzAlg}
\begin{align}
[M_{ab}, M_{cd}] &= \eta_{bc} M_{ad} - \eta_{ac} M_{bd} - \eta_{bd} M_{ac} + \eta_{ad} M_{bc} \\
[M_{ab}, \nabla_c] &= \eta_{bc} \nabla_a - \eta_{ac} \nabla_b \\
[M_{ab}, \nabla_\gamma^\pm] &= {(\sigma_{ab})_\gamma}^{\beta} \nabla_\beta^\pm, \qquad
[M_{ab}, \bar \nabla^\dgamma{}^\pm] = {(\bsigma_{ab})^\dgamma}_{\dbeta} \bar \nabla^\dbeta{}^\pm~.
\end{align}
\end{subequations}
The action of the dilatation and ${\rm U}(1)_R$ generators is given by
\begin{subequations}
\begin{alignat}{3}
[\bbD, \nabla_\alpha^\pm] &= \frac{1}{2} \nabla_\alpha^\pm~, &\qquad
[\bbD, \bar \nabla^\dalpha{}^\pm] &= \frac{1}{2} \bar \nabla^{\dalpha \pm}~, &\qquad
[\bbD, \nabla_a] &= \nabla_a~, \\
[\bbD, S_{\alpha}^\pm] &= -\frac{1}{2} S_{\alpha}^\pm, &\qquad
[\bbD, \bar S^{\dalpha \pm}] &= -\frac{1}{2} \bar S^{\dalpha \pm} &\qquad
[\bbD,K_a] &= -K_a~, \\
[\bbA, \nabla_\alpha^\pm] &= -i \nabla_\alpha^\pm~, &\qquad
[\bbA, \bar \nabla^\dalpha{}^\pm] &= +i \bar \nabla^\dalpha{}^\pm~, \\
[\bbA, S_{\alpha}^\pm] &= +i S_{\alpha}^\pm~, &\qquad
[\bbA, \bar S^{\dalpha \pm}] &= -i \bar S^{\dalpha \pm}~.
\end{alignat}
\end{subequations}
The special conformal and $S$-supersymmetry generators obey
\begin{subequations}\label{eq:SKalg}
\begin{align}
[K_a, \nabla_b] &= 2 \eta_{ab} \bbD - 2 M_{ab} ~,\\
\{S_\beta^\pm, \nabla_\alpha^\pm\} &= \pm 4 \eps_{\beta \alpha} \nabla^{\pm\pm}, \qquad\qquad
\{\bar S^{\dbeta \pm}, \bar \nabla^{\dalpha \pm}\} = \mp 4 \eps^{\dbeta \dalpha} \nabla^{\pm\pm}~, \\
\{S_\beta^\mp, \nabla_\alpha^\pm\} &= \pm (2 \eps_{\beta \alpha} \bbD - 2 M_{\beta \alpha}
     - i \eps_{\beta \alpha} \bbA) - 2 \eps_{\beta \alpha} \nabla^0 ~,\\
\{\bar S^{\dbeta \mp}, \bar \nabla^{\dalpha \pm}\} &= \mp (2 \eps^{\dbeta \dalpha} \bbD - 2 M^{\dbeta \dalpha}
     + i \eps^{\dbeta \dalpha} \bbA) + 2 \eps^{\dbeta \dalpha} \nabla^0 ~,\\
[K_a, \nabla_\alpha^\pm] &= i (\sigma_a)_{\alpha \dbeta} \,\bar S^{\dbeta \pm}~, \qquad\qquad
[K_a, \bar \nabla^\dalpha{}^\pm] = i (\bsigma_a)^{\dalpha \beta} \,S_{\beta}^\pm~, \\
[S_{\alpha}^\pm, \nabla_a] &= i (\sigma_a)_{\alpha \dbeta} \,\bar \nabla^{\dbeta \pm}~, \qquad\qquad
[\bar S^{\dalpha \pm}, \nabla_a] = i (\bsigma_a)^{\dalpha \beta} \,\nabla_{\beta}{}^\pm~,
\end{align}
\end{subequations}
and
\begin{align}
[\nabla^{\pm\pm}, S_{\ul\alpha}^{\pm}] = 0~, \qquad
[\nabla^{\mp\mp}, S_{\ul\alpha}^{\pm}] = S_{\ul\alpha}^{\mp}~, \qquad
[\nabla^0, S_{\ul\alpha}^{\pm}] = \pm S_{\ul\alpha}^{\pm}~.
\end{align}

Up to this point, we have only been discussing the algebra of the gauge generators
with themselves and with the covariant derivatives $\nabla_{\ul A}$.
These dictate how the connections transform under the corresponding symmetries.
(An explicit discussion of this can be found, for example, in \cite{Butter:CSG4d_2}.)
What remains is to specify the algebra of the covariant derivatives themselves,
corresponding to the torsion and curvatures on the supermanifold.
The various constraints imposed will dictate the supergeometry.

We begin by specifying the algebra of the
$\rm SU(2)$ covariant derivatives with the spinor derivatives:
\begin{subequations}\label{eq:algSU2}
\begin{alignat}{3}
[\nabla^{++}, \nabla^{--}] &= \nabla^0~, &\qquad
[\nabla^{0}, \nabla^{\pm\pm}] &= \pm 2 \nabla^{\pm\pm}~, \\
[\nabla^{\pm\pm}, \nabla_{\ul\alpha}^{\pm}] &= 0~, &\qquad
[\nabla^{\mp\mp}, \nabla_{\ul\alpha}^{\pm}] &= \nabla_{\ul\alpha}^{\mp}~, &\qquad
[\nabla^0, \nabla_{\ul\alpha}^{\pm}] &= \pm \nabla_{\ul\alpha}^{\pm}~.
\end{alignat}
\end{subequations}
These conditions imply that the $\rm SU(2)$ part of the
manifold is flat, possessing only (constant) torsion and no curvature,
and are necessary for the existence of a central gauge where
the $\rm SU(2)$ manifold (almost) decouples. In other words,
if we did not impose these constraints, then we would be introducing
new degrees of freedom that were absent before.
For the algebra of the spinor covariant derivatives, we impose the
analyticity constraint
\begin{align}\label{eq:AnalyticCommutes}
\{\nabla_{\ul \alpha}^\pm, \nabla_{\ul \beta}^\pm\} = 0~.
\end{align}
This is an integrability condition for the existence of analytic superfields,
which we will discuss shortly.
Now we find that the remainder of the dimension-1 curvatures can be written
\begin{align}\label{eq:defCurvSpinor}
\{\nabla_\alpha^\pm, \bnabla_{\dbeta}^\mp\} = \mp 2i \nabla_{\alpha \dbeta}~, \quad
\{\nabla_\alpha^\pm, \nabla_\beta^\mp\} = \pm 2 \eps_{\alpha \beta} \bar \cW~, \quad
\{\bnabla^{\dalpha \pm}, \bnabla^{\dbeta \mp}\} = \pm 2 \, \eps^{\dalpha \dbeta} \cW~.
\end{align}
The first equation of \eqref{eq:defCurvSpinor} is a conventional constraint
and serves to define $\nabla_{\alpha\dbeta} = (\sigma^a)_{\alpha \dbeta} \nabla_a$.
As a consequence, the vector
covariant derivative has vanishing algebra with the $\rm SU(2)$ derivatives,
$[\nabla^{\pm\pm}, \nabla_a] = [\nabla^0, \nabla_a] = 0$, and obeys the
other algebraic properties given in eqs. \eqref{eq:LorentzAlg} -- \eqref{eq:SKalg}.
The second and third equations involve a chiral primary operator $\cW$ and
its conjugate antichiral primary operator $\bar\cW$, which are constrained by
\begin{gather}
[\nabla^{\pm\pm}, \cW] = [\nabla^{0}, \cW] = [\bnabla_\dalpha^\pm, \cW] = 0~, \qquad
[\nabla^{\pm\pm}, \bar\cW] = [\nabla^{0}, \bar\cW] = [\nabla_\alpha^\pm, \bar \cW] = 0~, \eol
\{\nabla^{\gamma +}, [\nabla_\gamma^+, \cW]\} = \{\bnabla_\dgamma^+, [\bnabla^{\dgamma +}, \bar\cW]\}~.
\label{eq:Wconstraints}
\end{gather}
The solution corresponding to conformal superspace involves specifying $\cW$
in terms of a superfield $W_{\alpha\beta}$,
\begin{subequations}
\begin{align}
\cW &= \frac{1}{2} W^{\alpha \beta} M_{\beta \alpha}
	+ \frac{1}{4} \nabla^{\beta +} W_{\beta}{}^\alpha S_{\alpha}^-
	- \frac{1}{4} \nabla^{\beta -} W_{\beta}{}^\alpha S_{\alpha}^+
     + \frac{1}{4} \nabla^{\dalpha \beta} W_\beta{}^\alpha K_{\alpha \dalpha}~, \\
\bar \cW &= \frac{1}{2} \bar W_{\dalpha \dbeta} M^{\dbeta \dalpha}
     + \frac{1}{4} \bar \nabla_{\dbeta}^- \bar W^{\dbeta}{}_\dalpha \bar S^{\dalpha +}
	- \frac{1}{4} \bar \nabla_{\dbeta}^+ \bar W^{\dbeta}{}_\dalpha \bar S^{\dalpha -}
     + \frac{1}{4} \nabla_{\alpha \dbeta} \bar W^\dbeta{}_\dalpha K^{\dalpha \alpha}~.
\end{align}
\end{subequations}
These operators obey \eqref{eq:Wconstraints} provided
$W_{\alpha\beta}$ is primary and obeys the constraints
\begin{align}
\nabla^{\pm \pm} W_{\alpha\beta} = \nabla^0 W_{\alpha\beta} = \bar\nabla_{\dgamma}^\pm W_{\alpha\beta} = 0~, \qquad
\nabla^{\alpha\beta} W_{\alpha\beta} = \bar\nabla^{\dalpha \dbeta} \bar W_{\dalpha \dbeta}
\end{align}
where we have introduced the abbreviations
\begin{align}
\nabla^{\alpha\beta} := 2 \nabla^{(\alpha +} \nabla^{\beta) -} 
	= - 2 \nabla^{(\alpha -} \nabla^{\beta) +}~, \qquad
\bar\nabla^{\dalpha\dbeta} := 2 \bar\nabla^{(\dalpha +} \bar\nabla^{\dbeta) -}
	= - 2 \bar\nabla^{(\dalpha -} \bar\nabla^{\dbeta) +}~.
\end{align}
In other words, $W_{\alpha\beta}$ is a chiral primary superfield
inert under covariant $\rm SU(2)$ derivatives.

The dimension-3/2 curvatures can be written
\begin{align}
[\nabla_{\beta}^\pm, \nabla_{\alpha \dalpha}] = -2 \eps_{\beta \alpha} \bar\cW_\dalpha^\pm~, \qquad
[\bar\nabla_{\dbeta}^\pm, \nabla_{\alpha \dalpha}] = -2 \eps_{\dbeta \dalpha} \cW_\alpha^\pm~.
\end{align}
The operators $\cW_{\ul\alpha}^{\pm}$ are given by
$\cW_{\alpha}^{\pm} = -\frac{i}{2} [\nabla_{\alpha}^\pm,\cW]$
and
$\bar\cW_{\dalpha}^{\pm} = -\frac{i}{2} [\bar\nabla_{\dalpha}^\pm,\cW]$:
\begin{align}
\cW_\alpha^+ &=
	-\frac{i}{8} \nabla^{\dgamma \beta} \nabla_\alpha^+ W_\beta{}^\gamma K_{\gamma \dgamma}
	+ \frac{i}{16} (\nabla^+)^2 W_\alpha{}^\gamma S_\gamma^-
	+ \frac{i}{8} \nabla_\alpha^+ \nabla^{\beta -} W_\beta{}^\gamma S_\gamma^+
	+ \frac{1}{4} \nabla_\dgamma{}^\beta W_{\beta \alpha} \bar S^{\dgamma +}
	\eol & \quad
	- \frac{i}{4} \nabla^{\beta +} W^\gamma{}_\alpha M_{\gamma \beta}
	- \frac{i}{4} \nabla^{\beta +} W_{\beta \alpha} (\mathbb D - \frac{i}{2} \mathbb A - \nabla^0)
	+ \frac{i}{2} \nabla^{\beta -} W_{\beta \alpha} \nabla^{++}
	+ \frac{i}{2} W_{\alpha}{}^{\beta} \nabla_\beta^+~, \eol
\bar \cW_{\dalpha}^+ &=
	+ \frac{i}{8} \nabla^{\dbeta\gamma} \bar\nabla_\dalpha^{+} \bar W_\dbeta{}^\dgamma K_{\gamma\dgamma}
	+ \frac{i}{16} (\bar\nabla^+)^2 \bar W_\dalpha{}^\dgamma \bar S_{\dgamma}^-
	- \frac{i}{8} \bar\nabla_\dalpha^{+} \bar\nabla^{\dbeta-} \bar W_\dbeta{}^\dgamma \bar S_{\dgamma}^+
	+ \frac{1}{4} \nabla_\gamma{}^\dbeta \bar W_{\dbeta \dalpha} S^{\gamma +}
	\eol & \quad
	- \frac{i}{4} \bar\nabla_{\dbeta}^+ \bar W_{\dgamma\dalpha} \bar M^{\dgamma \dbeta}
	+ \frac{i}{4} \bar\nabla^{\dbeta+}\bar W_{\dbeta\dalpha}(\mathbb D + \frac{i}{2} \mathbb A - \nabla^0)
	- \frac{i}{2} \bar\nabla^{\dbeta-} \bar W_{\dbeta\dalpha} \nabla^{++}
	- \frac{i}{2} \bar W_{\dalpha}{}^\dbeta \bar\nabla_\dbeta^{+}~, \eol
\cW_{\ul\alpha}^- &= [\nabla^{--}, \cW_{\ul\alpha}^+]~.
\end{align}
Above we have introduced
$(\nabla^\pm)^2 := \nabla^{\gamma \pm} \nabla_\gamma^\pm$
and
$(\bnabla^\pm)^2 := \bnabla_\dgamma^{\pm} \bnabla^\dgamma{}^\pm$.
Note that these operators obey the rules
$[\nabla^{\pm\pm}, \cW_{\ul\alpha}^\pm] = 0$ and
$[\nabla^{\pm\pm}, \cW_{\ul\alpha}^\mp] = \cW_{\ul\alpha}^\pm$
as a consequence of \eqref{eq:Wconstraints}.

The dimension-2 curvatures $[\nabla_b, \nabla_a]$ are a bit more complicated.
Writing
\begin{align}
[\nabla_{\beta \dbeta}, \nabla_{\alpha \dalpha}] = -\cF_{\beta \dbeta\, \alpha \dalpha}
     = -2 \eps_{\dbeta \dalpha} \cF_{\sym{\beta \alpha}}
     + 2 \eps_{\beta \alpha} \cF_{\sym{\dbeta \dalpha}}~,
\end{align}
the anti-selfdual and selfdual components of $\cF_{ba}$ are given by
\begin{align}
\cF_{\sym{\beta \alpha}} = \frac{1}{4} \{\nabla_{(\beta}^+,[\nabla_{\alpha)}^-, \cW]\}~, \qquad
\cF_{\sym{\dbeta \dalpha}} = \frac{1}{4} \{\bar\nabla_{(\dbeta}^+,[\bar\nabla_{\dalpha)}^-, \bar \cW]\}.
\end{align}
The curvatures $\cF_{ba}$ must be invariant under the $\rm SU(2)$
derivatives, $[\nabla^{\pm\pm}, \cF_{ba}] = [\nabla^{0}, \cF_{ba}] = 0$.
The explicit expressions for $\cF_{ba}$
won't be of much use to us here, so we will not
discuss them explicitly. Instead, we collect them, along with the other
curvatures, in appendix \ref{app:Curvatures}.

We note that under the generalized $\widetilde{\phantom{a}}$ conjugation,
the derivatives transform as in \cite{GIOS}:
\begin{align}
\widetilde{\nabla_{\alpha}^\pm} = -\bar\nabla_{\dalpha}^\pm~, \qquad
\widetilde{\bar\nabla_{\dalpha}^\pm} = \nabla_{\alpha}^\pm~.
\end{align}

The superspace we have constructed here admits a full set of gauge transformations,
\begin{align}
\delta = \xi^{\ul A} \nabla_{\ul A} + \frac{1}{2} \l^{ab} M_{ba}
	+ \l \,\bbA + \L \,\bbD + \eps^A K_A~,
\end{align}
where each of the parameters may depend arbitrarily on the coordinates $z^{\ul M}$.

Now let us argue that we can always recover the central gauge of the previous section.
Because it is obvious that we can always start from the central gauge in constructing
$\cM^{4|8} \times \rm SU(2)$, we will only give a sketch of a proof.
As a consequence of the algebra \eqref{eq:algSU2}, one can always adopt
a gauge where $\nabla^{\pm\pm}$ and $\nabla^0$ are given by their forms in the
central gauge in terms of $v^{i\pm}$. This implies that the superspace vielbein
takes the form \eqref{eq:vielbeinCentral} and the other connections the form
\eqref{eq:connectionsCentral}. It is easy to prove that
$\Omega_M{}^{ab}$, $A_M$, $B_M$ and $F_M{}^a$ are independent of the $\rm SU(2)$
coordinates: one merely needs that the corresponding curvature components
$R_{\ul n M}$ all vanish in this gauge.
For the $S$-supersymmetry connection $F_M{}^{\ul\alpha \pm}$,
the vanishing of $R(S)_{\ul n M}{}^{\ul\alpha \pm}$ implies that
$F_M{}^{\ul\alpha \pm} = v_i^\pm F_M{}^{\ul \alpha i}$ as expected.
For the vielbein $E_M{}^A$, similar arguments imply that $E_M{}^a$ is
harmonic independent while $E_M{}^{\ul\alpha \pm} = E_M{}^{\ul\alpha i} v_i^\pm$.
Finally, a similar argument with $\cV_M{}^{\ul a}$ establishes that they
are given by $\cV_M{}^{\pm\pm} = \cV_M{}^{ij} v_i^\pm v_j^\pm$
and $\cV_M{}^{0} = \cV_M{}^{ij} v_i^+ v_j^-$.

\subsection{Consequences of analyticity}\label{sec:AnalyticSuperfields}
In this paper, we will not present specific actions (e.g. explicit models involving
hypermultiplets), so we will not have much need for an extended
discussion of the types of superfields possible in this superspace.
However, it is clear that if we wish to use the superspace
$\cM^{4|8} \times \rm SU(2)$ for projective multiplets like those discussed
in the introduction, then we must discuss (at least briefly) the consequences
of imposing analyticity on superfields.

Due to the integrability conditions \eqref{eq:AnalyticCommutes},
it is admissible to have primary analytic superfields $\Psi$,
\begin{align}
S_{\ul\alpha}^{\pm} \Psi = K_a \Psi = 0~, \qquad \nabla_{\ul\alpha}^+ \Psi = 0~.
\end{align}
Consistency with the algebra implies that $\Psi$ is a Lorentz scalar,
invariant under ${\rm U}(1)_R$, and obeys
\begin{align}\label{eq:AnalyticConsequences}
\nabla^0 \Psi = \bbD\Psi, \qquad \nabla^{++} \Psi = 0~.
\end{align}
The first condition tells us that
$\Psi$ must have a $\nabla^0$ charge equal to its conformal dimension.
For definiteness, let us denote both quantities by $n$.
The second condition ensures that,
in the central gauge, $\Psi$ is a holomorphic tensor on (an open domain of)
$\mathbb CP^1$.
These are exactly the same conditions (up to the
complexification discussed in section \ref{sec:SU2Geo.complex}) as those
for admissible projective multiplets $\cQ^{(n)}$ in the usual formulation of
projective superspace \cite{KLRT-M1, KLRT-M2, KT-M:DiffReps}.
These conditions also match those found for superconformal projective
multiplets in flat projective superspace \cite{Kuzenko:5Dallthat}.

\section{Superspace action principles on $\cM^{4|8} \times \rm SU(2)$}\label{sec:SuperActions}
The original supermanifold $\cM^{4|8}$ came equipped with two natural action principles,
involving respectively integrals over the full superspace and the
chiral superspace,
\begin{align}
\int \rd^4x\, \rd^4\q\, \rd^4\bar\q\, E\, \mathscr{L}~, \qquad
\int \rd^4x\, \rd^4\q\, \cE\, \mathscr{L}_c~.
\end{align}
Here $E$ and $\cE$ were defined respectively as
\begin{align}
E = \sdet E_M{}^A~, \qquad
\cE = \sdet
\begin{pmatrix}
E_m{}^a & E_{m}{}^\imath\,{}^\alpha{}_i \\
E_{\mu}{}^\imath\,{}^a & E_{\mu}{}^\imath\,{}^\alpha{}_i
\end{pmatrix}~;
\end{align}
the superspace Lagrangian $\mathscr{L}$ was required to be a conformal primary
scalar superfield of vanishing dilation and ${\rm U}(1)_R$ weight, inert
under ${\rm SU}(2)_R$,
\begin{align}
\bbD \mathscr{L} = \bbA \mathscr{L} = I^i{}_j \mathscr{L} = K_A \mathscr{L} = 0~;
\end{align}
and the chiral Lagrangian $\mathscr{L}_c$ was required to be a conformal
primary chiral scalar superfield, inert under ${\rm SU}(2)_R$, with certain
weights,
\begin{align}
\bbD \mathscr{L}_c = 2 \mathscr{L}_c~, \qquad
\bbA \mathscr{L}_c = 4i \mathscr{L}_c~, \qquad
\bar\nabla^\dalpha{}_i \mathscr{L}_c = I^i{}_j \mathscr{L}_c = K_A \mathscr{L}_c = 0~.
\end{align}
These properties of the respective Lagrangians can be proven, for example,
by applying the results of appendix \ref{app:Integration}.

Now that we have extended our superspace to $\cM^{4|8} \times \rm SU(2)$,
other possibilities emerge. The ones we will discuss below fall into
three classes: full superspace integrals involving integrals over both
$S^2$ and over a contour $\cC$; analytic superspace integrals over
a contour $\cC$; and chiral-analytic superspace integrals over
a contour $\cC$.

\subsection{Full superspace integrals}
It is straightforward to extend the full superspace action to include
an integral over $\rm SU(2) / \rm U(1)$. In the central basis, we can take
\begin{align}\label{eq:PHarmFullSuper} 
\int \rd^4x\, \rd^4\q\, \rd^4\bar\q\, E\, \int \rd v \mathscr{L}^0~, \qquad
\end{align}
where $\rd v$ is an abbreviation for the standard measure on the $S^2$,
\begin{align}
\rd v := \frac{i}{2\pi} \frac{\rd \z \wedge \rd \bar \z}{(1 + \z \bar \z)^2}~,
\end{align}
and $\mathscr{L}^0$ is assumed to have vanishing $D^0$ charge, vanishing
dilatation and ${\rm U}(1)_R$ weights, and to be globally
defined on $\rm SU(2)$, but otherwise to be unconstrained. In a generic
gauge, this action is written
\begin{align}\label{eq:PHarmFullSuperCentral} 
\int \rd^4x\, \rd^4\q\, \rd^4\bar\q\, \rd^2\z\, E^0\, \mathscr{L}^0~, \qquad
\end{align}
using the abbreviation
$\rd^2 \z:= \frac{i}{2\pi} \rd \z \wedge \rd \bar \z$
for the complex coordinates on the $S^2$. The rest of the usual $S^2$ measure is contained
in the full superspace measure
\begin{align}
E^0 = \sdet \begin{pmatrix}
E_M{}^A & E_M{}^{++} & E_M{}^{--} \\[0.2em]
E_\z{}^A & E_\z{}^{++} & E_\z{}^{--} \\[0.2em]
E_{\bar\z}{}^A & E_{\bar\z}{}^{++} & E_{\bar\z}{}^{--}
\end{pmatrix}~.
\end{align}

The full superspace action can also be extended to to involve an integral over a contour $\cC$.
The most natural choice is a purely holomorphic contour, given in the central gauge by\footnote{This
action principle is used as the universal action principle in the conventional formulation of
projective superspace \cite{KLRT-M1, KLRT-M2, KT-M:DiffReps}. We will discuss shortly why this form
is actually universal.}
\begin{align}\label{eq:ProjFullSuperCentral}
-\frac{1}{2\pi} \int \rd^4x\, \rd^4\q\, \rd^4\bar\q\, E\, \oint_\cC \cV^{++} \mathscr{L}^{--}~,
\end{align}
where $\mathscr{L}^{--}$ has vanishing dilatation and ${\rm U}(1)_R$ weights,
but is required to be holomorphic with $D^0$ charge $-2$,
\begin{align}
D^{++} \mathscr{L}^{--} = 0~, \qquad D^0 \mathscr{L}^{--} = -2 \mathscr{L}^{--}~.
\end{align}
Extending this to a generic gauge is straightforward. Letting $\tau$ be the coordinate
parametrizing the contour, we introduce the action
\begin{align}\label{eq:ProjFullSuper}
- \frac{1}{2\pi} \oint_\cC \rd \tau \int \rd^{4}x\, \rd^4\q\, \rd^4\bar\q\,  E^{++} \, \mathscr{L}^{--}~.
\end{align}
where
\begin{align}
E^{++} = \sdet \begin{pmatrix}
E_M{}^A & E_M{}^{++} \\[0.2em]
E_\tau{}^A & E_\tau{}^{++}
\end{pmatrix}~,
\end{align}
with $E_\tau{}^{\ul A}$ corresponding to the pullback of the one-form $E^{\ul A}$ to the
contour.\footnote{For example,
$E_\tau{}^{++} \equiv \dot \z\, E_\z{}^{++} + \dot {\bar \z}\, E_{\bar \z}{}^{++}$,
where $\dot {\phantom{a}}:= \rd / \rd \tau$.}
Applying the results of appendix \ref{app:Integration}, one can show that
$\mathscr{L}^{--}$ must be a covariantly holomorphic primary superfield
with vanishing dilatation and ${\rm U}(1)_R$ weights and $\nabla^0$ charge $-2$:
\begin{align}
0 = \nabla^{++} \mathscr{L}^{--} = K_A \mathscr{L}^{--} = \bbD \mathscr{L}^{--}
	= \bbA \mathscr{L}^{--}~, \qquad
\nabla^0 \mathscr{L}^{--} = -2 \mathscr{L}^{--}~.
\end{align}

Within projective superspace, the natural quantities are holomorphic on
$\rm SU(2) / U(1)$,
so the action principle \eqref{eq:ProjFullSuperCentral}
(or \eqref{eq:ProjFullSuper} in its generic form)
is more commonly encountered than \eqref{eq:PHarmFullSuper}
(or \eqref{eq:PHarmFullSuperCentral} in its generic form).
In fact, as we will shortly review, the action principle \eqref{eq:ProjFullSuperCentral} 
can also efficiently encapsulate the other relevant action principles
involving integrals over smaller superspaces.
Let us describe these other possibilities next.

\subsection{Analytic superspace integrals}
As discussed in the introduction, the natural action principle in projective superspace
involves a contour integral and a Grassmann integration over $\q^{\mu+} = v_\imath^+ \q^{\mu \imath}$
and $\bar\q^{\dmu +} = v_\imath^+ \bar\q^{\dmu \imath}$.
In flat projective superspace, such actions take the
form \cite{Siegel:ChiralActions, Kuzenko:SPH}
\begin{align}\label{eq:FlatProjAction}
-\frac{1}{2\pi} \oint_\cC v_{i}^+ \rd v^{i +} \int \rd^4x\, \rd^4\q^+\, \mathscr{L}^{++}
	= -\frac{1}{2\pi} \oint_\cC v_{i}^+ \rd v^{i +} \int \rd^4x\, (D^-)^4 \mathscr{L}^{++}~,
\end{align}
where $\mathscr{L}^{++}$ is a holomorphic analytic Lagrangian,
$D^{++} \mathscr{L}^{++} = D_{\ul\alpha}^+ \mathscr{L}^{++} = 0$.

The curved generalization of the analytic superspace integral \eqref{eq:FlatProjAction}
is naturally written
\begin{align}\label{eq:ProjAnalSuper}
- \frac{1}{2\pi} \oint_\cC \rd \tau \int \rd^4x\, \rd^4\q^+ \cE^{--} \mathscr{L}^{++}
\end{align}
where the measure is
\begin{align}
\cE^{--} = 
\sdet \begin{pmatrix}
E_m{}^a & E_m{}^{\ul \alpha +} & E_m{}^{++} \\[0.3em]
E_{\ul\mu +}{}^a & E_{\ul\mu +}{}^{\ul \alpha +} & E_{\ul\mu +}{}^{++} \\[0.3em]
E_\tau{}^a & E_\tau{}^{\ul \alpha +} & E_\tau{}^{++}
\end{pmatrix}~.
\end{align}
The action is invariant under all gauge transformations
provided $\mathscr{L}^{++}$ is a covariantly holomorphic,
analytic, conformal primary superfield,
\begin{align}
\nabla^{++} \mathscr{L}^{++} = \nabla_{\ul \alpha}^+ \mathscr{L}^{++} = K_A \mathscr{L}^{++} = 0~, 
\end{align}
with vanishing ${\rm U}(1)_R$ weight and equal dilatation and $\nabla^0$ weights,
\begin{align}
\bbA \mathscr{L}^{++} = 0~, \qquad
\bbD \mathscr{L}^{++} = \nabla^0 \mathscr{L}^{++} = 2 \mathscr{L}^{++}~.
\end{align}

The integral \eqref{eq:ProjAnalSuper} is the natural action principle in
projective superspace on $\cM^{4|8} \times \rm SU(2)$.  We will discuss its
component reduction in section \ref{sec:CompAction}. For now, we wish to establish the
relationship between analytic superspace actions \eqref{eq:ProjAnalSuper}
and full superspace actions \eqref{eq:ProjFullSuper}. Let us begin
by recalling two relationships between $\cN=1$ full superspace
and chiral superspace integrals, which are respectively written
\begin{align}
\int \rd^4x\, \rd^2\q\, \rd^2\bar\q\, E\, \mathscr{L}~, \qquad
\int \rd^4x\, \rd^2\q\, \cE \mathscr{L}_c~.
\end{align}
The first relationship is that any full superspace integral can be
written as a chiral superspace integral as
\begin{align}\label{eq:DtoF}
\int \rd^4x\, \rd^2\q\, \rd^2\bar\q\, E\, \mathscr{L}
	= -\frac{1}{4} \int \rd^4x\, \rd^2\q\, \cE \,\bar\nabla^2 \mathscr{L}
	= -\frac{1}{4} \int \rd^4x\, \rd^2\q\, \cE \,(\bar \cD^2 - 8 R) \mathscr{L}~.
\end{align}
We have written the chiral integrand in two ways, as both
$\bar\nabla^2 \mathscr{L}$ and as $(\bar \cD^2 - 8 R) \mathscr{L}$; the first
expression is appropriate for $\cN=1$ conformal superspace \cite{Butter:CSG4d}
while the second
involves the conventional formulation of $\cN=1$ Poincar\'e (old minimal)
superspace.\footnote{We use the conventions of \cite{WessBagger}. See also
\cite{GGRS, BK}, where different conventions are employed.}
The second relationship can be written
\begin{align}\label{eq:FtoDC}
\int \rd^4 x\, \rd^4\q\, \cE\, \mathscr{L}_c = -4 \int \rd^4 x \, \rd^4\q\, E\,\frac{X}{\bar \nabla^2 X} \, \mathscr{L}_c
	= -4 \int \rd^4 x \, \rd^4\q\, E\,\frac{X}{(\bar \cD^2 - 8 R) X} \, \mathscr{L}_c
\end{align}
where $X$ is a real primary superfield of dimension two. (The proof follows
by applying \eqref{eq:DtoF} to the right-hand side.)
In this expression,
$\bar\nabla^2 X$ is chiral and primary and so the second integrand
is primary as well as possessing the appropriate dilatation and ${\rm U}(1)_R$ weights.
The third integrand involves the same expression in Poincar\'e (old minimal) superspace.
This last expression is especially useful because we can adopt the Weyl gauge where
$X = 1$, in which case the above equality simplifies to
\begin{align}\label{eq:FtoDP}
\int \rd^4 x\, \rd^4\q\, \cE\, \mathscr{L}_c = \frac{1}{2}\int \rd^4 x \, \rd^4\q\, \frac{E}{R} \, \mathscr{L}_c~.
\end{align}
The two relationships \eqref{eq:DtoF} and \eqref{eq:FtoDP} can be understood as
inverses of each other.

It turns out that two analogous relationships can be constructed between full
superspace and analytic superspace, both over a contour $\cC$.
The first relationship we will establish is the analogue of \eqref{eq:DtoF},
\begin{align}\label{eq:FullToAnalyticProj}
- \frac{1}{2\pi} \oint_\cC \rd \tau \int \rd^{4}x\, \rd^4\q\, \rd^4\bar\q\, E^{++} \, \mathscr{L}^{--}
	= - \frac{1}{2\pi} \oint_\cC \rd \tau \int \rd^{4}x\, \rd^4\q^+\, \cE^{--} \, (\nabla^+)^4 \mathscr{L}^{--}~.
\end{align}
To prove this, we go to analytic gauge where
the covariant derivative $\nabla_{\ul\alpha -} \equiv \nabla_{\ul\alpha}^+$ is simply given by
$\nabla_{\ul \alpha -} = \pa / \pa \q^{\ul\alpha -}$. This is always
possible to do because of the constraints \eqref{eq:AnalyticCommutes}. 
This fixes the gauge up to $\q^{\ul\mu-}$-independent gauge transformations.
In this gauge,
$E^{++}$ is equal to $\cE^{--}$; the difference in apparent $\nabla^0$ charges of the two
quantities arises because in analytic gauge, any $\nabla^0$ gauge transformation must be accompanied
by a special diffeomorphism to maintain that gauge. The integral becomes
\begin{align}
- \frac{1}{32\pi} \oint_\cC \rd \tau \int \rd^{4}x\, \rd^4\q^+
	\pa^{\alpha}_- \pa_{\alpha -} \bar\pa_{\dalpha -} \bar\pa^{\dalpha}_- 
	\Big(\cE^{--} \mathscr{L}^{--} \Big)~.
\end{align}
Now we use the feature that $\cE^{--}$ is itself analytic in this gauge;
this follows from
\begin{align}
\pa_{\ul\alpha -} \cE^{--} = \cE^{--} (\pa_{\ul\alpha -} \cE_{\ul N}{}^{\ul B}) \cE_{\ul B}{}^{\ul N} (-)^N
	= \cE^{--}\, T_{\ul\alpha - \, \ul N}{}^{\ul B} \cE_{\ul B}{}^{\ul N} (-)^N = 0~.
\end{align}
As a result, we find
\begin{align}
- \frac{1}{32\pi} \oint_\cC \rd \tau \int \rd^{4}x\, \rd^4\q^+ \cE^{--}
	\pa^{\alpha}_- \pa_{\alpha -} \bar\pa_{\dalpha -} \bar\pa^{\dalpha}_- 
	\mathscr{L}^{--}~,
\end{align}
with the integrand equal to $(\nabla^+)^2 (\bar\nabla^+)^2 \mathscr{L}^{--}$ in this gauge.
Rewriting the result in a gauge-invariant way, we recover \eqref{eq:FullToAnalyticProj}.

In projective superspace, the expression analogous to \eqref{eq:FtoDC} is
\begin{align}\label{eq:ProjToFull}
- \frac{1}{2\pi} \oint_\cC \rd \tau \int \rd^4x\, \rd^4\q^+ \,\cE^{--} \mathscr{L}^{++}
	= - \frac{1}{2\pi} \oint_\cC \rd \tau \int \rd^4x\, \rd^4\q\, \rd^4\bar\q\, E^{++} \frac{X}{(\nabla^+)^4 X} \mathscr{L}^{++}~,
\end{align}
where $X$ is a real superfield of conformal dimension two and invariant under
the $\rm SU(2)$ derivatives. It is a straightforward
exercise to verify that $(\nabla^+)^4 X$ is a real conformal primary of dimension 4
and so the integrand on the right-hand side is a real primary superfield of vanishing weight.
The advantage of the right-hand side is that it can be formulated directly
in the central gauge. Indeed, an equivalent formulation of the right-hand side
appeared in \cite{KLRT-M1} (mirroring an identical construction in five dimensions
\cite{KT-M:5DSugra2}) where it was used to define analytic integration in the central gauge.
There the particular choice $X = W \bar W$ was made,
where $W$ was an abelian vector multiplet. Moving to the central gauge where
$E^{++} = E \cV_\tau^{++}$, one finds
\begin{align}\label{eq:ProjToFull2}
- \frac{1}{2\pi} \oint_\cC \rd \tau \int \rd^4x\, \rd^4\q^+ \,\cE^{--} \mathscr{L}^{++}
	= - \frac{1}{2\pi} \oint_\cC \cV^{++} \int \rd^4x\, \rd^4\q\, \rd^4\bar\q\, E
		\frac{16 \,W\bar W}{(\nabla^+)^2 W (\bar\nabla^+)^2 \bar W} \mathscr{L}^{++}~.
\end{align}
If one degauges conformal superspace to $\rm SU(2)$ superspace, then
$(\nabla^+)^2 W$ becomes $\big((\cD^+)^2 + 4 S^{++}\big) W$.
Adopting the super-Weyl gauge $W = 1$ leads to the final relation
\begin{align}\label{eq:ProjToFull3}
- \frac{1}{2\pi} \oint_\cC \rd \tau \int \rd^4x\, \rd^4\q^+ \cE^{--} \mathscr{L}^{++}
	= - \frac{1}{2\pi} \oint_\cC \cV^{++} \int \rd^4x\, \rd^4\q\, \rd^4\bar\q\, 
		\frac{E}{(S^{++})^2} \mathscr{L}^{++}~.
\end{align}
The expression on the right is a particularly elegant form of the
analytic action principle \cite{KLRT-M1}. Its advantage is that it permits
easy manipulation in the central gauge. The similarity between this result
and the $\cN=1$ analogue \eqref{eq:FtoDP} is especially striking.

\subsection{Chiral-analytic superspace}
The final superspace action principle we will discuss is a curious one 
because it involves an integration over 3/4 of the Grassmann variables.
Suppose one is given a complex conformal primary Lagrangian $\mathscr{L}^{0}$ which is
\emph{chiral-analytic},
\begin{align}
\bar\nabla_\dalpha^+ \mathscr{L}^{0} = 0~.
\end{align}
Such a Lagrangian would, in the analytic gauge, be independent of $\bar \q^{\dmu -}$.
Provided that the Lagrangian is holomorphic with certain weights,
\begin{align}
\nabla^{++} \mathscr{L}^0 = 0~, \qquad
\bbD \mathscr{L}^{0} = \mathscr{L}^{0}~, \qquad
\bbA \mathscr{L}^{0} = 2 \mathscr{L}^{0}~, \qquad
\nabla^0 \mathscr{L}^{0} = 0~,
\end{align}
then the following action is invariant:
\begin{align}
-\frac{1}{2\pi} \oint_\cC \rd \tau \int \rd^4x\, \rd^4 \q\, \rd^2 \bar \q^+\, \cE^{0} \mathscr{L}^{0} + \HC
\end{align}
where the measure is
\begin{align}
\cE^{0} =
\sdet \begin{pmatrix}
E_m{}^a & E_m{}^{\alpha \pm} & E_m{}^{\dalpha +} & E_m{}^{++} \\[0.3em]
E_{\mu \pm}{}^a & E_{\mu \pm}{}^{\alpha \pm} & E_{\mu \pm}{}^{\dalpha +} & E_{\mu \pm}{}^{++} \\[0.3em]
E_{\dmu +}{}^a & E_{\dmu +}{}^{\alpha \pm} & E_{\dmu +}{}^{\dalpha +} & E_{\dmu +}{}^{++} \\[0.3em]
E_\tau{}^a & E_\tau{}^{\alpha \pm} & E_\tau{}^{\dalpha +} & E_\tau{}^{++}
\end{pmatrix}~.
\end{align}
Such chiral-analytic actions are naturally higher-derivative, and have been discussed
recently in \cite{ButterKuzenko:HigherDerivs} in the context of curved
projective superspace, as well as \cite{ArAwBrEs} in the context of
flat harmonic superspace.

To evaluate such actions, one can convert them to analytic integrals by integrating
over the two $\q^{\mu-}$ coordinates:
\begin{align}
-\frac{1}{2\pi} \oint_\cC \rd \tau \int \rd^4x\, \rd^4 \q\, \rd^2 \bar \q^+\, \cE^{0} \mathscr{L}^{0}
	= 
\frac{1}{8\pi} \oint_\cC \rd \tau \int \rd^4x\, \rd^4 \q^+\, \cE^{--}
	(\nabla^+)^2 \mathscr{L}^{0}~.
\end{align}
One can check that the integrand $(\nabla^+)^2 \mathscr{L}^0$ satisfies all the
required properties of an analytic superspace Lagrangian.
Alternatively, one can lift a chiral-analytic superspace integral to full
superspace in the same way as eqs. \eqref{eq:ProjToFull} -- \eqref{eq:ProjToFull3}.
For example, using the antichiral field strength $\bar W$ of a vector multiplet, one has in the central gauge
\begin{align}\label{eq:CAtoFull}
-\frac{1}{2\pi} \oint_\cC \rd \tau \int \rd^4x\, \rd^4 \q\, \rd^2 \bar \q^+\, \cE^{0} \mathscr{L}^{0}
	= \frac{2}{\pi} \oint_\cC \cV^{++} \int \rd^4x\, \rd^4 \q\, \rd^4 \bar \q\, E
		\frac{\bar W}{(\bar\nabla^+)^2 \bar W}
		\mathscr{L}^{0}
\end{align}
or imposing the Weyl-U(1) gauge $\bar W=1$,
\begin{align}
-\frac{1}{2\pi} \oint_\cC \rd \tau \int \rd^4x\, \rd^4 \q\, \rd^2 \bar \q^+\, \cE^{0} \mathscr{L}^{0}
	= \frac{1}{2\pi} \oint_\cC \cV^{++} \int \rd^4x\, \rd^4 \q\, \rd^4 \bar \q\, E
		\frac{1}{S^{++}}
		\mathscr{L}^{0}~.
\end{align}
This formulation of the chiral-analytic projective superspace action appeared in
\cite{ButterKuzenko:HigherDerivs}.
Finally, we mention that one can convert a chiral-analytic integral to a chiral superspace
integral by integrating over $\q^{\ul\mu+}$.
This is easiest to construct in the central gauge:
\begin{align}
-\frac{1}{2\pi} \oint_\cC \rd \tau \int \rd^4x\, \rd^4 \q\, \rd^2 \bar \q^+\, \cE^{0} \mathscr{L}^{0}
	&= \frac{1}{8\pi} \int \rd^4x\, \rd^4 \q\, \cE\,\oint_\cC \cV^{++} (\bar\nabla^-)^2 \mathscr{L}^{0}~.
\end{align}
The simplest proof of this is to convert the full superspace integral on
the right-hand side of \eqref{eq:CAtoFull} to a chiral superspace integral
while remaining in central gauge.

\section{Component reduction of analytic superspace action}\label{sec:CompAction}
Our goal in this section is to perform the component reduction of the general
analytic superspace action
\begin{align}
S = - \frac{1}{2\pi} \oint_\cC \rd \tau \int \rd^4x\, \rd^4\q^+ \cE^{--} \mathscr{L}^{++}~.
\end{align}
That is, we will perform the four Grassmann integrals explicitly (using a certain
gauge) and then give the result of the action in the so-called central gauge.

We begin by noting that the action can be interpreted as
evaluated at $\q^{\ul\mu-} = 0$.
Along this submanifold, it is possible to adopt a
gauge where $\nabla_{\ul \alpha +} = \pa / \pa \q^{\ul \alpha+}$,
corresponding to
\begin{align}
\cE^{--} = \sdet
\begin{pmatrix}
E_m{}^a & E_m{}^{++} & E_m{}^{\ul \alpha +} \\[0.2em]
E_\tau{}^a & E_\tau{}^{++} & E_\tau{}^{\ul \alpha +} \\[0.2em]
0 & 0 & \delta_{\ul\mu}{}^{\ul\alpha}
\end{pmatrix}
= \sdet \begin{pmatrix}
E_m{}^a & E_m{}^{++} \\[0.2em]
E_\tau{}^a & E_\tau{}^{++}
\end{pmatrix}
\equiv e^{++}
\end{align}
so our goal is to evaluate
\begin{align}
S = -\frac{1}{2\pi} \, \frac{1}{16} \int \rd^4x \oint_\cC \rd \tau \, 
	(\pa_+)^2 (\bar\pa_+)^2 (e^{++} \mathscr{L}^{++})~.
\end{align}

At this stage, we emphasize that $\q^{\ul\mu\pm}$-independent gauge
transformations are still permitted in the gauge
$\nabla_{\ul \alpha +} = \pa / \pa \q^{\ul \alpha+}$.
In other words, the gauge of the \emph{component fields} at $\q^{\ul\mu\pm}=0$
remains completely unfixed. Naturally, one expects the resulting action
should take its simplest form if we adopt the central gauge at $\q^{\ul\mu\pm}=0$,
and we will do this at the very end. However, it is not
easy to impose central gauge at the component level
\emph{prior} to taking the $\q^{\ul \mu+}$ derivatives,
so we will remain in a more general gauge for the time being.

To organize the calculation,
it is convenient to write the integrand in a way which emphasizes that it is
a five-form. That is, the action can be written as
\begin{align}
S = -\frac{1}{2\pi} \, \frac{1}{16} \int_{\cM^4 \times \cC}
	(\pa_+)^2 (\bar\pa_+)^2 (\widehat e^{++} \cL^{--})~.
\end{align}
where $\widehat e^{++}$ is the volume five-form
\begin{align}
\widehat e^{++}
	= \rd x^0 \wedge \rd x^1 \wedge \rd x^2 \wedge \rd x^3 \wedge \rd \tau \, e^{++}
	= \frac{1}{4!} \eps_{abcd}\, E^a \wedge E^b \wedge E^c \wedge E^d \wedge E^{++}~.
\end{align}
Taking the $\q^{\ul\mu+}$ derivatives of this five-form proves to be significantly
simpler than the corresponding calculation with the determinant.
Expanding out the action, one finds
\begin{align}
S = - \frac{1}{2\pi} \int_{\cM^4 \times \cC} \cJ
\end{align}
where $\cJ$ is a five-form given by
\begin{align}\label{eq:defJ}
\cJ &= \widehat e^{++}\, (\nabla^-)^4 \mathscr{L}^{++}
- \frac{1}{8} \pa^\alpha_+ \widehat e^{++}\, \nabla_\alpha^- (\bar \nabla^-)^2 \mathscr{L}^{++}
- \frac{1}{8} \bar\pa_{\dalpha +} \widehat e^{++}\, \bar\nabla^{\dalpha -} (\nabla^-)^2 \mathscr{L}^{++}
\eol & \quad
+ \frac{1}{16} (\pa_+)^2 \widehat e^{++}\, (\bar\nabla^-)^2 \mathscr{L}^{++}
+ \frac{1}{16} (\bar\pa_+)^2 \widehat e^{++}\, (\nabla^-)^2 \mathscr{L}^{++}
+ \frac{1}{8} \pa_{\alpha+} \bar\pa_{\dalpha +} \widehat e^{++} [\nabla^{\alpha -}, \bar\nabla^{\dalpha -}] \mathscr{L}^{++}
\eol & \quad
- \frac{1}{8} \pa^\alpha_+ (\bar \pa_+)^2 \widehat e^{++}\, \nabla_\alpha^- \mathscr{L}^{++}
- \frac{1}{8} \bar\pa_{\dalpha +} (\pa_+)^2 \widehat e^{++}\, \bar\nabla^{\dalpha -} \mathscr{L}^{++}
+ \frac{1}{16} (\pa_+)^2 (\bar\pa_+)^2 \widehat e^{++}\, \mathscr{L}^{++}~.
\end{align}
In the above expression, we have replaced $\pa_{\ul\alpha+} \rightarrow \nabla_{\ul\alpha+}$
for all the derivatives acting upon the analytic Lagrangian $\mathscr{L}^{++}$.
This is allowed because after projecting to $\q^{\ul \mu+} = \q^{\ul\mu -} = 0$
(which is implicitly assumed above)
the result holds in a general component gauge.
To recover the explicit expression for $\cJ$, one must evaluate each 
of the $\q^{\ul\mu+}$ derivatives of $\widehat e^{++}$. This can be done
systematically, although the resulting formulae grow quite complicated
as the number of spinor derivatives increases. The results are given
in eqs. \eqref{eq:dVolForm} -- \eqref{eq:d4VolForm} of appendix \ref{app:CompDetails},
where the details of
the calculation are also included. We emphasize that upon using eqs.
\eqref{eq:dVolForm} -- \eqref{eq:d4VolForm},
the result for $\cJ$ is given in a general component gauge.

Some comments should now be made about the nature of this five-form:
\begin{itemize}
\item It is invariant under all gauge transformations,
up to an exact form. This is a direct consequence of its origin
from a gauge-invariant superspace action, but it can be checked explicitly.
A straightforward calculation shows, for example, that $\cJ$ transforms
under $S$-supersymmetry,
$\delta = \eta^{\ul\alpha+} S_{\ul\alpha}^- - \eta^{\ul\alpha -} S_{\ul\alpha}^+$,
into an exact form involving $\eta^{\ul\alpha+}$. Therefore,
strictly speaking, $\cJ$ is \emph{not} a conformal primary five-form,
although its integral is invariant.

\item Viewed as a five-form in superspace, $\cJ$ is closed.
In principle, this can also be established by an explicit computation
but is a direct consequence of its construction. Under an arbitrary
diffeomorphism on $\cM^{4|8} \times \rm SU(2)$, $\cJ$ transforms as
a form,
\begin{align}
\delta_\xi \cJ = \rd (\imath_\xi \cJ) + \imath_\xi \rd \cJ~.
\end{align}
The first term vanishes upon integration over the bosonic manifold
$\cM^4 \times \cC$, while the second must vanish for arbitrary
$\xi$ because the original action
was invariant under diffeomorphisms of all types.
This implies that $\cJ$ is closed.
\end{itemize}
These two features are indicative of the superform approach to
supersymmetric invariants \cite{CaDaFr:Vol2.Ecto},
known within the superspace literature as the ectoplasm method
\cite{Gates:Ecto, GaGrKWSi:Ecto} (see also \cite{Hasler:1996wc}).
However, the usual ectoplasm construction usually assumes that
the superform $\cJ$ is expanded entirely in terms of the supervielbein.
In our case, this would mean
$\cJ = \frac{1}{5!} E^{\ul A_1} \wedge \cdots \wedge E^{\ul A_5} \cJ_{\ul A_5 \cdots \ul A_1}$
for the five-form $\cJ$. In contrast, the five-form we have found
above, given by \eqref{eq:defJ} upon substituting
\eqref{eq:dVolForm} -- \eqref{eq:d4VolForm}, does not generically
have this usual form. It is in fact a rather complicated expansion
involving the explicit appearance of even the $S$-supersymmetry and
special conformal connections.

A dramatic simplification of $\cJ$ occurs if we now adopt the central gauge for the 
$\q^{\ul\mu \pm} = 0$ components of the connections. We leave the details
again to appendix \ref{app:CompDetails} and merely summarize that the
action can then be written
$S = \int \rd^4x \, e\, \cL$
where the Lagrangian $\cL$ involves a contour integral with two distinct
integrands,
\begin{align}\label{eq:CompAction}
\cL &= -\frac{1}{2\pi} \oint_\cC \cV^{++} \cL^{--}
	+ \frac{1}{2\pi} \oint_\cC \cV^{--} \cL^{++}
\end{align}
where
\begin{align}\label{eq:L--}
\cL^{--} &= \frac{1}{16} (\nabla^-)^2 (\bar\nabla^-)^2 \mathscr{L}^{++}
	- \frac{i}{8} (\bar\psi_m^- \bsigma^m)^\alpha \nabla_\alpha^- (\bar\nabla^-)^2 \mathscr{L}^{++}
	- \frac{i}{8} (\psi_m^- \sigma^m)_\dalpha \bar\nabla^{\dalpha -} (\nabla^-)^2 \mathscr{L}^{++}
	\eol & \quad
	+ \frac{1}{4} \Big(
	(\psi_n^- \sigma^{nm})^\alpha \bar\psi_m{}^\dalpha{}^-
	+ \psi_n{}^\alpha{}^-  (\bsigma^{nm}\bar\psi_m^-)^\dalpha
	- i \cV_m^{--} (\sigma^m)_{\alpha \dalpha} \Big) [\nabla_\alpha^{-}, \bar\nabla_\dalpha^{-}]  \mathscr{L}^{++}
	\eol & \quad
	+ \frac{1}{4} (\psi_m^- \sigma^{mn} \psi_n^-) (\nabla^-)^2 \mathscr{L}^{++}
	+ \frac{1}{4} (\bar\psi_m^- \bsigma^{mn} \bar\psi_n^-) (\bar \nabla^-)^2 \mathscr{L}^{++}
	\eol & \quad
	- \Big(
	\frac{1}{2} \eps^{mnpq} (\psi_m^- \sigma_n \bar\psi_p^-) \psi_q^{\alpha -}
	- 2 (\psi_m^- \sigma^{mn})^\alpha \cV_n^{--} \Big) \nabla_\alpha^- \mathscr{L}^{++}
	\eol & \quad
	+ \Big(
	\frac{1}{2} \eps^{mnpq} (\bar\psi_m^- \bsigma_n \psi_p^-) \bar\psi_{q\dalpha}^{-}
	- 2 (\bar\psi_m^- \bsigma^{mn})_\dalpha \cV_n^{--} \Big) \bar\nabla^{\dalpha -} \mathscr{L}^{++}
	\eol & \quad
	+ 3 \eps^{mnpq} (\psi_m^- \sigma_n \bar\psi_p^-) \cV_q^{--} \mathscr{L}^{++}
\end{align}
and
\begin{align}\label{eq:L++}
\cL^{++}
	&= -\Big[3 D 
	+ 4 f_a{}^a 
	- 4 (\bar \psi_m^- \bsigma^{mn} \hat{\bar\phi}_n^+) 
	+ 4 (\psi_m^{-} \sigma^{mn} \hat\phi_n^{+}) 
	- 3\, \eps^{mnpq}
		 (\psi_m^{-} \sigma_n \bar\psi_p^{-}) \cV_q^{++}
	\Big] \mathscr{L}^{++}
	\eol & \quad
	+ \Big[\frac{3}{2} \chi^{\alpha +}
	- i (\bar \phi_m^{+} \bsigma^m)^\alpha 
	+ 2 (\psi_m^- \sigma^{mn})^\alpha \cV_n^{++} \Big] \nabla_\alpha^- \mathscr{L}^{++}
	\eol & \quad
	- \Big[\frac{3}{2} \chi_\dalpha^{+}
	- i (\phi_m^{+} \sigma^m)_\dalpha 
	+ 2 (\bar\psi_m^- \bsigma^{mn})_\dalpha \cV_n^{++}\Big] \bar\nabla^{\dalpha -} \mathscr{L}^{++}
	\eol & \quad
	- \frac{i}{4} \cV_m^{++} (\bsigma^m)^{\dalpha \alpha} [\nabla_\alpha^-, \bar\nabla_\dalpha^-] \mathscr{L}^{++}~.
\end{align}
The component fields appearing above are defined in \cite{Butter:CSG4d_2}
and correspond to the matter content of $\cN=2$ conformal supergravity.
These consist of (i) five fundamental connections --
the vierbein $e_m{}^a$, the gravitini $\psi_m{}^{\alpha}{}_i$,
the $\rm SU(2)_R$ and $\rm U(1)_R$ connections $\cV_m{}^i{}_j$
and $A_m$, and the dilatation connection $b_m$;
(ii) covariant auxiliary fields $W_{ab}$, $\chi_{\alpha i}$, and $D$; and
(iii) composite connections $\omega_m{}^{ab}$, $\phi_m{}^{\alpha i}$ and $f_m{}^a$,
given in terms of the other fields, which are
associated respectively with Lorentz, $S$-supersymmetry
and special conformal gauge symmetries.
In the expression for $\cL^{++}$, we have used the symbol $\hat \phi_m{}^{\ul\alpha+}$
to denote the gravitino-dependent part of the $S$-supersymmetry connection. It is
given by
\begin{align}
\hat \phi_m{}_\alpha{}^j &:= \phi_{m}{}_\alpha{}^j + \frac{i}{4} (\sigma_{m} \bar\chi^{j})_\alpha
	= \frac{i}{2} \Big(\sigma^{pn} \sigma_m - \frac{1}{3} \sigma_m \bsigma^{pn} \Big)_{\alpha \dbeta}
          \Big(\cD_p \bpsi_{n}{}^{\dbeta j}
               + \frac{i}{4} \bar W_{ab} (\bsigma^{ab} \bsigma_p \psi_{n}{}^j)^\dbeta \Big)~, \eol
\hat \bphi_{m}{}^\dalpha{}_j &:= \bphi_{m}{}^\dalpha{}_j + \frac{i}{4} (\bsigma_{m} \chi_{j})^\dalpha =
     \frac{i}{2} \Big(\bsigma^{pn} \bsigma_m - \frac{1}{3} \bsigma_m \sigma^{pn} \Big)^{\dalpha \beta}
          \Big(\cD_p \psi_{n\, \beta j} -
               \frac{i}{4} W_{ab} (\sigma^{ab} \sigma_p \bar\psi_{n j})_\beta \Big)~.
\end{align}
Note that $\cL^{++}$ vanishes in the rigid limit.

We point out also that the five-form $\cJ$ associated with $\cL$ still does
not possess the usual ectoplasmic form even in the central gauge because
of the explicit
appearance of the composite connections $f_m{}^a$ and $\phi_m{}^{\alpha i}$.

Introducing
\begin{align}
\omega^{--} \equiv \int \rd^4x\, e\, \cL^{--}~, \qquad
\omega^{++} \equiv \int \rd^4x\, e\, \cL^{++}~,
\end{align}
it is a straightforward exercise to demonstrate that $\omega$ is closed as
a one-form on $\rm SU(2)$,
\begin{align}\label{eq:DomegaToDL}
D^{++} \omega^{--} = D^{--} \omega^{++} \quad \Longleftrightarrow \quad
e \, D^{++} \cL^{--} = e\, D^{--} \cL^{++} + \text{total $x$-derivative}~.
\end{align}
This is a direct consequence of our construction, but it can also be
checked explicitly.

The importance of two distinct integrands can be attributed to the fact that
$\cL^{--}$ is not holomorphic, even up to a total derivative.
The presence of the $\cL^{++}$ term is quite necessary in order
for the full action to be invariant under all of the component gauge transformations.
These include not only $S$-supersymmetry and $Q$-supersymmetry but also $\rm SU(2)$
diffeomorphisms that leave us in the central basis.
Recall that these act as
\begin{align}
\delta_\l = -\l^{++} D^{--} + \l^0 D^0 + \l^{--} D^{++}
\end{align}
where $\l^{\pm\pm}$ and $\l^0$ are given by \eqref{eq:KillingLambda}, now with
$\l^i{}_j$ potentially depending on $x$. Invariance under $\delta_\l$ can actually
be used to uniquely determine $\cL^{--}$ and $\cL^{++}$ starting from the
leading term in $\cL^{--}$.

At this stage, we should mention that the action \eqref{eq:CompAction} is actually
invariant under another group of transformations -- arbitrary
diffeomorphisms on the $\rm SU(2)$ manifold,
\begin{align}
\delta v^{i+} = -\xi^{++} v^{i-} + \xi^0 v^{i+}~, \qquad
\delta v_{i}^- = \xi^{--} v_i^+ - \xi^0 v_i^{-}~,
\end{align}
where $\xi^{\pm\pm}$ and $\xi^0$ are $x$-independent but otherwise
arbitrary. This implies an invariance of the action under
small deformations of the contour $\cC$.

The component action \eqref{eq:CompAction}
can be compared with the original expression (4.13) in \cite{KT-M:DiffReps}
(where $\rm SU(2)$ superspace was used) as well as the later result (4.13) in \cite{BuNo:CompRed}
(using conformal superspace). Both expressions
involve only the first contour integral with $\cL^{--}$.
This earlier formulation of projective superspace
can be interpreted in our language as involving a complex $\rm SU(2)$ manifold
(i.e. an $\rm SL(2, \mathbb C)$ manifold)
as discussed in section \ref{sec:SU2Geo.complex}. This involves making
a certain complexification of the harmonic variables $v^{i\pm}$,
\begin{align}
\begin{pmatrix}
v^{i+} \\
v_i^-
\end{pmatrix}
\longrightarrow
\begin{pmatrix}
v^i \\
u_i / (v, u)
\end{pmatrix} 
\end{align}
where the coordinate $u_i$ is \emph{not} the complex conjugate of $v^i$.
Then it is possible to choose a contour in $\rm SL(2, \mathbb C)$
where $v^i$ varies with $u_i$ fixed,
with the requirement that $(v,u)$ be nonzero along the contour. In such a case,
$\cV^{--} = 0$ on the $\rm SL(2, \mathbb C)$ manifold
and so the second contour integral vanishes automatically
even through $\cL^{++}$ is nonzero.
Moreover, if we take the rigid limit with non-constant $u_i$,
it is easy to see that $\cL^{++}$ vanishes even though $\cV^{--}$ is nonzero.
Thus we recover both the original flat space formulation of \cite{KLR, Siegel:ChiralActions}
with arbitrary $u_i$ as well as the curved formulation of
\cite{KT-M:DiffReps} with fixed $u_i$.

We emphasize that the original derivation of $\cL^{--}$ in \cite{KT-M:DiffReps}
was based on a very similar observation to \eqref{eq:DomegaToDL}. The method
there was to construct $\cL^{--}$ iteratively by first specifying the
leading term, analogous to $(\nabla^-)^4 \mathscr{L}^{++}$, and then to add in
the terms needed to ensure that $\cL^{--}$ was independent of the fixed
coordinate $u_i$, up to a total contour derivative (analogous to $D^{--} \cL^{++}$)
and a total spacetime derivative. More explicitly, let us consider the
complexified version of the expression \eqref{eq:L--} for $\cL^{--}$ in the central gauge,
\begin{align}\label{eq:cL--Leading}
\cL^{--} = (\nabla^-)^4 \mathscr{L}^{++}(v) + \cdots
	= \frac{1}{16} \frac{u_i u_j u_k u_l}{(v,u)^4} \nabla^{ij} \bar\nabla^{kl} \mathscr{L}^{++}(v) + \cdots
\end{align}
Following the same argument as \cite{KT-M:DiffReps}, the action must
be invariant under constant shifts $\delta u_i$, which can be parametrized as
\begin{align}\label{eq:delta_u}
\delta u_i = \alpha u_i + \beta v_i~,
\end{align}
in terms of $x$-independent parameters $\alpha$ and $\beta$.
(This is possible since $v_i$ and $u_i$ are linearly independent along the contour.)
The parameters $\alpha$ and $\beta$ must depend on the contour coordinate $\tau$ in
order for $\delta u_i$ to be $\tau$-independent, but the precise relationship
will not concern us here. The important feature is that
$\delta v_i^- = \beta v_i^+ / (v,u)$ and so the transformation
\eqref{eq:delta_u} can be interpreted as the $\rm SL(2, \mathbb C)$ diffeomorphism
$\delta = \xi^{--} D^{++}$ with $\xi^{--} = \beta / (v,u)$.
This acts only on $v_i^-$. It follows that
\begin{align}\label{eq:deltaL--Comp}
\delta \cL^{--} = \xi^{--} D^{++} \cL^{--}~.
\end{align}
Now in order for this to vanish under the contour integral, it must be that
\eqref{eq:DomegaToDL} holds for some choice of function $\cL^{++}$.
This allows one to iteratively determine all contributions to
$\cL^{--}$ starting from the leading term \eqref{eq:cL--Leading}.
This uniquely specifies $\cL^{--}$ and $\cL^{++}$ in \eqref{eq:L--}
and \eqref{eq:L++}.
Now assuming that $\cL^{--}$ has been so constructed, one has
\begin{align}
\delta \cL^{--} = \xi^{--} D^{--} \cL^{++} + \text{total $x$ derivative}~.
\end{align}
Using $D^{--} \delta u_i = 0$, one can prove $D^{--} \beta \propto D^{--} \xi^{--} = 0$,
and so one recovers
\begin{align}
\delta \cL^{--} = D^{--} (\xi^{--} \cL^{++}) + \text{total $x$ derivative}~.
\end{align}
The remaining contour can then be discarded and invariance under \eqref{eq:delta_u}
confirmed.

A natural question to ask is what happens if we keep an $\rm SL(2, \mathbb C)$
manifold but allow $u_i$ to vary along the contour, as
in \cite{KLR, Siegel:ChiralActions}.
We may still demand the invariance of the
action under \eqref{eq:delta_u}, but now \emph{there is no need for any constraint
to be imposed on $\alpha$ or $\beta$}.
We find as before \eqref{eq:deltaL--Comp}.
This leads (using $\delta \cV^{++} = \delta (v_i \,\rd v^i) = 0$) to
\begin{align}\label{eq:deltaL--Action}
\delta \oint_\cC \cV^{++} \cL^{--}
	= \oint_\cC \cV^{++} \xi^{--} D^{--} \cL^{++} + \text{total $x$ derivative}~,
\end{align}
which does not vanish automatically. But now the second contour integral is not zero,
so we must analyze its variation.
This involves calculating $\delta \cV^{--}$ using the expression for
the complexified vielbeins \eqref{eq:VielbeinsComplex}. The result is
$\delta \cV^{--} = \rd \xi^{--} + 2 \xi^{--} \cV^0$,
which is the same expression as \eqref{eq:deltacV} found on the real
$\rm SU(2)$ manifold. This leads to
\begin{align}\label{eq:deltaL++Action}
\delta \oint_\cC \cV^{--} \cL^{++}
	&= \oint_\cC (\delta \cV^{--} \cL^{++} + \cV^{--} \xi^{--} D^{++} \cL^{++}) \eol
	&= \oint_\cC (\rd \xi^{--} \cL^{++} + 2 \xi^{--} \cV^0 \cL^{++}
		+ \xi^{--} \cV^{--} D^{++} \cL^{++})
\end{align}
and the difference between \eqref{eq:deltaL--Action} and \eqref{eq:deltaL++Action} is,
after rewriting $\cV^{\ul a} D_{\ul a} \cL^{++} = \rd \cL^{++}$,
\begin{align}
- \delta \oint_\cC \cV^{++} \cL^{--}
	+ \delta \oint_\cC \cV^{--} \cL^{++}
	= \oint_\cC (\rd \xi^{--} \cL^{++} + \xi^{--} \rd \cL^{++})
	= 0~.
\end{align}
(In the above equation, we discarded the total $x$ derivative.)

This is a happy state of affairs. The expression \eqref{eq:CompAction}, which
we derived using a real $\rm SU(2)$ manifold in the central gauge,
proves to generalize to an $\rm SL(2, \mathbb C)$ manifold in the central gauge,
\emph{no matter the behavior of $u_i$ along the contour}, so long as
$(v,u) \neq 0$ is maintained.
In practice, one expects the calculation either with constant $u_i$ or
with $u_i = \bar v_i$ to be convenient: both correspond to 
special cases of a more general formulation involving
the auxiliary manifold $\rm SL(2, \mathbb C)$.
That we can make arbitrary shifts \eqref{eq:delta_u} ensures that one can
analytically continue from $u_i = \bar v_i$ to $u_i = \text{constant}$ (and back again)
without any difficulty. This ensures that the formulation presented
here and the conventional formulation \cite{KLRT-M1, KLRT-M2, KT-M:DiffReps}
are equivalent.

\section{Conclusion}

In this paper we have constructed curved projective
superspace using the supermanifold $\cM^{4|8} \times \rm SU(2)$.
This approach generalizes previous work
\cite{KLRT-M1, KLRT-M2, KT-M:DiffReps} in four dimensions,
which we have interpreted as the central gauge of 
the superspace $\cM^{4|8} \times {\rm SL}(2, \mathbb C)$, 
the complexified version of the superspace taken here.
This approach to curved projective superspace can straightforwardly be
extended to dimensions two through six using the existing body of
work \cite{KT-M:5DSugra1, KT-M:5DSugra2, KT-M:5DSugra3, T-M:2DSugra,KLT-M:3DSugra, LT-M:6DSugra}.

In particular, a recent paper \cite{Linch:Superforms} has explored superforms
in $6D$ curved superspace \cite{LT-M:6DSugra},
motivated partly by an attempt
to construct the component form associated with the $6D$
projective superspace action principle.
It seems to us that an interpretation of
$6D$ projective superspace along the lines we have taken here
should be possible.
We reiterate here that the five-form $\cJ$
corresponding to the component Lagrangian of the $4D$
analytic projective
superspace action, which we gave implicitly in
\eqref{eq:defJ} upon substituting eqs. \eqref{eq:dVolForm} -- \eqref{eq:d4VolForm},
rather curiously does not possess the standard form
$\cJ = \frac{1}{5!} E^{\ul A_1} \wedge \cdots \wedge E^{\ul A_5}
\cJ_{\ul A_5 \cdots \ul A_1}$
of an expansion purely in terms of the supervielbeins.
It is plausible that this is a source of the difficulties
observed in \cite{Linch:Superforms}. Another intriguing feature of
\cite{Linch:Superforms} was its use of pure spinor Lorentz harmonics to
drastically simplify the study of the complex of differential forms;
perhaps a curved superspace which implements such
Lorentz harmonics directly within the superfields
could have powerful applications.

To keep our construction as simple as possible,
we have avoided introducing a Yang-Mills connection on 
$\cM^{4|8} \times \rm SU(2)$, but there is no barrier to doing so.
This was already discussed in the conventional formulation \cite{KLRT-M1, KLRT-M2},
and the extension to the formulation here is completely straightforward.
Similarly, we have not discussed the various possible actions
one can construct involving covariantly arctic, antarctic,
tensor and vector multiplets.
These have been discussed elsewhere in the
conventional approach; see \cite{Kuzenko:2008qz}
where the vector multiplet action and off-shell
supergravity-matter actions with a tensor multiplet compensator
were constructed in curved superspace.
Their construction in the general gauge is similarly
straightforward.

What then is the benefit of this new extended formulation?
From our point of view, a main advantage is that it transparently
admits the existence of an analytic gauge where (at least locally)
$\nabla_{\ul\alpha}^+ = \pa / \pa \q^{\ul\alpha-}$
and $D^{++} = v_{i}^+ \pa / \pa v_i^{-}$. This is possible only
if one can make arbitrary diffeomorphisms on the supermanifold,
just as in harmonic superspace. In such a gauge,
covariantly analytic superfields are characterized simply by their
independence of $v_i^{-}$ and $\q^{\ul\alpha-}$.
It is well-known in harmonic superspace that the analytic gauge
(known as the \emph{analytic basis} in the harmonic context)
plays a critical role when one constructs the supergravity
prepotentials \cite{Galperin:1987ek}.
It seems likely that analytic gauge should help resolve
the problem of finding supergravity prepotentials in
projective superspace, a partial solution of which
was presented in \cite{KT-M:5DSugra1}. Perhaps the
harmonic and projective approaches could even be related to each
other, as was the case with the gauge prepotentials
\cite{Kuzenko:HarmProj, JainSiegel:HarmProj}.
We intend to revisit this subject in the near future.

\section*{Acknowledgements}
The author is grateful to both Sergei Kuzenko and
Gabriele Tartaglino-Mazzucchelli for valuable suggestions
and comments on the manuscript as well as William Linch and
Evgeny Ivanov for valuable correspondence and suggestions.
This work was supported in part by the ERC Advanced Grant no. 246974,
{\it ``Supersymmetry: a window to non-perturbative physics''}
and by the European Commission
Marie Curie International Incoming Fellowship grant no.
PIIF-GA-2012-627976.	

\appendix

\section{Curvatures of conformal superspace on $\cM^{4|8} \times \rm SU(2)$}\label{app:Curvatures}

\subsection{Torsion}
The torsion two-forms are defined by
\begin{subequations}
\begin{align}
T^a &:= \rd E^a + E^b \wedge \Omega_b{}^a + E^a \wedge B~, \\
T^{\alpha \pm} &:= \rd E^{\alpha \pm} + \frac{1}{2} E^{\alpha \pm} \wedge B
     - i E^{\alpha \pm} \wedge A + E^{\beta \pm} \wedge \Omega_\beta{}^\alpha
     + i  E^b \wedge F_{\dgamma}^\pm \,(\bsigma_b)^{\dgamma \alpha}~,
	\label{eq:Talpha}
\\
T^{\dalpha \pm} &:= \rd E^{\dalpha \pm} + \frac{1}{2} E^{\dalpha \pm} \wedge B
     + i E^{\dalpha \pm} \wedge A - E^\dbeta{}^\pm \wedge \Omega_\dbeta{}^{\dalpha}
     - i  E^b \wedge F_\gamma{}^{\pm} \,(\bsigma_b)^{\dalpha \gamma}~, \\
T^{\pm\pm} &:= \rd E^{\pm\pm}
     + 4 E^{\ul\beta \pm} \wedge F_{\ul\beta}^\pm ~, \\
T^{0} &:= \rd E^{0}
     + 2 E^{\ul\beta +} \wedge F_{\ul\beta}^-
     + 2 E^{\ul\beta -} \wedge F_{\ul\beta}^+~.
\end{align}
\end{subequations}
The non-vanishing components of the torsion tensor can be grouped by dimension:
\begin{itemize}
\item Dimension 0
\begin{subequations}
\begin{alignat}{2}
T_{\gamma \pm\, \dbeta \mp}{}^a &= \pm 2i (\sigma^a)_{\gamma \dbeta}~, \eol
T_{\pm\pm \,\ul\beta\mp}{}^{\ul\alpha\pm} &= \mp\delta_{\ul\beta}{}^{\ul\alpha}~, &\qquad
T_{0 \,\ul\beta\pm}{}^{\ul\alpha\pm} &= \pm \delta_{\ul\beta}{}^{\ul\alpha}~,  \eol
T_{0 \,\pm\pm}{}^{\pm\pm} &= \pm 2~, &\qquad
T_{--\, ++}{}^0 &= 1~.
\end{alignat}
\item Dimension 1
\begin{align}
T_{\dgamma \pm\, \beta \dbeta}{}^{\alpha \pm} = i \eps_{\dgamma \dbeta} W_\beta{}^\alpha~, \qquad
T_{\gamma \pm\, \beta \dbeta}{}^{\dalpha \pm} = -i \eps_{\gamma \beta} \bar W_\dbeta{}^\dalpha~.
\end{align}
\item Dimension 3/2
\begin{alignat}{2}
T_{\gamma \dgamma \,\beta \dbeta\,}{}^{\alpha \pm} &=
          \frac{1}{2} \eps_{\dgamma \dbeta} \nabla^{\alpha \pm} W_{\gamma \beta}~, &\qquad
T_{\gamma \dgamma \,\beta \dbeta\,}{}^{\dalpha \pm} &=
          \frac{1}{2} \eps_{\gamma \beta} \bar\nabla^{\dalpha \pm} \bar W_{\dgamma \dbeta}~, \eol
T_{\beta \pm\, \alpha \dalpha}{}^{\pm \pm} &= -i \eps_{\beta \alpha} \bar\nabla^{\dphi\pm} \bar W_{\dphi \dalpha}~, &\qquad
T_{\dbeta \pm\, \alpha \dalpha}{}^{\pm \pm} &= i \eps_{\dbeta \dalpha} \nabla^{\phi\pm} W_{\phi \alpha}~, \eol
T_{\beta \pm\, \alpha \dalpha}{}^{0} &= -\frac{i}{2} \eps_{\beta \alpha} \bar\nabla^{\dphi\mp} \bar W_{\dphi \dalpha}~, &\qquad
T_{\dbeta \pm\, \alpha \dalpha}{}^{0} &= \frac{i}{2} \eps_{\dbeta \dalpha} \nabla^{\phi\mp} W_{\phi \alpha}
\end{alignat}
\item Dimension 2
\begin{align}
T_{\beta \dbeta\,\alpha \dalpha}{}^{\pm\pm} =
     - \frac{1}{4} \eps_{\dbeta \dalpha} \nabla^{\gamma\pm} \nabla_\gamma^\pm W_{\beta \alpha}
     + \frac{1}{4} \eps_{\beta \alpha} \bar \nabla_{\dgamma}^\pm \bar\nabla^{\dgamma \pm} \bar W_{\dbeta \dalpha}~, \eol
T_{\beta \dbeta\,\alpha \dalpha}{}^{0} =
     - \frac{1}{4} \eps_{\dbeta \dalpha} \nabla^{\gamma+} \nabla_\gamma^- W_{\beta \alpha}
     + \frac{1}{4} \eps_{\beta \alpha} \bar \nabla_{\dgamma}^+ \bar\nabla^{\dgamma -} \bar W_{\dbeta \dalpha}~.
\end{align}
\end{subequations}
\end{itemize}

Some subtleties arise when one compares these equations to those in
\cite{Butter:CSG4d_2}. For example, there one finds (relabeling
$\Phi^j{}_i \rightarrow \cV^j{}_i$)
\begin{align}\label{eq:TalphaCentral}
T^\alpha{}_i &= \rd E^\alpha{}_i
     + E^\alpha{}_j \wedge \cV^j{}_i
	+ \frac{1}{2} E^\alpha{}_i \wedge B
     - i E^\alpha{}_i \wedge A + E^\beta{}_i \wedge \Omega_\beta{}^\alpha
     + i  E^b \wedge F_{\dgamma i} \,\bsigma_b^{\dgamma \alpha}~.
\end{align}
There is an apparent discrepancy in the second term, which is absent in 
the corresponding equation for $T^{\alpha \pm}$. This is because here the
tensor $\cV^j{}_i$ is no longer interpreted as part of the vielbein and so
the formal definition of the torsion two-form differs. However,
what does not differ is the actual equation one finds for
$\rd E^\alpha{}_i$. From \cite{Butter:CSG4d_2}, one finds the \emph{constraint}
\begin{align}
T^\alpha{}_i = -i E^b \wedge E_{\dgamma i}\, (\bar \sigma_b)^{\dgamma \beta} W_\beta{}^\alpha
	- \frac{1}{8} E^b \wedge E^c \, (\sigma_{cb})^{\gamma \beta} \nabla^\alpha{}_i W_{\gamma \beta}~,
\end{align}
which should be equated to \eqref{eq:TalphaCentral} to give a constraint on
$\rd E^\alpha{}_i$. In our framework here, we have instead
\begin{align}\label{eq:TalphaConstraint}
T^{\alpha \pm}
	&= -i E^b \wedge E_\dbeta^\pm (\bar \sigma_b)^{\dbeta \beta} W_\beta{}^\alpha
	- \frac{1}{8} E^b \wedge E^c (\sigma_{cb})^{\gamma \beta} \nabla^{\alpha \pm} W_{\gamma \beta}
	\eol & \qquad
	\mp E^{\alpha \mp} \wedge E^{\pm\pm}
	\pm E^{\alpha \pm} \wedge E^0~.
\end{align}
This should be equated with \eqref{eq:Talpha} to find a constraint for $\rd E^{\alpha \pm}$.
In the central basis, the two equations for $\rd E^{\alpha}{}_i$ are identical.
The ``additional'' terms in the second line of \eqref{eq:TalphaConstraint} are the same
as the terms ``missing'' in \eqref{eq:Talpha}; this swapping amounts merely to
a redefinition of the torsion two-form. Moreover, this redefinition does not change
the values of the tangent space components $T_{C B}{}^{A}$, so the same algebra
of covariant derivatives holds in both approaches.

A similar alteration happens in the definitions of $T^{\pm\pm}$ and $T^0$
when compared with the $\rm SU(2)$
curvature $R(\cV)^i{}_j$ given in \cite{Butter:CSG4d_2}. Nevertheless, the
values of $T_{CB}{}^{\pm\pm}$ and $T_{CB}{}^0$ are identical in the central basis
to $R(\cV)_{CB}{}^{ij} v_i^\pm v_j^\pm$ and $R(\cV)_{CB}{}^{ij} v_i^+ v_j^-$.

This swapping of terms between the constraints on
and the definition of the torsion tensor occurs also when one 
compares the curvature $R(P)_{nm}{}^a$ from the tensor calculus formulation of
conformal supergravity with the torsion tensor $T_{nm}{}^a$. These differ by a term
proportional to $\psi_m{}_j \sigma^a \bar \psi_n{}^j$. In the component formulation,
this bilinear appears in the definition of $R(P)_{nm}{}^a$ (which is set to zero).
In the supergravity formulation, it appears in the constraint equation from the nonzero
component $T_{\gamma}{}^k{}_{\dbeta j}{}^a = 2i \delta^k_j (\sigma^a)_{\gamma \dbeta}$.
However, the curvature $[\nabla_b, \nabla_a]$ is the same in both approaches,
as is the equation for $\rd e^a$, which is used to determine the spin connection.

\subsection{Lorentz curvature}
The conformal Lorentz curvature two-form is
\begin{align}
R^{ba} &= \rd \Omega^{ba} + \Omega^{bc} \wedge \Omega_c{}^a - 2 E^{[b} \wedge F^{a]}
     + 4 E^{\beta}{}^- \wedge F^{\alpha +} \,(\sigma^{ba})_{\alpha\beta}
     - 4 E^{\beta}{}^+ \wedge F^{\alpha -} \,(\sigma^{ba})_{\alpha\beta}
	\eol & \qquad
     + 4 E^{\dbeta}{}^+ \wedge F^{\dalpha-} \,(\bsigma^{ba})_{\dalpha\dbeta}
     - 4 E^{\dbeta}{}^- \wedge F^{\dalpha+} \,(\bsigma^{ba})_{\dalpha\dbeta}
\end{align}
and may be canonically decomposed as
\begin{align}
R_{DC \,\beta\dbeta\, \alpha \dalpha} =
2 \eps_{\dbeta \dalpha} R_{DC \beta \alpha}
- 2 \eps_{\beta \alpha} R_{DC \dbeta \dalpha}.
\end{align}
It is simplest to express the curvature results in terms of these components.
We group the non-vanishing components by dimension.
\begin{itemize}
\begin{subequations}
\item {Dimension 1}
\begin{gather}
R_{\delta + \gamma - \,\dbeta \dalpha} = -2 \eps_{\delta \gamma} \bar W_{\dbeta \dalpha}~, \qquad
R_{\ddelta + \dgamma - \,\beta \alpha} = 2 \eps_{\ddelta \dgamma} W_{\beta \dalpha}
\end{gather}
\item {Dimension 3/2}
\begin{gather}
R_{\delta\mp}{}_{\,\gamma \dgamma\,}{}_{\dbeta \dalpha} =
     \mp \frac{i}{2} \eps_{\delta \gamma} \bar\nabla_\dbeta^\pm \bar W_{\dalpha \dgamma}
     \mp \frac{i}{2} \eps_{\delta \gamma} \bar\nabla_\dalpha^\pm \bar W_{\dbeta \dgamma}~, \\
R_{\ddelta\mp}{}_{\,\gamma \dgamma\,}{}_{\beta \alpha} =
     \mp\frac{i}{2} \eps_{\ddelta \dgamma} \nabla_\beta^\pm W_{\alpha \gamma}
     \mp\frac{i}{2} \eps_{\ddelta \dgamma} \nabla_\alpha^\pm W_{\beta \gamma}
\end{gather}
\item{Dimension 2}
\begin{align}
R_{\delta \ddelta\,\gamma \dgamma\,\beta \alpha} &=
     + \frac{1}{4} \eps_{\ddelta \dgamma} \nabla_{\beta \alpha} W_{\delta \gamma}
     - \frac{1}{8} \eps_{\ddelta \dgamma}
          (\eps_{\delta \beta} \eps_{\gamma \alpha} + \eps_{\delta \alpha} \eps_{\gamma \beta})
          \nabla_{\phi \rho} W^{\rho \phi}
     + \eps_{\delta \gamma} \bar W_{\ddelta \dgamma} W_{\beta \alpha} \\
R_{\delta \ddelta \,\gamma \dgamma\, \dbeta \dalpha} &=
     - \frac{1}{4} \eps_{\delta \gamma} \bar \nabla_{\dbeta \dalpha} \bar W_{\ddelta \dgamma}
     + \frac{1}{8} \eps_{\delta \gamma}
          (\eps_{\ddelta \dbeta} \eps_{\dgamma \dalpha} + \eps_{\ddelta \dalpha} \eps_{\dgamma \dbeta})
          \bar \nabla_{\dphi \dot\rho} \bar W^{\dot\rho \dphi}
     - \eps_{\ddelta \dgamma} W_{\delta \gamma} \bar W_{\dbeta \dalpha}
\end{align}
\end{subequations}
\end{itemize}

\subsection{Dilatation and ${\rm U}(1)_R$ curvatures}
The conformal field strengths for dilatations and chiral rotations are
\begin{align}
R(\bbD) &= \rd B + 2 E^a \wedge F_a
	- 2 E^{\ul\alpha}{}^- \wedge F_{\ul\alpha}^+
	+ 2 E^{\ul\alpha}{}^+ \wedge F_{\ul\alpha}^-~, \\
R(\bbA) &= \rd A
	+ i E^\alpha{}^- \wedge F_\alpha^+
	- i E^\alpha{}^+ \wedge F_\alpha^-
	- i E^\dalpha{}^- \wedge F_\dalpha^+
	+ i E^\dalpha{}^+ \wedge F_\dalpha^-~.
\end{align}
We group the non-vanishing components by dimension.
\begin{itemize}
\begin{subequations}
\item {Dimension 3/2}
\begin{gather}
R(\bbD)_{\beta\mp}{}_{\,\alpha \dalpha} = \pm\frac{i}{2} \eps_{\beta \alpha} \bar\nabla^{\dphi \pm}
     \bar W_{\dphi \dalpha} ~, \qquad
R(\bbD)_{\dbeta\mp}{}_{\,\alpha \dalpha} = \mp\frac{i}{2} \eps_{\dbeta \dalpha} \nabla^{\phi \pm} W_{\phi \alpha}~,\\
R(\bbA)_{\beta\mp}{}_{\,\alpha \dalpha} = \mp\frac{1}{4} \eps_{\beta \alpha} \bar\nabla^{\dphi \pm}
     \bar W_{\dphi \dalpha} ~, \qquad
R(\bbA)_{\dbeta\mp}{}_{\,\alpha \dalpha} = \mp\frac{1}{4} \eps_{\dbeta \dalpha} \nabla^{\phi \pm} W_{\phi \alpha}
\end{gather}
\item{Dimension 2}
\begin{align}
R(\bbD)_{\beta \dbeta\,\alpha \dalpha} &=
    \frac{1}{8} \eps_{\dbeta \dalpha} (\nabla_\beta{}^\phi W_{\phi \alpha}
          + \nabla_\alpha{}^\phi W_{\phi \beta})
     - \frac{1}{8} \eps_{\beta \alpha} (\bar\nabla_{\dbeta \dphi} \bar W^\dphi{}_\dalpha
          + \bar\nabla_{\dalpha \dphi} \bar W^\dphi{}_\dbeta) \\
R(\bbA)_{\beta \dbeta\,\alpha \dalpha} &=
    - \frac{i}{16} \eps_{\dbeta \dalpha} (\nabla_\beta{}^\phi W_{\phi \alpha}
          + \nabla_\alpha{}^\phi W_{\phi \beta})
     - \frac{i}{16} \eps_{\beta \alpha} (\bar\nabla_{\dbeta \dphi} \bar W^\dphi{}_\dalpha
          + \bar\nabla_{\dalpha \dphi} \bar W^\dphi{}_\dbeta)
\end{align}
\end{subequations}
\end{itemize}

\subsection{Special superconformal curvatures}
The special superconformal curvatures $R(K)^A$,
consisting of $S$-supersymmetry $R(S)^{\ul\alpha\pm}$
and special conformal curvatures $R(K)^a$, are defined by
\begin{align}
R(K)^a &= \rd F^a - F^b \wedge \Omega_b{}^a - F^a \wedge B
	\eol & \qquad
     - 2i F^{\alpha +} \wedge F^\dalpha{}^- \,(\sigma^a)_{\alpha \dalpha}
     + 2i F^{\alpha -} \wedge F^\dalpha{}^+ \,(\sigma^a)_{\alpha \dalpha}~,\\
R(S)^{\alpha \pm} &= \rd F^{\alpha \pm}
     - \frac{1}{2} F^{\alpha \pm} \wedge B
     + i F^{\alpha \pm} \wedge A + F^{\beta \pm} \wedge \Omega_\beta{}^\alpha
     \eol & \qquad
     \mp F^{\alpha \pm} \wedge E^0
     \pm F^{\alpha \mp} \wedge E^{\pm\pm}
     - i F^b \wedge E_{\dalpha}{}^\pm \,(\bsigma_b)^{\dalpha \alpha}~, \\
R(S)^{\dalpha\pm} &= \rd F^{\dalpha\pm}
     - \frac{1}{2} F^{\dalpha\pm} \wedge B
     - i F^{\dalpha\pm} \wedge A
     - F^{\dbeta\pm} \wedge \Omega_\dbeta{}^\dalpha
     \eol & \qquad
     \mp F^{\dalpha \pm} \wedge E^0
     \pm F^{\dalpha \mp} \wedge E^{\pm\pm}
     + i F^b \wedge E_\alpha{}^{\pm} \,(\bsigma_b)^{\dalpha \alpha}~.
\end{align}

The non-vanishing components of
$R(K)_{\ul C \ul B}{}_{\alpha \dalpha} = R(K)_{\ul C \ul B}{}^a (\sigma_a)_{\alpha\dalpha}$
are given by
\begin{subequations}
\begin{align}
R(K)_{\gamma+\, \beta -\, \alpha\dalpha} &= \eps_{\gamma\beta}
	\nabla_{\alpha \dphi} \bar W^{\dphi}{}_\dalpha~, \qquad
R(K)_{\dgamma+\, \dbeta -\, \alpha\dalpha} = -\eps_{\dgamma\dbeta} 
	\nabla_\dalpha{}^{\phi} W_{\phi\alpha}~, \\
R(K)_{\gamma\mp}{}{}_{\,\beta \dbeta\,\alpha\dalpha}
     &= \pm\frac{i}{2} \eps_{\gamma \beta} \bar\nabla^\pm_\dbeta \nabla_{\alpha \dphi} \bar W^\dphi{}_\dalpha,\qquad
R(K)_{\dgamma\mp}{}{}_{\,\beta \dbeta\,\alpha\dalpha}
     = \pm\frac{i}{2} \eps_{\dgamma \dbeta} \nabla^\pm_\beta \nabla_{\dalpha}{}^\phi W_{\phi \alpha}~, \\
R(K)_{\gamma \dgamma\, \beta \dbeta\, \alpha \dalpha}
     &=  -\frac{1}{8} \eps_{\dgamma \dbeta} \nabla_\dalpha{}^\phi \nabla_{\gamma \beta} W_{\phi \alpha}
          - \frac{1}{8} \eps_{\gamma \beta} \nabla_\alpha{}^\dphi \bar\nabla_{\dgamma \dbeta} \bar W_{\dphi \dalpha}
          \eol & \quad
          + \frac{1}{4} \eps_{\dgamma \dbeta} \nabla_\gamma{}^\dphi (\bar W_{\dphi \dalpha} W_{\beta \alpha})
          + \frac{1}{4} \eps_{\dgamma \dbeta} \nabla_\beta{}^\dphi (\bar W_{\dphi \dalpha} W_{\gamma \alpha})
          \eol & \quad
          + \frac{1}{4} \eps_{\gamma \beta} \nabla_\dgamma{}^\phi (W_{\phi \alpha} \bar W_{\dbeta \dalpha})
          + \frac{1}{4} \eps_{\gamma \beta} \nabla_\dbeta{}^\phi (W_{\phi \alpha} \bar W_{\dgamma \dalpha})
          \eol & \quad
          + \frac{i}{4} \eps_{\dgamma \dbeta} (\bar\nabla^{\dphi-}\bar W_{\dphi \dalpha})
               (\nabla_{(\gamma}^+ W_{\beta) \alpha})
          - \frac{i}{4} \eps_{\dgamma \dbeta}(\bar\nabla^{\dphi+} \bar W_{\dphi \dalpha})
               (\nabla_{(\gamma}^- W_{\beta) \alpha})
          \eol & \quad
          + \frac{i}{4} \eps_{\gamma \beta} (\nabla^{\phi +} W_{\phi \alpha})
               (\bar \nabla_{(\dgamma}^- \bar W_{\dbeta) \dalpha})
          - \frac{i}{4} \eps_{\gamma \beta} (\nabla^{\phi-} W_{\phi \alpha})
               (\bar \nabla_{(\dgamma}^+ \bar W_{\dbeta) \dalpha})~.
\end{align}
\end{subequations}

We give the non-vanishing components of $R(S)_{\ul C \ul B}{}^{\ul\alpha \pm}$ grouped
by dimension.
\begin{itemize}
\begin{subequations}
\item Dimension 3/2
\begin{align}
R(S)_{\dgamma +\, \dbeta -}{}^{\alpha \pm} = \frac{1}{2} \eps_{\dgamma \dbeta}
	\nabla^{\phi\pm} W_\phi{}^\alpha~, \qquad
R(S)_{\gamma +\, \beta -}{}^{\dalpha \pm} = \frac{1}{2} \eps_{\gamma \beta}
	\bar\nabla^{\dphi\pm} \bar W_\dphi{}^\dalpha
\end{align}
\item Dimension 2
\begin{alignat}{2}
R(S)_{\gamma \pm\, \beta \dbeta\,}{}^{\alpha \pm} &= \frac{1}{2} \eps_{\gamma \beta}
	\nabla^{\dphi \alpha} \bar W_{\dphi \dbeta}~, &\quad
R(S)_{\dgamma \pm\, \beta \dbeta\,}{}^{\dalpha \pm} &= \frac{1}{2} \eps_{\dgamma \dbeta}
	\nabla^{\dalpha \phi} W_{\phi \beta}~, \\
R(S)_{\dgamma \pm\, \beta \dbeta\,}{}^{\alpha \pm} &=
	\pm \frac{i}{4} \eps_{\dgamma \dbeta} \nabla_\beta^\mp \nabla^{\phi \pm} W_\phi{}^\alpha~, &\quad
R(S)_{\gamma \pm\, \beta \dbeta\,}{}^{\dalpha \pm} &=
	\pm \frac{i}{4} \eps_{\gamma \beta} \bar\nabla_\dbeta^\mp \bar\nabla_\dphi^{\pm} \bar W^{\dphi \dalpha}~, \\
R(S)_{\dgamma \mp\, \beta \dbeta\,}{}^{\alpha \pm} &=
	\pm \frac{i}{8} \eps_{\dgamma \dbeta} (\nabla^\pm)^2 W_\beta{}^\alpha~, &\quad
R(S)_{\gamma \mp\, \beta \dbeta\,}{}^{\dalpha \pm} &=
	\pm \frac{i}{8} \eps_{\gamma \beta} (\bar\nabla^\pm)^2 \bar W_\dbeta{}^\dalpha
\end{alignat}
\item Dimension 5/2
\begin{align}
R(S)_{\gamma \dgamma \,\beta \dbeta \,}{}^{\alpha\pm} &=
     \pm\frac{1}{16} \eps_{\dgamma \dbeta}
          (\nabla^\pm)^2 \nabla_{(\gamma}^\mp W_{\beta)}{}^\alpha
     + \frac{1}{4} \eps_{\gamma \beta}
          \left(
               i \nabla^{\dphi\alpha} \bar\nabla_{(\dgamma}^\pm \bar W_{\dbeta) \dphi}
               + \bar W_{\dgamma \dbeta}  \nabla^{\phi \pm} W_{\phi}{}^\alpha
          \right)~, \\
R(S)_{\gamma \dgamma \,\beta \dbeta \,}{}^{\dalpha \pm} &=
     \mp \frac{1}{16} \eps_{\gamma \beta}
		(\bar \nabla^\pm)^2 \bar \nabla_{(\dgamma}^\mp \bar W_{\dbeta)}{}^\dalpha
     - \frac{1}{4} \eps_{\dgamma \dbeta}
          \left(
               i \nabla^{\dalpha \phi} \nabla_{(\gamma}^\pm W_{\beta) \phi}
               + W_{\gamma \beta}  \bar \nabla_\dphi^{\pm} \bar W^{\dphi\dalpha}
          \right)
\end{align}
\end{subequations}
\end{itemize}

\section{Integration over submanifolds}\label{app:Integration}
In this appendix, we briefly review some elements of integration theory
over submanifolds. A complementary discussion can be found in \cite{KT-M:DiffReps}.

Let $\cM$ be a supermanifold of dimension $D$ with local coordinates $z^M$, $M=1, \ldots, D$.
We denote the grading of a coordinate $z^M$ by $(-)^M$.
The manifold possesses a vielbein $E_M{}^A$ and we can
introduce an integral over a Lagrangian $\mathscr{L}$ in the usual way as
\begin{align}
S = \int \rd^D z\, E\, \mathscr{L}~.
\end{align}
Provided that $\mathscr{L}$ transform as a scalar field under diffeomorphisms,
$\delta_\xi \mathscr{L} = \xi^M \pa_M \mathscr{L}$, the action $S$ is invariant.
If the manifold possesses an additional local symmetry group $\cH$ with generators $X_{\ul a}$,
under which the vielbein transforms as
\begin{align}\label{eq:dHViel}
\delta_\cH E_M{}^A = E_M{}^B g^{\ul c} f_{\ul c B}{}^A~,
\end{align}
with structure constants $f_{\ul c B}{}^A$ (see the discussion in e.g. \cite{Butter:CSG4d_2})
then the action $S$ is invariant provided $\mathscr{L}$ transforms as
\begin{align}
\delta_\cH \mathscr{L} = -(-)^A g^{\ul b} f_{\ul b A}{}^A\, \mathscr{L}~.
\end{align}

Now suppose we are given a submanifold $\mathfrak{M}$ of dimension $d$ with local
coordinates $\mathfrak{z}^m$, $m=1, \ldots, d$. We have in mind a situation where
the original coordinates $z^M$ can be decomposed (at least in the vicinity of
$\mathfrak{M}$) as $z^M = (\mathfrak{z}^m, y^\mu)$ with the submanifold
$\mathfrak{M}$ corresponding to the surface with $y^\mu = 0$.
We make no assumptions about whether $\mathfrak{z}^m$ and $y^\mu$ are bosonic
or fermionic; in fact, we are interested in cases where both consist of
bosonic and fermionic coordinates.
We decompose the vielbein and its inverse as
\begin{align}
E_M{}^A =
\begin{pmatrix}
\cE_m{}^a & E_m{}^\alpha \\
E_\mu{}^a & E_\mu{}^\alpha
\end{pmatrix}~, \qquad
E_A{}^M =
\begin{pmatrix}
E_a{}^m & E_a{}^\mu \\
E_\alpha{}^m & \phi_\alpha{}^\mu
\end{pmatrix}~, \qquad
\end{align}
with the assumption that both $\cE_m{}^a$ and $\phi_\alpha{}^\mu$ are invertible,
with inverses $\cE_a{}^m$ and $\phi_\mu{}^\alpha$, respectively.
This allows one to compactly specify all the remaining components of the vielbein
and its inverse in terms of these quantities, and
$E_m{}^\alpha$ and $E_\alpha{}^m$:
\begin{align}
E_M{}^A &=
\left(\begin{array}{c|c}
\cE_m{}^a & E_m{}^\alpha \\ \hline
- \phi_\mu{}^\beta E_\beta{}^n \cE_n{}^a & \phi_\mu{}^\alpha - \phi_\mu{}^\beta E_\beta{}^n E_n{}^\alpha
\end{array}\right)~, \eol
E_A{}^M &=
\left(\begin{array}{c|c}
\cE_a{}^m - \cE_a{}^n E_n{}^\beta E_\beta{}^m & -\phi_\mu{}^\beta E_\beta{}^n E_n{}^a \\ \hline
E_\alpha{}^m & \phi_\alpha{}^\mu
\end{array}\right)~.
\end{align}
No assumptions need to be made about $E_m{}^\alpha$ or $E_{\alpha}{}^m$.
One can check that
\begin{align}
E \equiv \sdet E_M{}^A = \sdet \cE_m{}^a \sdet \phi_\mu{}^\alpha
	= \frac{\sdet \cE_m{}^a}{\sdet \phi_\alpha{}^\mu}~,
\end{align}
although we won't make use of this feature.

Now consider the action $\cS$ over the submanifold $\mathfrak{M}$ with
Lagrangian $\cL$:
\begin{align}\label{eq:submanifoldS}
\cS = \int \rd^d \mathfrak{z}\, \cE\, \cL~, \qquad \cE = \sdet \cE_m{}^a~.
\end{align}
This is invariant under $\mathfrak{z}^m$ diffeomorphisms provided $\cL$
transforms as a scalar function. If we impose $f_{\ul c \beta}{}^a=0$,
then \eqref{eq:dHViel} implies $\delta_H \cE = (-)^a g^{\ul b} f_{\ul b a}{}^a$.
So a set of sufficient conditions for $\cH$-invariance is
\begin{align}
\delta_\cH \cL = -(-)^a g^{\ul b} f_{\ul b a}{}^a\,\cL~, \qquad
f_{\ul c \beta}{}^a = 0~.
\end{align}

It turns out that $\cS$ can also be made invariant under diffeomorphisms generated by $\xi^\mu$.
The easiest way to see this is to note that because $E_\alpha{}^\mu \equiv \phi_\alpha{}^\mu$
is invertible, it is possible to construct a one-to-one relation between any diffeomorphism in $\xi^\mu$ 
and a covariant diffeomorphism generated by $\xi'^\alpha = \xi^\mu \phi_\mu{}^\alpha$ modulo a certain
diffeomorphism in $\mathfrak{z}^m$ and an $\cH$ gauge transformation. Recall that
a covariant diffeomorphism is given by
\begin{align}
\delta_\xi = \xi^A \nabla_A = \xi^A E_A{}^M \pa_M - \xi^A H_A{}^{\ul b} X_{\ul b} ~,
\end{align}
where $H_M{}^{\ul a}$ is the connection associated with the group $\cH$.
Taking $\xi^A = (0, \xi'^\alpha) = (0, \xi^\mu \phi_\mu{}^\alpha)$, one finds
\begin{align}
\xi'^\alpha \nabla_\alpha &= \xi'^\alpha \phi_\alpha{}^\mu \pa_\mu + \xi'^\alpha E_\alpha{}^m \pa_m
	- \xi'^\alpha H_\alpha{}^{\ul b} X_{\ul b} \eol
	&= \xi^\mu \pa_\mu + \xi^\mu \phi_\mu{}^\alpha E_\alpha{}^m \pa_m
	- \xi^\mu \phi_\mu{}^\alpha H_\alpha{}^{\ul b} X_{\ul b}~.
\end{align}
Since we have already established invariance under $\mathfrak{z}^m$ diffeomorphisms
and $\cH$ gauge transformations, we need only check covariant diffeomorphisms
generated by arbitrary $\xi^\alpha$. This will establish invariance under the full set
of diffeomorphisms.
To prove invariance under covariant diffeomorphisms with parameter $\xi^\alpha$, observe that
\begin{align}
\delta \cE_m{}^a = \cE_m{}^b \xi^\gamma T_{\gamma b}{}^a + E_m{}^\beta \xi^\gamma T_{\gamma \beta}{}^a~.
\end{align}
We will restrict our attention to situations where
$T_{\gamma\beta}{}^a = 0$
so only the first term in $\delta \cE_m{}^a$ contributes.
Noting that $\delta \cL = \xi^\alpha \nabla_\alpha \cL$, it follows that
the remaining sufficient conditions for invariance of
the action \eqref{eq:submanifoldS} are
\begin{align}
\nabla_\alpha \cL = -(-)^b T_{\alpha b}{}^b \, \cL~, \qquad T_{\gamma \beta}{}^a = 0~.
\end{align}

\section{Component action derivation}\label{app:CompDetails}
In this appendix, we describe how to derive the component action of
\begin{align}
S = - \frac{1}{2\pi} \oint_\cC \rd \tau \int \rd^4x\, \rd^4\q^+ \cE^{--} \mathscr{L}^{++}~.
\end{align}
The integral can be understood as evaluated at $\q^{\ul \mu -} = 0$,
since these Grassmann variables do not appear in the measure.
To evaluate the action, it helps to exploit the $\q^{\ul \mu +}$-dependent
parts of our gauge transformations (including covariant diffeomorphisms)
to fix the gauge\footnote{This is just the superspace analogue of
Riemann normal coordinates for the Grassmann coordinates \cite{McArthur:SNC}.
For an extensive discussion of using normal coordinates to derive
component actions, see \cite{KT-M:DiffReps}.}
$\nabla_{\ul\alpha +} = \pa/\pa \q^{\ul\alpha +}$.
Now the analytic superspace vielbein is given by
\begin{align}\label{eq:CompEvalGauge}
\cE_{\ul M}{}^{\ul A} \Big\vert_{\q^- = 0} = 
\begin{pmatrix}
E_m{}^a & E_m{}^{++} & E_m{}^{\ul \alpha +} \\[0.2em]
E_\tau{}^a & E_\tau{}^{++} & E_\tau{}^{\ul \alpha +} \\[0.2em]
0 & 0 & \delta_{\ul\mu}{}^{\ul\alpha}
\end{pmatrix}
=
\begin{pmatrix}
e_m{}^a & \cV_m{}^{++} & \frac{1}{2} \psi_m{}^{\ul \alpha +} \\[0.2em]
e_\tau{}^a & \cV_\tau{}^{++} & \frac{1}{2} \psi_\tau{}^{\ul \alpha +} \\[0.2em]
0 & 0 & \delta_{\ul\mu}{}^{\ul\alpha}
\end{pmatrix}~.
\end{align}
In the last equality, we have relabeled the components of the one-forms
$\cE^{\ul A}$ by $e^a$, $\cV^{++}$, and $\frac{1}{2}\psi^{\ul \alpha+}$ to
simplify the notation that will follow.\footnote{A precise notation would
reserve these labels for the component projections $E^{\ul A}\vert_{\q=0}$,
but in practice it is often convenient to use the same labels for the component
expansions as well as the full superfields.}
Its determinant $\cE^{--}$ is equal in this gauge to $e^{++}$ given by
\begin{align}
e^{++} = \det
\begin{pmatrix}
e_m{}^a & \cV_m{}^{++} \\
e_\tau{}^a & \cV_\tau{}^{++}
\end{pmatrix}~.
\end{align}
This determinant is over the five-by-five component vielbein describing
both the base manifold with coordinates $x^m$ and the $\rm SU(2)$ contour
with coordinate $\tau$.

The easiest way to evaluate the component action is to rewrite $S$ as
\begin{align}
S = -\frac{1}{2\pi} \, \frac{1}{16} \int_{\cM^4 \times \cC}
	(\pa_+)^2 (\bar\pa_+)^2 (\widehat e^{++} \cL^{--})~.
\end{align}
where $\widehat e^{++}$ is the volume five-form
\begin{align}
\widehat e^{++}
	= \rd x^0 \wedge \rd x^1 \wedge \rd x^2 \wedge \rd x^3 \wedge \rd \tau \, e^{++}
	= \frac{1}{4!} \eps_{abcd}\, e^a \wedge e^b \wedge e^c \wedge e^d \wedge \cV^{++}~.
\end{align}
In order to evaluate successive spinor derivatives of $\widehat e^{++}$, one must
work out the rules for spinor differentiation of the one-forms $e^a$ and $\cV^{++}$
in the gauge where $\nabla_{\ul\alpha +} = \pa_{\ul\alpha+}$.
These can be derived by using the relations for the corresponding
curvatures $T^a$ and $T^{++}$. For example, from the definition
of $T^a$, one can show that
\begin{align}
T_{\nu+ M}{}^a &= - (-)^M \,T_{\nu+ \dbeta -}{}^a \, E_M{}^{\dbeta -} = \pa_{\nu+} e_M{}^a \quad \implies \quad
\pa_{\nu+} e^a = -i (\sigma^a)_{\nu\dbeta} \bar \psi^{\dbeta -}~.
\end{align}
Similar relations can be used to define the spinor derivative of any one-form.
The ones we will need are
\begin{alignat}{2}
\pa_{\alpha +} e^a &= -i (\sigma^a)_{\alpha \dbeta} \bar \psi^{\dbeta -}~, &\quad
\pa_{\alpha +} \cV^{++} &= 2 \phi_{\alpha}^+ + \frac{3i}{2} e^b (\sigma_b)_{\alpha\dbeta} \bar\chi^{\dbeta +}~, \eol
\pa_{\alpha +} \psi^{\beta -} &= -2 \delta_\alpha{}^\beta \cV^{--}~, &\quad
\pa_{\alpha +} \bar\psi^{\dbeta -} &= 0~, \eol
\pa_{\alpha +} \phi^{\beta +} &= - 2 e^c R(S)_{c\alpha+}{}^{\beta +}~, &\quad
\pa_{\alpha +} \bar\phi^{\dbeta +} &= \frac{3}{2} \psi_\alpha^- \bar \chi^{\dbeta +}
	- 2 e^c R(S)_{c\alpha+ }{}^{\dbeta +}
	+ 2 i f^b (\sigma_b)_\alpha{}^\dbeta~, \eol
\pa_{\alpha +} \cV^{--} &= 0~, & \quad
\pa_{\alpha +} f^b &= -e^c R(K)_{c \alpha+}{}^b - \frac{1}{2} \psi^{\gamma-} R(K)_{\gamma - \alpha +}{}^b~,
\end{alignat}
as well as their complex conjugates,
\begin{alignat}{2}
\bar\pa_{\dalpha +} e^a &= i (\sigma^a)_{\beta\dalpha} \psi^{\beta -}~, &\quad
\bar\pa_{\dalpha +} \cV^{++} &= 2 \phi_{\dalpha}^+ - \frac{3i}{2} e^b (\sigma_b)_{\beta\dalpha} \chi^{\beta +}~, \eol
\bar\pa_{\dalpha +} \bar\psi^{\dbeta -} &= -2 \delta_\dalpha{}^\dbeta \cV^{--}~, &\quad
\bar\pa_{\dalpha +} \psi^{\beta -} &= 0~, \eol
\bar\pa_{\dalpha +} \bar\phi^{\dbeta +} &= - 2 e^c R(S)_{c\dalpha+}{}^{\dbeta+}~, &\quad
\bar\pa_{\dalpha +} \phi^{\beta +} &= \frac{3}{2} \psi_\dalpha^- \chi^{\beta +}
	- 2 e^c R(S)_{c\dalpha+ }{}^{\beta+}
	- 2 i f^b (\sigma_b)_\dalpha{}^\beta~, \eol
\bar\pa_{\dalpha +} \cV^{--} &= 0~, & \quad
\bar\pa_{\dalpha +} f^b &= -e^c R(K)_{c \dalpha+}{}^b - \frac{1}{2} \bar\psi^{\dgamma-} R(K)_{\dgamma - \dalpha +}{}^b~.
\end{alignat}
As with the other connections, we label the superfield connections $F^A$ by their
component names, $F^A = (f^a, \frac{1}{2} \phi^{\ul\alpha\pm})$.

Applying these rules and using the explicit expressions for the curvatures $R(K)$
and $R(S)$ where needed, one can derive all the spinor derivatives of $\widehat e^{++}$.
Suppressing the explicit $\wedge$ symbol from now on, we find
\begin{align}\label{eq:dVolForm}
\pa^\alpha_+ \,\widehat e^{++}
	&= \eps_{abcd} \,e^a  e^b  e^c 
	\Big(
	\frac{i}{6} \bar \psi_\dbeta^-  \cV^{++} \,(\bar \sigma^d)^{\dbeta \alpha}
	+ \frac{1}{12} e^d  \phi^{\alpha +}
	\Big)~, \eol
\bar\pa_{\dalpha +} \,\widehat e^{++}
	&= \eps_{abcd} \,e^a  e^b  e^c 
	\Big(
	\frac{i}{6} \psi^\beta{}^-  \cV^{++} \,(\sigma^d)_{\beta \dalpha}
	+ \frac{1}{12} e^d  \phi_\dalpha^+
	\Big)~.
\end{align}
The second spinor derivatives are
\begin{align}
(\pa_+)^2 \widehat e^{++}
	&=
	- 2i \,e^a  e^b  \bar\psi^{\dalpha -}  \bar\psi^{\dbeta -}
		 \cV^{++} \, (\bsigma_{ab})_{\dalpha\dbeta}
	+ \frac{2i}{3} \,e^a  e^b  e^c  \bar\psi^\dbeta{}^- 
		\phi^{\beta +} \, \eps_{abcd} \,(\sigma^d)_{\beta \dbeta}
	\eol & \qquad
	- \frac{1}{4} e^a  e^b  e^c  e^d  \bar\psi_\dbeta^{-}\,
		\bar\chi^{\dbeta +} \eps_{abcd}~, \eol
(\bar \pa_+)^2 \widehat e^{++}
	&=
	2i \,e^a  e^b  \psi^{\alpha -}  \psi^{\beta -}
		 \cV^{++} \, (\sigma_{ab})_{\alpha\beta}
	+ \frac{2i}{3} \,e^a  e^b  e^c  \psi^\beta{}^- 
		\bar \phi^{\dbeta +} \, \eps_{abcd} \,(\sigma^d)_{\beta \dbeta}
	\eol & \qquad	
	- \frac{1}{4} e^a  e^b  e^c  e^d  \psi^{\beta -}\,
		\chi_\beta^+ \eps_{abcd}~, \eol
\pa_{\alpha +} \bar\pa_{\dalpha +} \widehat e^{++}
	&= -\frac{i}{3} \eps_{abcd} e^a  e^b  e^c 
		\Big(\cV^{--}  \cV^{++} (\sigma^d)_{\alpha \dalpha}
		+ \psi^{\beta -}  \phi_\alpha^+ (\sigma^d)_{\beta \dalpha}
		+ \bar\psi^{\dbeta -}  \bar\phi_\dalpha^+ (\sigma^d)_{\alpha \dbeta}
	\Big)
	\eol & \quad
	+ \frac{i}{6} e^a  e^b  e^c  e^d  f^f (\sigma_f)_{\alpha \dalpha} \eps_{abcd}
	+ \frac{1}{2} e^a  e^b  \psi^{\beta -}  \bar \psi^{\dbeta -}  \cV^{++}
		(\sigma^c)_{\alpha \dbeta} (\sigma^d)_{\beta \dalpha} \eps_{abcd}~.
\end{align}
The terms with three spinor derivatives are
\begin{align}
\pa^\alpha_{+} (\bar\pa_+)^2 \,\widehat e^{++}
	&= \eps_{abcd} \,e^a  e^b  e^c  e^d  \Big[
	\frac{1}{2} \cV^{--}\, \chi^{\alpha +}
	+ \frac{1}{4} \psi^{\alpha -}\, D
	-\frac{1}{6} \psi^{\beta -} \Big(T_{de}{}^{0} - R(\bbD)_{de}\Big)
		(\sigma^{de})_\beta{}^\alpha
	\Big]
	\eol & \quad
	- \eps_{abcd}\, e^a  e^b  e^c  \Big[
		\frac{4i}{3} \bar \phi_\dbeta^{+}  \cV^{--}\, (\bsigma^d)^{\dbeta\alpha}
		- \frac{4}{3} \psi^{\beta -}  f^e \,(\sigma^d \bsigma_e)_\beta{}^\alpha
		+ i \bar \psi_\dbeta^-  \psi^\beta{}^- \, \chi_\beta^+
			\, (\bsigma^d)^{\dbeta \alpha}
	\Big]
	\eol & \quad
	+ e^a  e^b  \Big[
	8 i \,\psi^\beta{}^-  \cV^{--}  \cV^{++}\, (\sigma_{ab})_\beta{}^\alpha
	- 2 \,\bar \psi_{\dalpha}^-  \psi^\beta{}^+  \bar\phi^{\dbeta +}
		\eps_{abcd} (\bar\sigma^c)^{\dalpha \alpha} (\sigma^d)_{\beta \dbeta}
	\eol & \qquad \qquad
	+ 4 i \,\psi^\beta{}^-  \psi^\gamma{}^-  \phi^\alpha{}^+\,
		(\sigma_{ab})_{\beta \gamma}
	\Big]
	+ 4 \,e^a  \bar \psi^\dbeta{}^-  \psi^\beta{}^- 
		\psi^\alpha{}^-  \cV^{++}\, (\sigma_a)_{\beta \dbeta}~, \eol
\bar\pa_{\dalpha+} (\pa_+)^2 \,\widehat e^{++}
	&= \eps_{abcd} \,e^a  e^b  e^c  e^d  \Big[
	- \frac{1}{2} \cV^{--}\, \chi_\dalpha^{+}
	- \frac{1}{4} \bar\psi_\dalpha^{-}\, D
	+ \frac{1}{6} \bar\psi_{\dbeta}^- \Big(T_{de}{}^{0} - R(\bbD)_{de}\Big)
		(\bsigma^{de})^\dbeta{}_\dalpha
	\Big]
	\eol & \quad
	+ \eps_{abcd}\, e^a  e^b  e^c  \Big[
		\frac{4i}{3} \phi^\beta{}^{+}  \cV^{--}\, (\sigma^d)_{\beta\dalpha}
		- \frac{4}{3} \bar\psi_\dbeta^{-}  f^e \,(\bsigma^d \sigma_e)^\dbeta{}_\dalpha
		- i \psi^{\beta -}  \bar\psi_\dbeta^- \, \bar\chi^{\dbeta +}
			\, (\sigma^d)_{\beta \dalpha}
	\Big]
	\eol & \quad
	+ e^a  e^b  \Big[
	8 i \,\bar\psi_\dbeta^-  \cV^{--}  \cV^{++}\, (\bsigma_{ab})^\dbeta{}_\dalpha
	- 2 \,\psi^{\beta-}  \bar\psi^{\dbeta +}  \phi^{\gamma +}
		\eps_{abcd} (\bar\sigma^c)_{\beta \dalpha} (\sigma^d)_{\gamma \dbeta}
	\eol & \qquad \qquad
	- 4 i \,\bar\psi^\dbeta{}^-  \bar\psi^\dgamma{}^-  \bar\phi_\dalpha^+\,
		(\bsigma_{ab})_{\dbeta \dgamma}
	\Big]
	- 4 \,e^a  \bar \psi^\dbeta{}^-  \psi^\beta{}^- 
		\bar\psi_\dalpha^-  \cV^{++}\, (\sigma_a)_{\beta \dbeta}~.
\end{align}
The highest term involves four spinor derivatives:
\begin{align}\label{eq:d4VolForm}
(\pa_+)^2 (\bar\pa_+)^2 \widehat e^{++}
	&= 2D \, \eps_{abcd} \,e^a  e^b  e^c  e^d  \cV^{--}
	\eol & \quad
	+ e^a  e^b  e^c  \Big[
		8 \,\psi^\beta{}^-  \bar \psi^\dbeta{}^- \, (\sigma_c)_{\beta \dbeta}
		\Big(T_{ab}{}^{0} - R(\bbD)_{ab} \Big)
		+ \frac{32}{3} f^d  \cV^{--} \eps_{abcd}
		\eol & \qquad
		+ 4 i \,\bar \psi_\dalpha^-  \cV^{--} \, \chi_\alpha^{+}
			(\bsigma^d)^{\dalpha \alpha} \eps_{abcd}
		- 4i \,\psi^{\alpha -}  \cV^{--}\, \bar \chi^{\dalpha +}
			(\sigma^d)_{\alpha \dalpha}\, \eps_{abcd}
	\Big]
	\eol & \quad
	+ e^a  e^b \Big[
		32 i \,\bar \psi^{\dbeta -}  \bar \phi^{\dgamma+}  \cV^{--}
			\, (\bsigma_{ab})_{\dbeta \dgamma}
		+ 32 i \,\psi^{\beta -}  \phi^{\gamma+}  \cV^{--}
			\, (\sigma_{ab})_{\beta \gamma}
		\eol & \qquad
		+ 12 i \,\bar \psi^{\dbeta -}  \bar \psi^{\dgamma -} 
			\psi^{\alpha -}\, \chi_\alpha^+\, (\bsigma_{ab})_{\dbeta \dgamma}
		- 12 i \,\psi^{\beta -}  \psi^{\gamma -} 
			\bar\psi_\dalpha^{-}\, \bar\chi^{\dalpha +}\, (\sigma_{ab})_{\beta \gamma}
		\eol & \qquad
		+ 32 \, \psi^{\beta -}  \bar \psi^{\dbeta -}  f_a\,
			(\sigma_b)_{\beta \dbeta}
	\Big]
	\eol & \quad
	+ 16 e^a \psi^{\alpha -}  \bar \psi^{\dalpha -} 
	\Big[
		\psi^{\beta -}  \phi_\beta^{+}
		- \bar \psi_\dbeta^-  \bar \phi^{\dbeta +}
		+ 3\, \cV^{--} \cV^{++}\, 
	\Big] (\sigma_a)_{\alpha \dalpha}~.
\end{align}
In the above expressions, we note that the curvatures
$T_{ab}{}^0$ and $R(\bbD)_{ab}$ were actually found by spinor differentiation
of covariant fields such as $\chi^{\alpha +}$ and $\bar\chi^{\dalpha +}$
that appeared at lower dimensions, using the explicit expressions for
$T^0$ and $R(\bbD)$ in terms of $W_{\alpha\beta}$ and $\bar W_{\dalpha\dbeta}$.

The component action can then be written as
\begin{align}
S = - \frac{1}{2\pi} \oint_\cC \rd \tau \int \rd^4x\, \rd^4\q^+ \cE^{--} \mathscr{L}^{++}
	= - \frac{1}{2\pi} \int_{\cM^4 \times \cC} \cJ
\end{align}
where $\cJ$ is a five-form given by \eqref{eq:defJ}.
The full expression for the five-form $\cJ$ is quite complicated in the general
component gauge.
In practice, one should always analyze component actions in the central gauge. Recall in this gauge
\begin{align}
e^a = \rd x^m\, e_m{}^a~, \qquad \psi{}^{\alpha +} = \rd x^m \, \psi_m{}^{\alpha +}~, \quad\text{etc.}
\end{align}
while only the connections $\cV^{\pm\pm}$ and $\cV^{0}$ possess a $\rd \tau$ component,
\begin{align}
\cV^{\pm\pm} = \rd x^m\, \cV_m{}^{\pm\pm} + \rd \tau \, \cV_\tau{}^{\pm\pm}~, \quad
\cV^{0} = \rd x^m\, \cV_m{}^{0} + \rd \tau \, \cV_\tau{}^{0}~.
\end{align}
Because the integral selects out only the component $\cJ$ involving
$\rd x^0  \rd x^1 \rd x^2 \rd x^3 \rd \tau$, only
those components of $\cJ$ involving at least one of $\cV^{\pm\pm}$
and $\cV^0$ can contribute. Now one can
make a dramatic simplification by going to the central gauge:
\begingroup
\allowdisplaybreaks
\begin{align*}
[\pa^\alpha_+ \,\widehat e^{++}]_{\rm CG}
	&\sim \frac{i}{6} \eps_{abcd} \,e^a  e^b  e^c  \bar \psi_\dbeta^-  \cV^{++} \,(\bar \sigma^d)^{\dbeta \alpha}~, \\
[\bar\pa_{\dalpha +} \,\widehat e^{++}]_{\rm CG}
	&\sim \frac{i}{6} \eps_{abcd} \,e^a  e^b  e^c 
		\psi^\beta{}^-  \cV^{++} \,(\sigma^d)_{\beta \dalpha}~, \\
[(\pa_+)^2 \widehat e^{++}]_{\rm CG}
	&\sim
	- 2i \,e^a  e^b  \bar\psi^{\dalpha -}  \bar\psi^{\dbeta -}
		 \cV^{++} \, (\bsigma_{ab})_{\dalpha\dbeta}~, \\
[(\bar \pa_+)^2 \widehat e^{++}]_{\rm CG}
	&\sim
	2i \,e^a  e^b  \psi^{\alpha -}  \psi^{\beta -}
		 \cV^{++} \, (\sigma_{ab})_{\alpha\beta}~, \\
[\pa_{\alpha +} \bar\pa_{\dalpha +} \widehat e^{++}]_{\rm CG}
	&\sim e^a  e^b  \Big(
	\frac{1}{2} \psi^{\beta -}  \bar \psi^{\dbeta -}  \cV^{++}
		(\sigma^c)_{\alpha \dbeta} (\sigma^d)_{\beta \dalpha}
	-\frac{i}{3} e^c 
		\cV^{--}  \cV^{++} \, (\sigma^d)_{\alpha \dalpha}
	\Big)\eps_{abcd}~, \\
[\pa^\alpha_{+} (\bar\pa_+)^2 \,\widehat e^{++}]_{\rm CG}
	&\sim \frac{1}{2} \eps_{abcd} \,e^a  e^b  e^c  e^d  \cV^{--}\, \chi^{\alpha +}
	- \frac{4i}{3} \eps_{abcd}\, e^a  e^b  e^c
		\bar \phi_\dbeta^{+}  \cV^{--}\, (\bsigma^d)^{\dbeta\alpha}
	\eol & \quad
	+ 8 i e^a  e^b 
		\psi^\beta{}^-  \cV^{--}  \cV^{++}\, (\sigma_{ab})_\beta{}^\alpha
	+ 4 \,e^a  \bar \psi^\dbeta{}^-  \psi^\beta{}^- 
		\psi^\alpha{}^-  \cV^{++}\, (\sigma_a)_{\beta \dbeta}~, \\
[\bar\pa_{\dalpha+} (\pa_+)^2 \,\widehat e^{++}]_{\rm CG}
	&\sim - \frac{1}{2} \eps_{abcd} \,e^a  e^b  e^c  e^d  
	\cV^{--}\, \chi_\dalpha^{+}
	+ \frac{4i}{3} \eps_{abcd}\, e^a  e^b  e^c 
		\phi^\beta{}^{+}  \cV^{--}\, (\sigma^d)_{\beta\dalpha}
	\eol & \quad
	+ 8 i e^a  e^b  \bar\psi_\dbeta^-  \cV^{--}  \cV^{++}\, (\bsigma_{ab})^\dbeta{}_\dalpha
	- 4 \,e^a  \bar \psi^\dbeta{}^-  \psi^\beta{}^- 
		\bar\psi_\dalpha^-  \cV^{++}\, (\sigma_a)_{\beta \dbeta}~, \\
[(\pa_+)^2 (\bar\pa_+)^2 \widehat e^{++}]_{\rm CG}
	&\sim e^a  e^b  e^c  \Big(2\, e^d \cV^{--} D
		+ 4 i \bar \psi_\dalpha^-  \cV^{--} \, \chi_\alpha^{+}
			(\bsigma^d)^{\dalpha \alpha}
		\eol & \qquad \qquad
		- 4i \psi^{\alpha -}  \cV^{--}\, \bar \chi^{\dalpha +}
			(\sigma^d)_{\alpha \dalpha}\,\Big)\eps_{abcd} 
	\eol & \quad
	+ \frac{32}{3} e^a  e^b  e^c  f^d  \cV^{--} \eps_{abcd}
	+ 48\, e^a 
		 \psi^{\alpha -}  \bar\psi^{\dalpha -}  \cV^{--}
			 \cV^{++}\, (\sigma_a)_{\alpha \dalpha}
	\eol & \quad
	+ 32 i e^a  e^b \,\bar \psi^{\dbeta -}  \bar \phi^{\dgamma+}  \cV^{--}
		\, (\bsigma_{ab})_{\dbeta \dgamma}
	+ 32 i e^a  e^b \,\psi^{\beta -}  \phi^{\gamma+}  \cV^{--}
		\, (\sigma_{ab})_{\beta \gamma}~.
\end{align*}
\endgroup
Converting the five-form into its corresponding integral density gives
\begin{align}
\int_{\cM^4 \times \cC} \cJ = \oint_\cC \rd \tau \int \rd^4x\, e\, \Big( \cV_\tau^{++} \cL^{--} - \cV_\tau^{--} \cL^{++} \Big)
\end{align}
where
\begin{align}
\cL^{--} &= \frac{1}{16} (\nabla^-)^2 (\bar\nabla^-)^2 \cL^{++}
	- \frac{i}{8} (\bar\psi_m^- \bsigma^m)^\alpha \nabla_\alpha^- (\bar\nabla^-)^2 \cL^{++}
	- \frac{i}{8} (\psi_m^- \sigma^m)_\dalpha \bar\nabla^{\dalpha -} (\nabla^-)^2 \cL^{++}
	\eol & \quad
	+ \frac{1}{4} \Big(
	(\psi_n^- \sigma^{nm})^\alpha \bar\psi_m{}^\dalpha{}^-
	+ \psi_n{}^\alpha{}^-  (\bsigma^{nm}\bar\psi_m^-)^\dalpha
	- i \cV_m^{--} \sigma^m_{\alpha \dalpha} \Big) [\nabla_\alpha^{-}, \bar\nabla_\dalpha^{-}]  \cL^{++}
	\eol & \quad
	+ \frac{1}{4} (\psi_m^- \sigma^{mn} \psi_n^-) (\nabla^-)^2 \cL^{++}
	+ \frac{1}{4} (\bar\psi_m^- \bsigma^{mn} \bar\psi_n^-) (\bar \nabla^-)^2 \cL^{++}
	\eol & \quad
	- \Big(
	\frac{1}{2} \eps^{mnpq} (\psi_m^- \sigma_n \bar\psi_p^-) \psi_q^{\alpha -}
	- 2 (\psi_m^- \sigma^{mn})^\alpha \cV_n^{--} \Big) \nabla_\alpha^- \cL^{++}
	\eol & \quad
	+ \Big(
	\frac{1}{2} \eps^{mnpq} (\bar\psi_m^- \bsigma_n \psi_p^-) \bar\psi_{q\dalpha}^{-}
	- 2 (\bar\psi_m^- \bsigma^{mn})_\dalpha \cV_n^{--} \Big) \bar\nabla^{\dalpha -} \cL^{++}
	\eol & \quad
	+ 3 \eps^{mnpq} (\psi_m^- \sigma_n \bar\psi_p^-) \cV_q^{--} \cL^{++}
\end{align}
and
\begin{align}
\cL^{++}
	&= -\Big[3 D 
	+ \frac{3i}{2} (\bar \psi_m^- \bsigma^m \chi^+) 
	- \frac{3i}{2} (\psi_m^{-} \sigma^m \bar\chi^+) 
	+ 4 f_a{}^a 
	\eol & \qquad \qquad
	- 4 (\bar \psi_m^- \bsigma^{mn} \bar\phi_n^+) 
	+ 4 (\psi_m^{-} \sigma^{mn}\phi_n^{+}) 
	- 3\, \eps^{mnpq}
		 (\psi_m^{-} \sigma_n \bar\psi_p^{-}) \cV_q^{++}
	\Big] \cL^{++}
	\eol & \quad
	+ \Big[\frac{3}{2} \chi^{\alpha +}
	- i (\bar \phi_m^{+} \bsigma^m)^\alpha 
	+ 2 (\psi_m^- \sigma^{mn})^\alpha \cV_n^{++} \Big] \nabla_\alpha^- \cL^{++}
	\eol & \quad
	- \Big[\frac{3}{2} \chi_\dalpha^{+}
	- i (\phi_m^{+} \sigma^m)_\dalpha 
	+ 2 (\bar\psi_m^- \bsigma^{mn})_\dalpha \cV_n^{++}\Big] \bar\nabla^{\dalpha -} \cL^{++}
	\eol & \quad
	- \frac{i}{4} \cV_m^{++} (\bsigma^m)^{\dalpha \alpha} [\nabla_\alpha^-, \bar\nabla_\dalpha^-] \cL^{++}~.
\end{align}

\end{document}